\documentclass[12pt,a4paper,amsmath,amssymb,aps]{article}

\usepackage{graphicx}
\usepackage{setspace}
\usepackage{fancyhdr}
\usepackage{multirow}
\usepackage{color}
\usepackage{dcolumn}
\usepackage{bm}
\usepackage{subfigure}
\usepackage{booktabs}
\usepackage{rotating}
\usepackage{longtable}

\newcommand{\bra}[1] {\langle #1 |}
\newcommand{\ket}[1] {| #1 \rangle}

\newcommand{\me}[3]{\bra{#1} #2 \ket{#3}}

\begin{document}

\title{The nitrogen-vacancy colour centre in diamond}

\author{Marcus W. Doherty,$^{a,b}$ Neil B. Manson,$^b$ Paul Delaney,$^c$ \\
Fedor Jelezko,$^{d}$ J\"org Wrachtrup$^e$ and Lloyd C.L. Hollenberg$^a$\footnote{Corresponding author. E-mail address: lloydch@unimelb.edu.au. Tel.: +61 3 8344 4210. Fax.: +61 3 9347 4783}\\
\\
\small$^a$ School of Physics, University of Melbourne, VIC 3010, Australia \\
\small$^b$ Laser Physics Centre, Research School of Physics and Engineering,\\
\small Australian National University, ACT 0200, Australia\\
\small $^c$ School of Mathematics and Physics, Queen's University Belfast,\\
\small  Northern Ireland BT7 1NN, United Kingdom \\
\small$^d$ Institut f$\mathrm{\ddot{u}}$r Quantenoptik, Universit$\mathrm{\ddot{a}}$t Ulm, Ulm D-89073, Germany \\
\small$^e$ 3$^{\mathrm{rd}}$ Institute of Physics and Research Center SCOPE,\\
\small University Stuttgart, Pfaffenwaldring 57, D-70550 Stuttgart, Germany}

\date{\today}
\maketitle

\begin{abstract}
The nitrogen-vacancy (NV) colour centre in diamond is an important physical system for emergent quantum technologies, including quantum metrology, information processing and communications, as well as for various nanotechnologies, such as biological and sub-diffraction limit imaging, and for tests of entanglement in quantum mechanics. Given this array of existing and potential applications and the almost 50 years of NV research, one would expect that the physics of the centre is well understood, however, the study of the NV centre has proved challenging, with many early assertions now believed false and many remaining issues yet to be resolved. This review represents the first time that the key empirical and \textit{ab initio} results have been extracted from the extensive NV literature and assembled into one consistent picture of the current understanding of the centre. As a result, the key unresolved issues concerning the NV centre are identified and the possible avenues for their resolution are examined. 
\end{abstract}

\section{Introduction}
\label{chapter:introduction}

The detection of single negatively charged nitrogen-vacancy (NV$^-$) colour centres in 1997 \cite{gruber97} marks a critical point in the evolution of diamond based quantum technologies. Although observations of ensembles of NV$^-$ centres were routinely performed prior to 1997, the detection of single centres soon enabled demonstrations of photostable single photon generation \cite{drab99,brouri00,kurt00}, which highlighted the NV$^-$ centre for implementation in quantum optical networks, as well as demonstrations of optical preparation and readout of the centre's electronic spin \cite{jelezko04,jelezko04b}, which identified the NV$^-$ centre as a possible solid state spin qubit suitable for quantum information processing and quantum metrology devices. Following these demonstrations, the growth of research into the NV$^-$ centre and the development of applications of the centre have been incredibly rapid and a number of important milestones have been reached. Although the NV centre is also found in the neutral charge state (NV$^0$), NV$^0$ has not yet been employed in any demonstrations akin to NV$^-$ and there exists only very speculative arguments \cite{gali09} concerning its suitability as a spin qubit (due to no reported observations of optical readout). Consequently, the motivations for studying NV$^0$ have thus far been restricted to obtaining insight into NV$^-$.

The NV$^-$ centre has been employed in milestone room temperature demonstrations of quantum registers built upon the NV$^-$ electronic spin and proximal N and $^{13}$C nuclear spins \cite{dutt07,neumann08}. At low temperature ($<10$ K), spin-photon entanglement between the ground state spin of a single NV$^-$ centre and the polarisation of an emitted photon has also been demonstrated \cite{togan10}. In terms of scalability, demonstrations of NV$^-$-NV$^-$ spin coupling \cite{neumann10} have been performed and important steps towards photonic coupling have been made \cite{santori10}. Beyond quantum information processing, there exist important applications in room-temperature nanoscale magnetometry \cite{balasubramanian09,degen08,taylor08,balasubramanian08,maze08,maertz10,meriles10,steinert10,laroui10,hall10b,zhao11}, bio-magnetometry \cite{mcguinness11}, electrometry \cite{dolde11} and decoherence microscopy \cite{cole09,hall09,hall10} using the NV$^-$ centre. Adding to the already impressive list of metrology applications, the NV$^-$ centre has also been recently proposed as a sensitive nanoscale thermometer \cite{acosta10,toyli12}.

Since the demonstrations of the NV$^-$ centre as a solid-state spin qubit and a nanoscale multi-sensor, the centre has been used to perform experiments addressing fundamental issues in quantum and condensed matter theory. These experiments include Bell state violation \cite{waldherr11b}, magnetic resonance beyond the rotating frame approximation \cite{fuchs09}, detection of the Meissner effect \cite{bouchard11} and quantum process tomography \cite{howard06}. Recently, the NV$^-$ centre has been proposed as an ideal system to study mesoscopic quantum entanglement between an NV$^-$ spin qubit and the zero-point vibrations of a magnetic micromechanical cantilever, or alternatively, the use of a cantilever to entangle NV$^-$ spin qubits separated by mesoscopic distances \cite{wrachtrup09,rabl09}. As a photostable single photon source, the NV$^-$ centre has been utilised to demonstrate quantum cryptography \cite{beveratos02,alleaume04} and the generation of single surface plasmons on silver nanowires \cite{kolesov09}. The ability to tune the optical transition of the NV$^-$ centre using strain and electric fields has also led to the centre being proposed for tests of cavity quantum electrodynamics \cite{santori10,yang10,buckley10} and photonic devices such as Q-switching photonic modules \cite{greentree06,devitt07}.

In the almost 5 decades of NV research, many properties of both charge states have been established. However, the understanding of these properties has not been without contention and significant revision and there remain a number of properties that are not well understood. The notable past issues concerning the NV$^-$ centre have been the identification of its charge state, the spin multiplicity of the ground electronic state as well as the intermediate electronic states involved in the non-radiative decay from the optically excited electronic state to the ground electronic state that enables the optical preparation and readout of the centre's electronic spin. The ongoing issues concerning the NV$^-$ centre include the detail of the aforementioned non-radiative decay, anomalous aspects of the centre's vibronic structure, the Jahn-Teller effect in the centre's optical transition and temperature variations of the centre's properties. Other important aspects of the NV centre that are not well understood are the factors that govern the relative charge state concentrations, the mechanism of photoconversion between the charge states and the associated spectral diffusion of the NV$^-$ optical transition. The theoretical methods that have been employed to understand the NV centre's properties have been the semi-empirical molecular model of deep level defects in semiconductors \cite{togan10,loubser77, loubser78,doherty11,manson06,reddy87,lenef96,goss97,martin99b,manson07,maze11} and \textit{ab initio} calculations \cite{gali09,mainwood94,goss96,luszczek04,pushkarchuk05,larrson08,gali08,zyubin08,lin08,hossain08,gali09b,gali09c,delaney10,delaney10b,weber10,ma10,gali11,zhang11}.
These theoretical methods have complementary strengths and weaknesses, which necessitate their combined application to address the ongoing issues.  Clearly, the improved understanding gained from the resolution of these ongoing issues will enable enhanced implementation of the NV centre in its many impressive applications.

Both the rapid expansion of the field of NV research and the need to resolve the ongoing issues motivate this detailed review of the properties of the NV centre in which we identify and define ongoing issues and discuss avenues to their resolution. In addition we aim to provide a comprehensive and self-consistent reference for those both new and old to NV research, which is free of any past interpretations that are now believed false. In the following subsections, the properties of the NV centre will be firstly introduced in order to orientate those new to the centre and to also facilitate the subsequent brief introduction to the quantum technology applications of the centre that have motivated its intense investigation. The final subsection presents the scope of this review, where the domain of the NV literature considered in this review is defined and the method of the review is discussed.

\subsection{Introduction to the NV centre}
\label{section:introductiontotheNVcentre}

\begin{figure}[hbtp]
\begin{center}
\mbox{
\subfigure[]{\includegraphics[width=0.4\columnwidth] {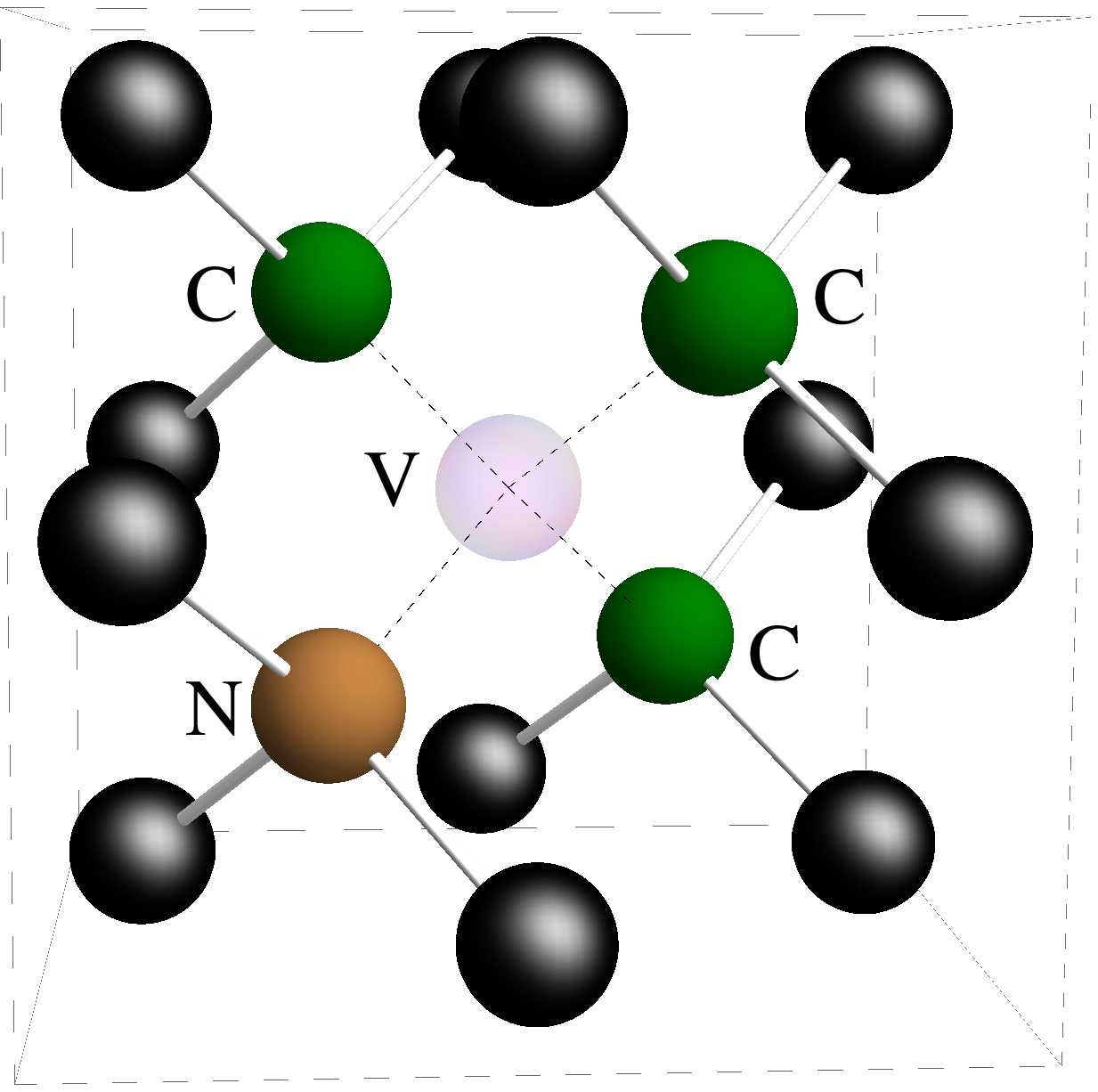}}
\subfigure[]{\includegraphics[width=0.6\columnwidth] {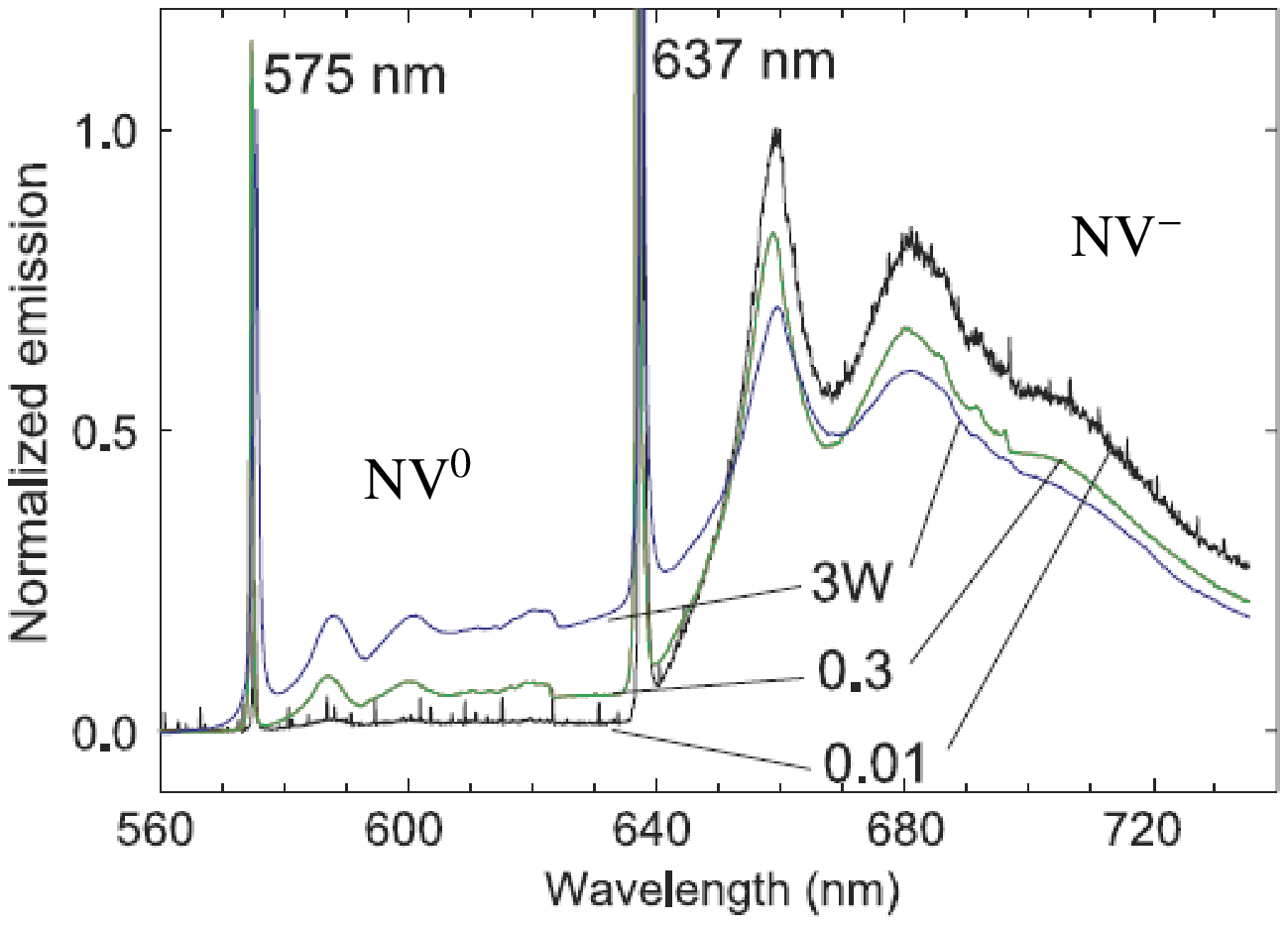}}}
\caption[The structure of the NV centre and example emission spectra of NV$^0$ and NV$^-$]{(a) Schematic of the nitrogen-vacancy centre and diamond lattice depicting the vacancy (transparent), the nearest neighbour carbon atoms to the vacancy (green), the substitutional nitrogen atom (brown), and the next-to-nearest carbon neighbours to the vacancy (black). (b) Normalised emission spectra of an ensemble of NV$^-$ and NV$^0$ centres at 10 K for different excitation (532 nm) powers as indicated \cite{manson05}. The photoconversion between the charge states is apparent in the increase of the emission of NV$^0$ relative to NV$^-$ with increasing excitation power. }
\label{fig:introtoNV}
\end{center}
\end{figure}

\begin{figure}[hbtp]
\begin{center}
\mbox{
\subfigure[]{\includegraphics[width=0.33\columnwidth] {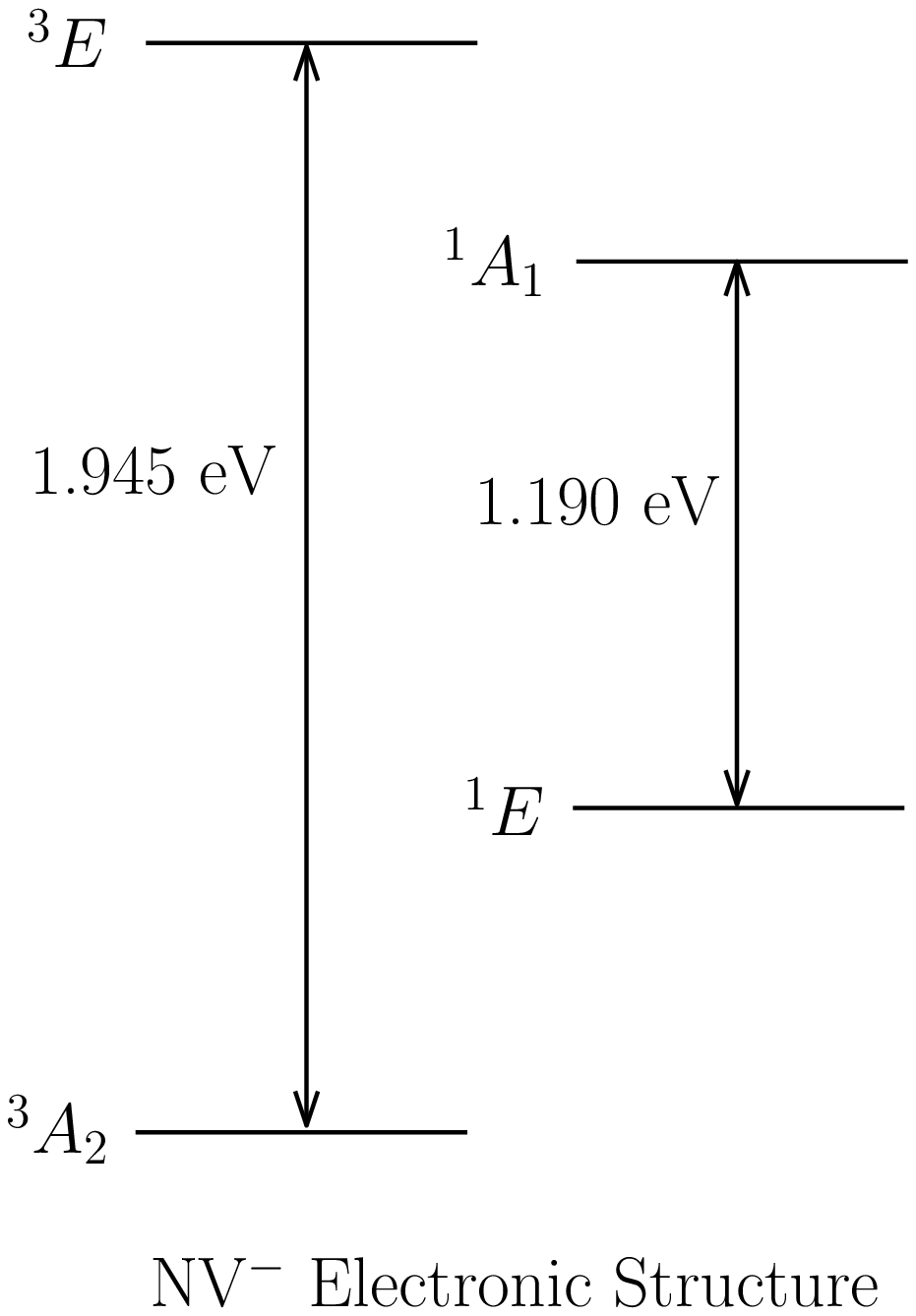}}
\subfigure[]{\includegraphics[width=0.33\columnwidth] {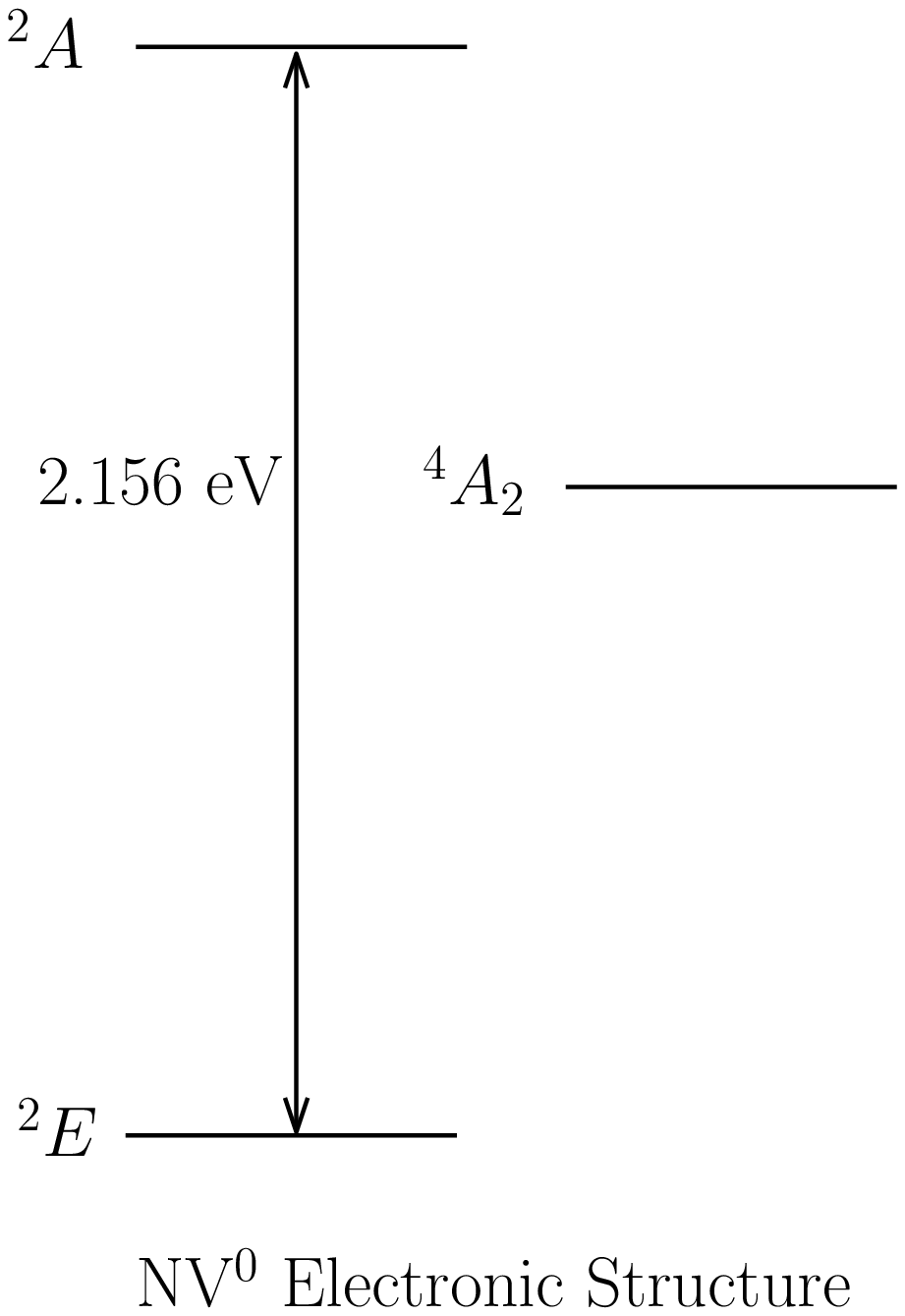}}
}
\caption[The NV defect levels within the diamond bandgap and known aspects of the electronic structures of NV$^0$ and NV$^-$.]{Schematics of (a) known aspects of the NV$^-$ electronic structure \cite{doherty11} and (b) known aspects of the NV$^0$ electronic structure \cite{felton08}. The observed optical \cite{duPreez, davies79} and infrared \cite{rogers08} zero phonon lines (ZPLs) are as indicated. Note that the relative energies of the triplet ($^3$$E$, $^3$$A_2$) and singlet states ($^1$$A_1$, $^1$$E$) of NV$^-$ and the doublet ($^2$$A$, $^2$$E$) and quartet states ($^4$$A_2$) of NV$^0$ are currently unknown.}
\label{fig:introelectronicstructure}
\end{center}
\end{figure}

The NV centre is a point defect in diamond with $C_{3v}$ symmetry consisting of a substitutional nitrogen - lattice vacancy pair orientated along the [111] crystalline direction \cite{davies76} (refer to figure \ref{fig:introtoNV}). The centre may be found as an `in grown' product of the chemical vapour deposition (CVD) diamond synthesis process \cite{vlasov00} or as a product of radiation damage and annealing \cite{duPreez} or ion implantation and annealing \cite{meijer05} in bulk and nanocrystalline diamond. The centre is known to exist in negative (NV$^-$) \cite{loubser77, loubser78} and neutral (NV$^0$) \cite{mita96} charge states. The identifying features of NV$^-$ and NV$^0$ are their optical zero phonon lines (ZPLs) at 1.945 eV (637 nm) \cite{duPreez} and 2.156 eV (575 nm) \cite{davies79}, respectively, and associated vibronic bands that extend from their ZPLs to higher/ lower energy in absorption/emission (see figure \ref{fig:introtoNV} for examples of the emission vibronic bands). An additional infrared ZPL at 1.190 eV (1042 nm) that is only observable during optical illumination (with energy $>$1.945 eV) has been associated with the NV$^-$ centre \cite{rogers08}. The sharp ZPLs and well defined vibronic bands of NV$^0$ and NV$^-$ indicate that the optical transitions occur between discrete defect levels that are deep within the diamond bandgap, such that the continua of valence or conduction band levels are not involved in the optical transitions. The NV centre is thus known as a deep-level defect in diamond \cite{loubser77, loubser78}. A schematic of the known aspects of the electronic structures of NV$^-$ and NV$^0$ are depicted in figure \ref{fig:introelectronicstructure}.

\begin{figure}[hbtp]
\begin{center}
\mbox{
\subfigure[]{\includegraphics[width=0.47\columnwidth] {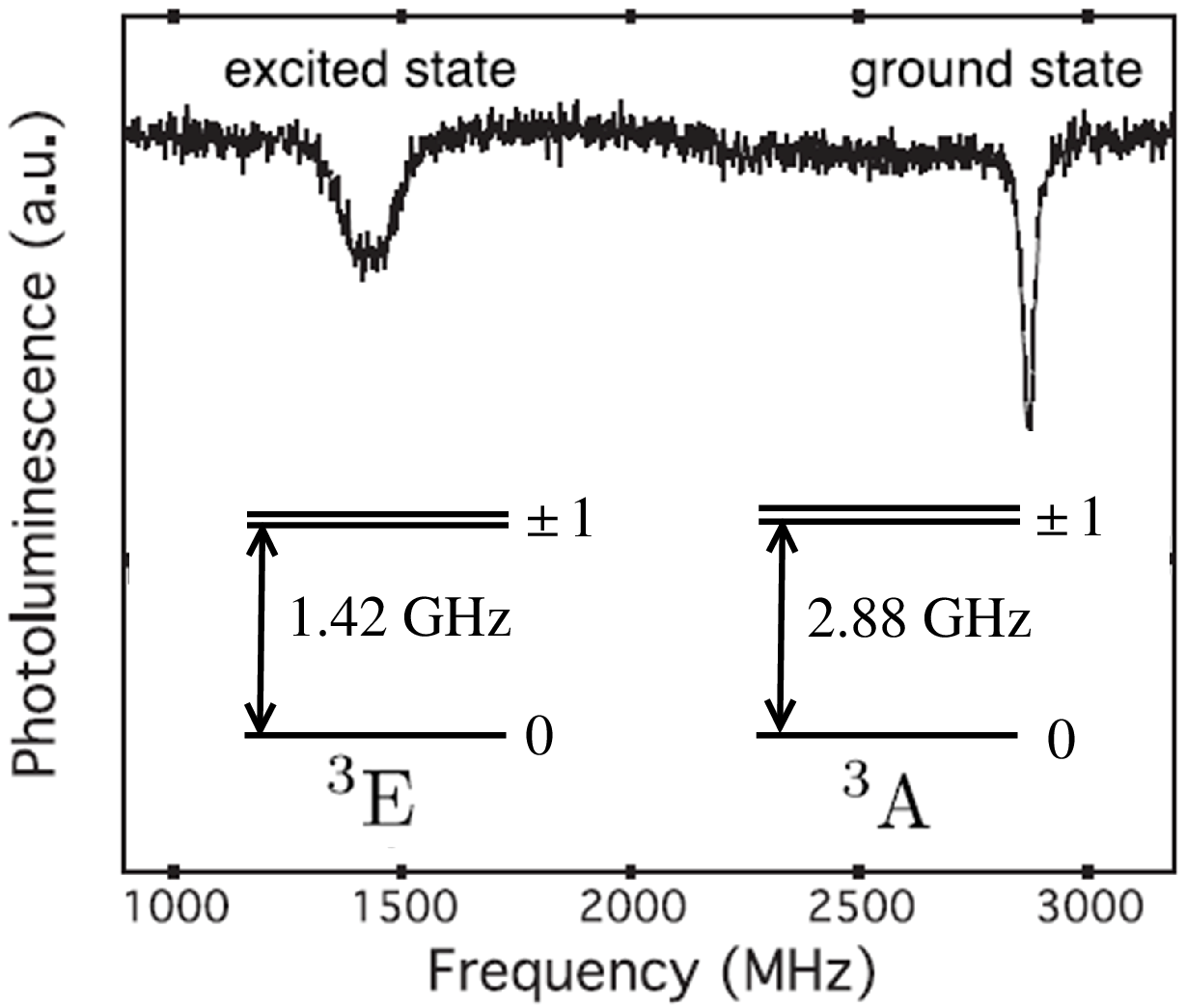}}
\subfigure[]{\includegraphics[width=0.35\columnwidth] {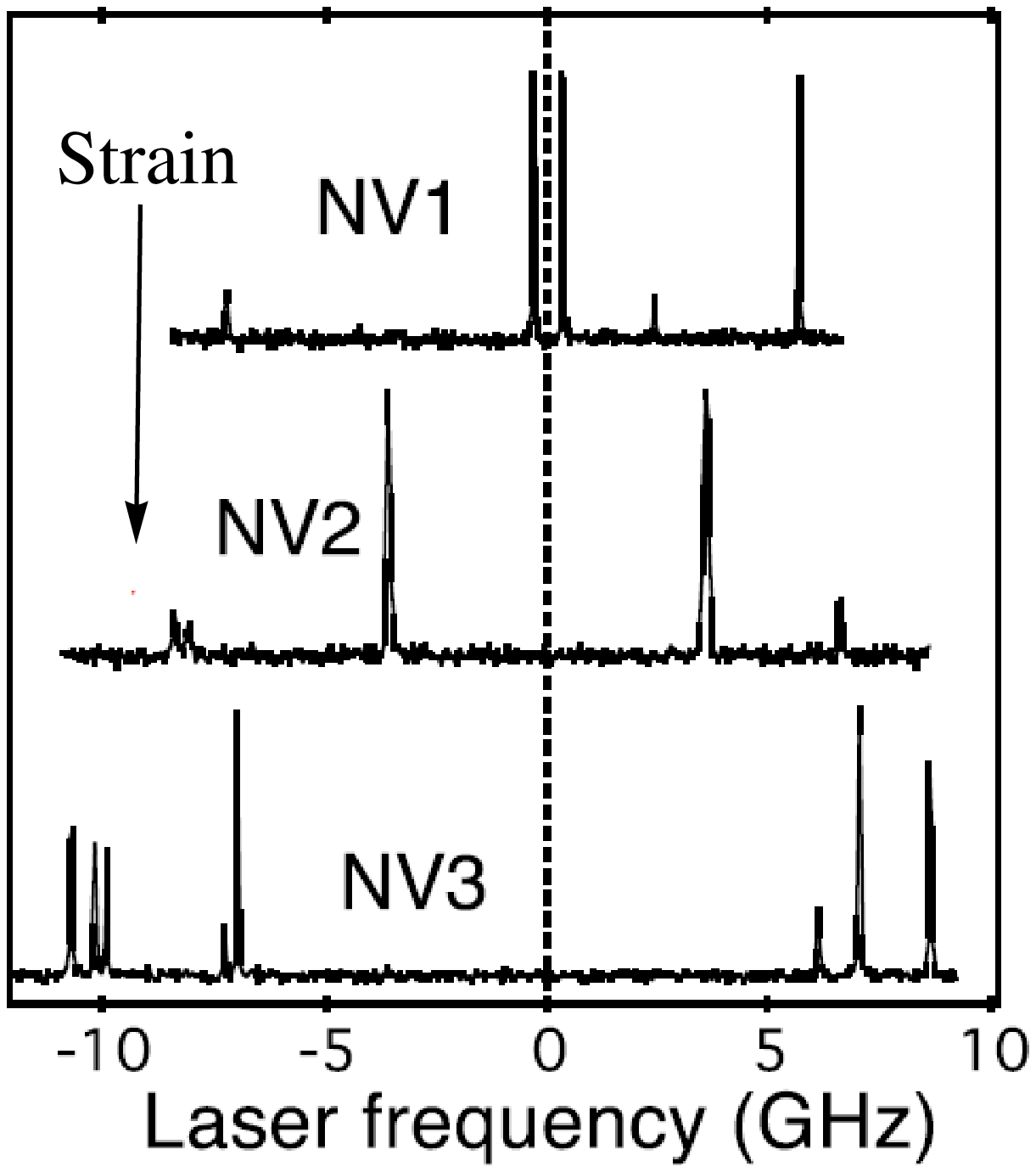}}}
\caption[Room temperature ODMR and low temperature excitation spectra of single NV$^-$ centres]{(a) Room temperature ODMR spectra \cite{neumann09} of a single NV$^-$ centre exhibiting the zero field splittings of the $m_s=0$ and $m_s=\pm1$ spin sub-levels of $^3A_2$ ($\sim$ 2.88 GHz \cite{loubser77, loubser78}) and $^3E$ ($\sim$1.42 GHz \cite{fuchs08}). (b) Low temperature (4 K) excitation spectra \cite{batalov09} of different single centres exhibiting the highly strain dependent NV$^-$ optical ZPL fine structure.}
\label{fig:introODMRspectra}
\end{center}
\end{figure}

The NV$^-$ centre may also be identified by a zero field magnetic resonance at $\sim$2.88 GHz \cite{loubser77, loubser78} (see figure \ref{fig:introODMRspectra}). This magnetic resonance occurs between the $m_s = 0$ and $m_s = \pm1$ spin sub-levels of the spin triplet ground state $^3A_2$ and can be detected by either conventional electron paramagnetic resonance (EPR) \cite{loubser77, loubser78} or optically detected magnetic resonance (ODMR) \cite{vanOort88} techniques. A further ODMR at $\sim$1.42 GHz \cite{fuchs08,neumann09} is observed at room temperature (see figure \ref{fig:introODMRspectra}) and is attributed to a zero field splitting of the spin triplet excited state $^3E$ that is analogous to that of $^3A_2$. The magnetic resonances of $^3A_2$ and $^3E$ behave as would be expected of triplet spins in a trigonal crystal field and are characterised by the approximately isotropic electron g-factors $\sim$2.0028 \cite{loubser77, loubser78} and $\sim$ 2.01 \cite{fuchs08} respectively. The magnetic resonances of $^3A_2$ \cite{vanOort90, vanOort91} and $^3E$ \cite{fuchs08, neumann09} also exhibit weak interactions with strain and electric fields.

\begin{figure}[hbtp]
\begin{center}
\mbox{
\subfigure[]{\includegraphics[width=0.32\columnwidth] {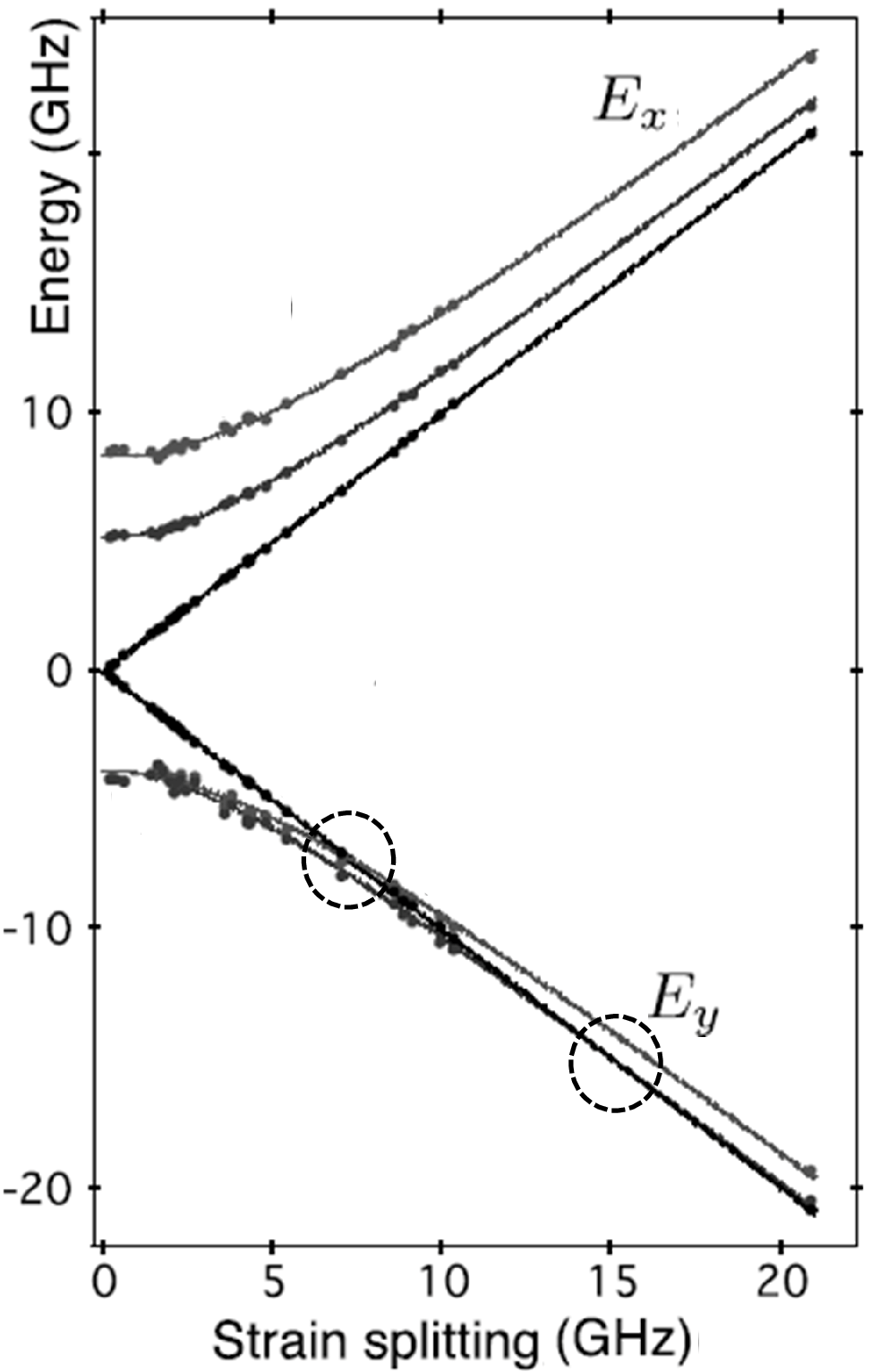}}
\subfigure[]{\includegraphics[width=0.55\columnwidth] {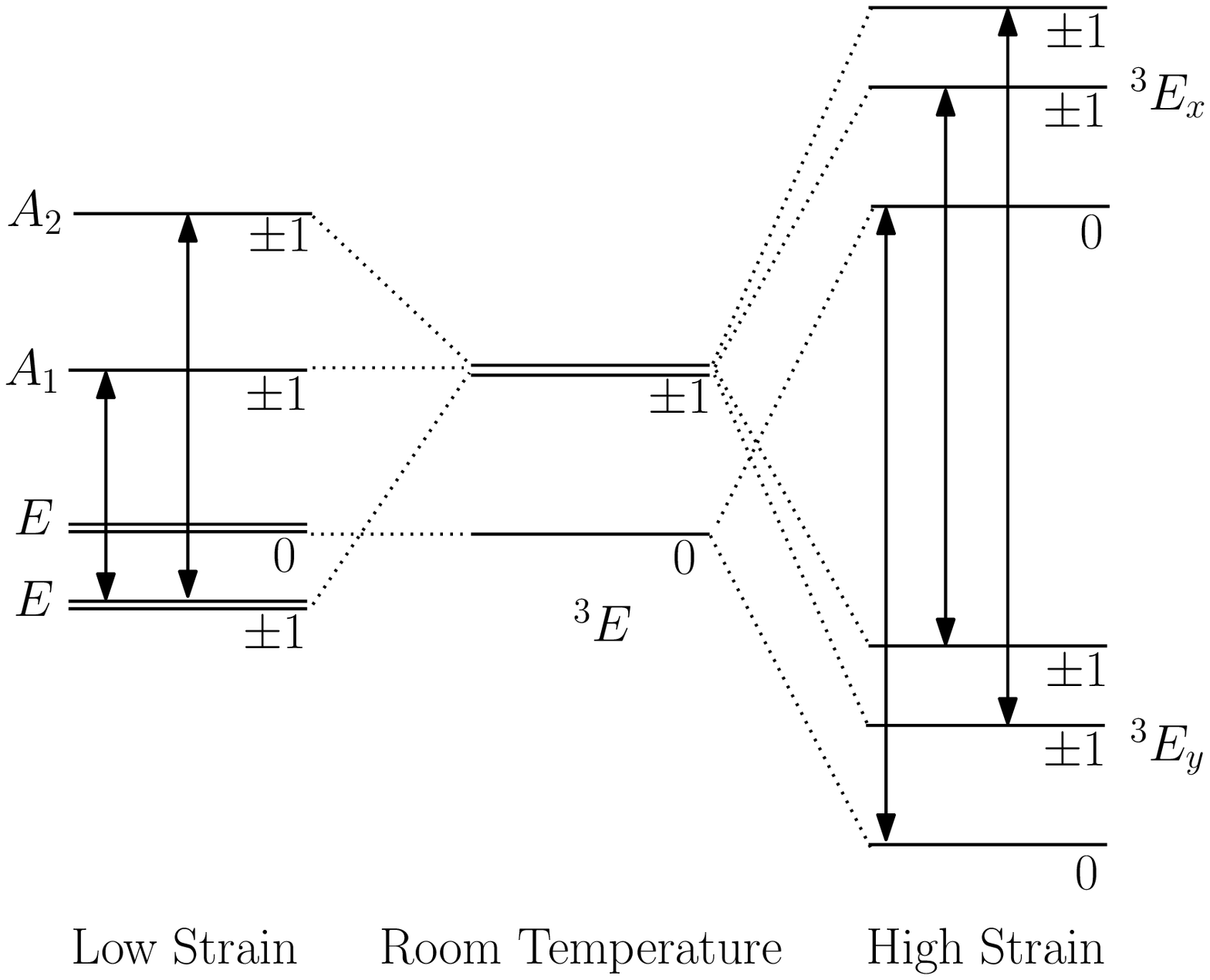}}}
\caption[The observed low temperature strain dependence of the $^3E$ fine structure and the proposed orbital averaging process]{(a) The strain dependence of the $^3E$ fine structure observed at low temperature (4 K) \cite{batalov09}. The level anti-crossings of the lower branch ($E_y$) are circled. (b) Schematic of the observed low temperature $^3E$ fine structure at low (left) and high (right) strain together with the observed room temperature fine structure (centre) \cite{manson06}. The spin-projection ($m_s = 0$, $\pm1$) of each fine structure level is as indicated. The phonon transitions that are proposed to give rise to orbital averaging are indicated by arrows \cite{doherty11}.}
\label{fig:intro3Efinestructure}
\end{center}
\end{figure}

\begin{figure}[hbtp]
\begin{center}
\mbox{
\subfigure[]{\includegraphics[width=0.45\columnwidth] {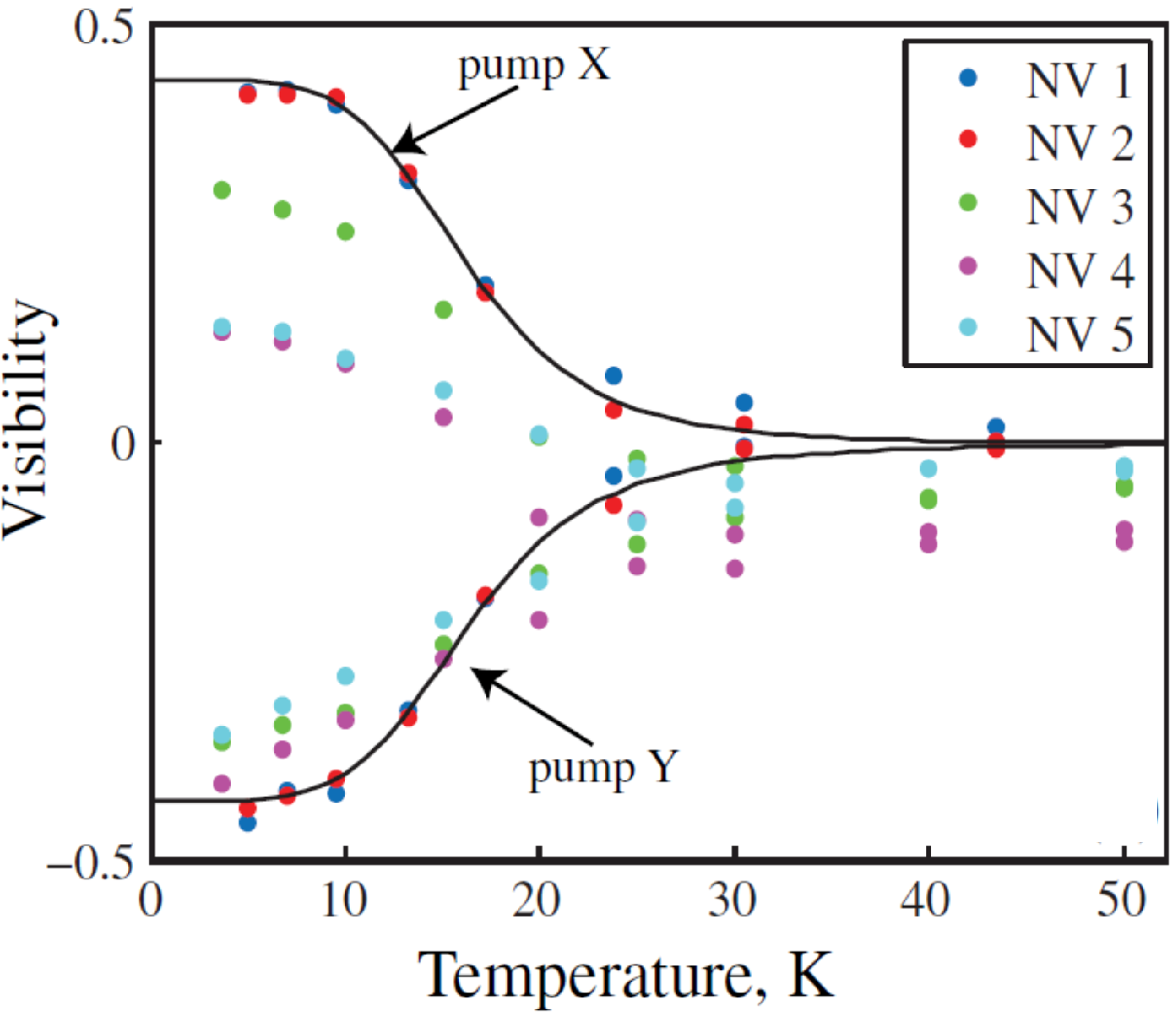}}
\subfigure[]{\includegraphics[width=0.275\columnwidth] {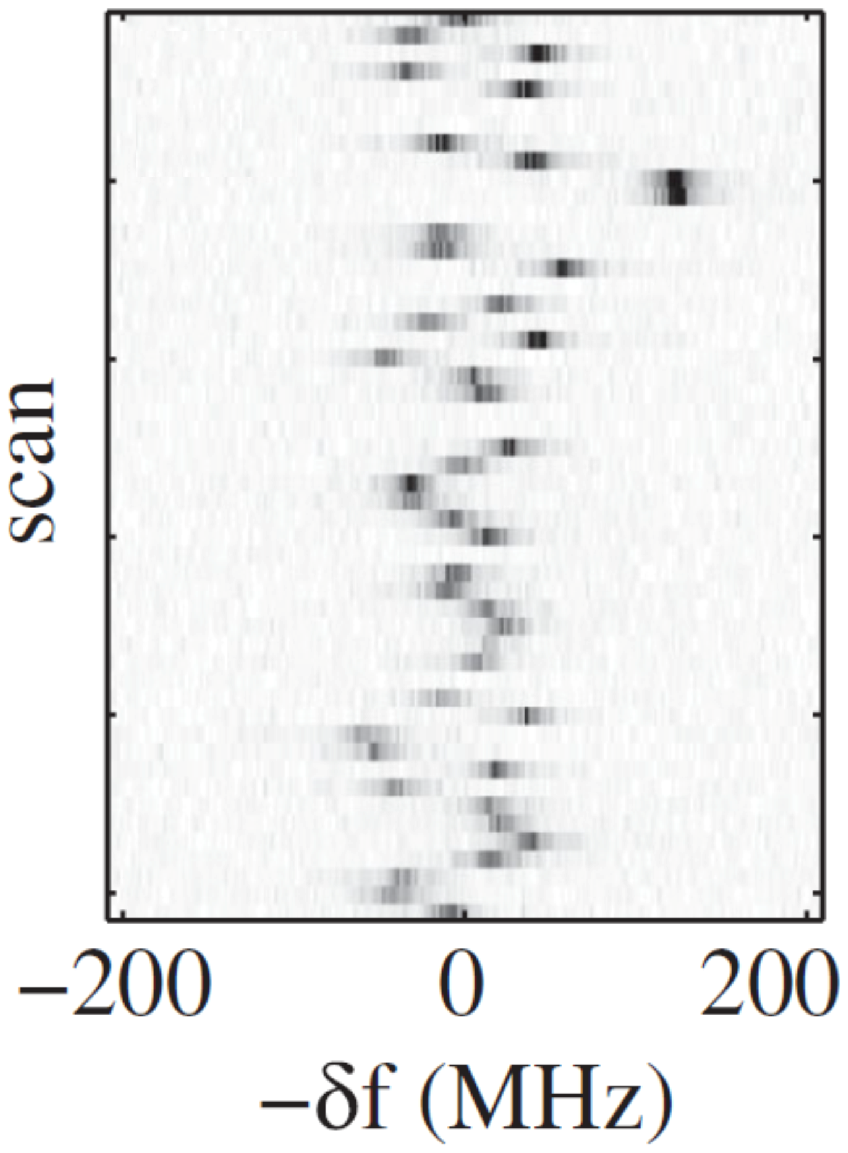}}}
\caption[Spectral diffusion of the NV$^-$ optical ZPL and its temperature dependent polarisation]{(a) Temperature dependence of the fluorescence polarisation visibility $(I_X-I_Y)/(I_X+I_Y)$ for different single NV$^-$ centres (assorted coloured dots) and different excitation polarisations (pump X/Y as indicated) \cite{fu09}. The departure from 100$\%$ visibility signifies phonon transitions between the $^3E$ fine structure states that act to depolarise the fluorescence \cite{fu09}. (b) A series of resonant excitation spectra scans of a single NV$^-$ centre at 10 K demonstrating the spectral diffusion of the optical ZPL (dark spots) in terms of detuning from 1.945 eV over time \cite{fu09}.}
\label{fig:depolarisationdiffusionintro}
\end{center}
\end{figure}

At low temperatures ($<$10 K), the excitation spectrum  of the NV$^-$ optical ZPL (refer to figure \ref{fig:introODMRspectra}) reveals a fine structure of the ZPL that is highly dependent on strain \cite{batalov09} and electric fields \cite{tamarat08}. The fine structure of the ZPL implies a fine structure of $^3E$ that is much more complicated than that suggested by the associated ODMR at room temperature. The low temperature fine structure of $^3E$ and its observed dependence on strain  \cite{batalov09} is depicted in figure \ref{fig:intro3Efinestructure} (analogous results have been obtained using electric fields \cite{tamarat08}). It is clear that the presence of strain and/or an electric field splits the $^3E$ fine structure into two branches; the lower branch undergoes two level anti-crossings (as marked in figure \ref{fig:intro3Efinestructure}) that mix the lower fine structure states to a degree dependent on the strain/electric field, whereas the upper branch does not experience any level anti-crossings and the upper fine structure states remain approximately unmixed \cite{tamarat08}. The mixing of the lower fine structure states results in optical spin-flip transitions when the lower branch is excited, whilst the unmixed upper fine structure states maintain optical spin selection rules ($\Delta S=0$, $\Delta m_s=0$) and result in optical spin-conserving transitions when the upper branch is excited \cite{tamarat08}. The collapse of the low temperature $^3E$ fine structure into a single broad ODMR at room temperature is thought to be due to phonon mediated orbital averaging over the $^3E$ fine structure levels \cite{rogers09}. This orbital averaging process is depicted in figure \ref{fig:intro3Efinestructure} and has the consequence that individual centres with different $^3E$ fine structures at low temperature (due to differing crystal strains) will have near identical $^3E$ fine structures at room temperature. In addition to their role in orbital averaging, phonons are also believed responsible for the temperature dependent depolarisation and broadening of the optical ZPL (see figure \ref{fig:depolarisationdiffusionintro}) \cite{fu09}. A thorough understanding of the phonon processes in $^3E$ is yet to be obtained.

\begin{figure}[hbtp]
\begin{center}
\mbox{
\subfigure[]{\includegraphics[width=0.4\columnwidth] {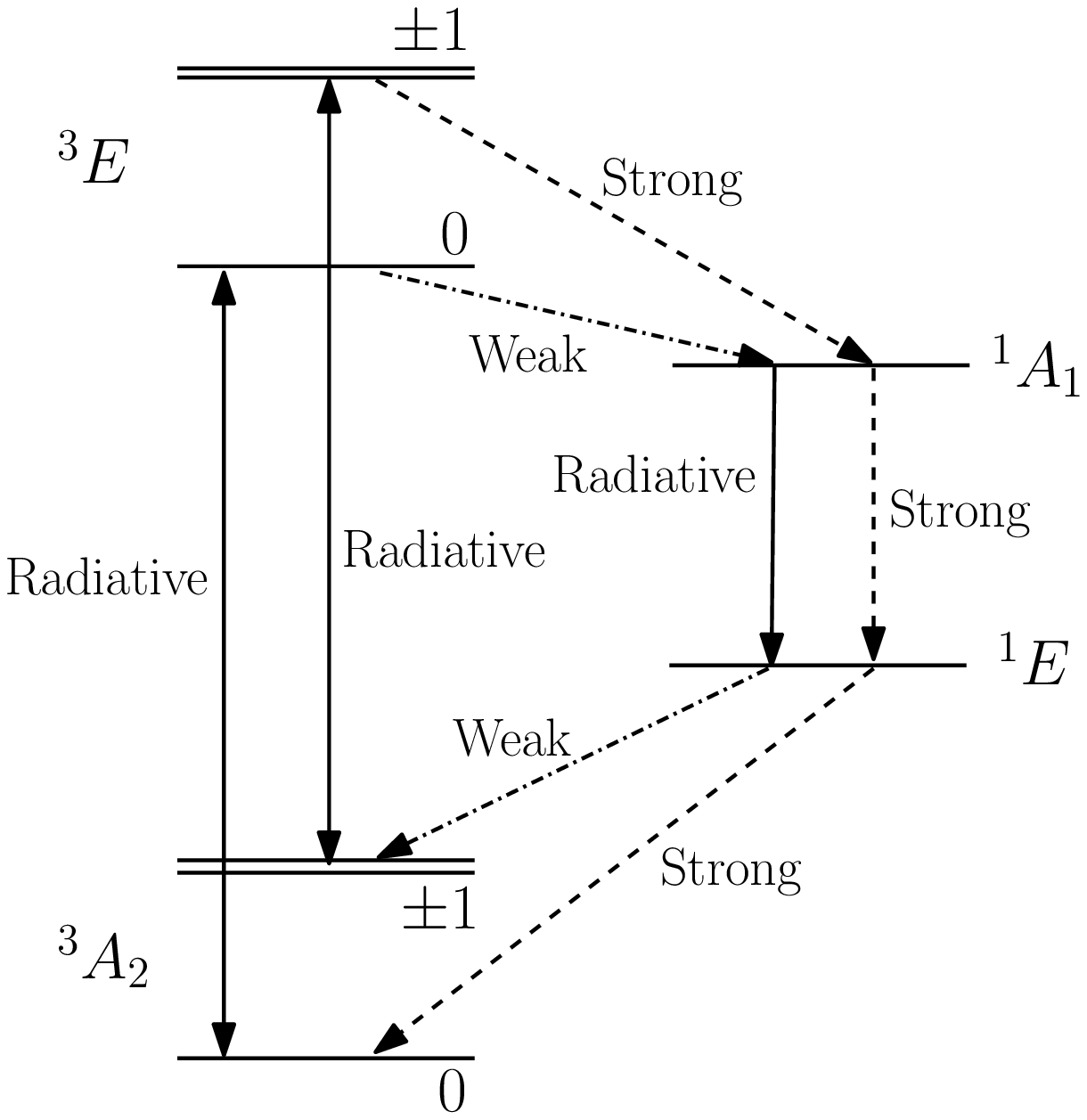}}
\subfigure[]{\includegraphics[width=0.6\columnwidth] {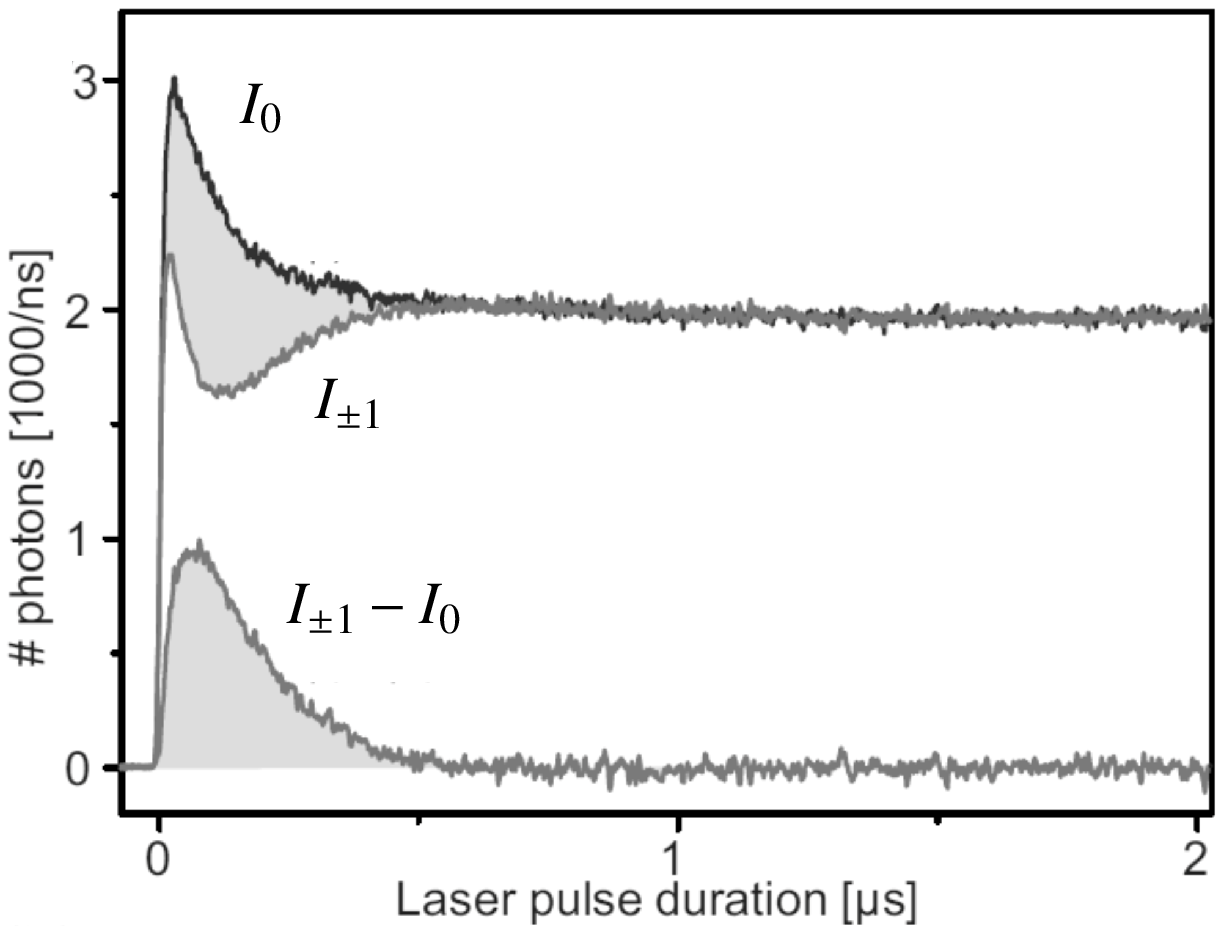}}}
\caption[Radiative and non-radiative transitions of NV$^-$ that lead to optical spin-polarisation and readout and example time-dependent fluorescence of pure spin states]{(a) Schematic of the NV$^-$ electronic structure including the $^3A_2$ and $^3E$ (room temperature) fine structure \cite{doherty11}. The optical and infrared transitions are denoted by solid arrows and the weak ($-\cdot-$) and strong (- - -) non-radiative transitions to, between and from the intermediate singlet states are denoted by dashed arrows. (b) Examples of the time-dependent fluorescence $I_{m_s}$ for pure initial spin states ($m_s=0$,$\pm1$) \cite{steiner10}. The data was collected from a single centre using off-resonant optical excitation and accumulated over many measurement repetitions. The contrast in the fluorescence of the pure spin states is demonstrated in the lower half of the plot.}
\label{fig:introopticaldynamics}
\end{center}
\end{figure}

The optical detection of the magnetic resonances of the NV$^-$ centre is enabled by the differing fluorescence of the $m_s=0$ and $\pm1$ spin projections \cite{vanOort88}. As demonstrated in figure \ref{fig:introopticaldynamics}, the spin dependence of the NV$^-$ fluorescence can be explained by different non-radiative decay pathways for the $m_s=0$ and $\pm1$ spin projections from $^3E$ to $^3A_2$ via intermediate dark states (believed to be $^1A_1$ and $^1E$) \cite{manson06,martin00,harrison04}. As these non-radiative decay pathways compete with fluorescent decay, the spin projection with the weakest non-radiative decay out of $^3E$ will appear brightest and vice versa. The observed optical enhancement of the $^3A_2$ EPR signal strength implies that the probability of finding the centre in a particular spin projection is polarised by optical excitation, a process known as optical spin-polarisation \cite{loubser77, loubser78}. The combination of a number of observations has led to the consensus that the $m_s=0$ spin projection is the bright projection and also the projection into which the centre is optically polarised \cite{manson90,he92,manson92,he93a}. It is currently believed that the non-radiative decay from the $^3E$ spin sub-levels to the dark states is slower for $m_s=0$ than for $m_s=\pm1$ and that the non-radiative decay out of the dark states to the $^3A_2$ spin sub-levels is similar for both spin sub-levels, thereby leading to a net polarisation into the $m_s=0$ sub-level \cite{togan10,manson06,nizovtsev03,batalov08,robledo10}. Through observing the time-dependence of the centre's fluorescence (refer to figure \ref{fig:introopticaldynamics}), it is clear that the process of optical spin-polarisation takes a few optical cycles; the initial fluorescence is representative of the initial spin state of the centre and spin-polarisation is achieved once the optical steady state has been reached. Consequently, the optical dynamics of the NV$^-$ centre can be used to optically readout the centre's spin state and to simultaneously spin-polarise the centre into the $m_s=0$ spin projection \cite{harrison04}. Specifically, optical readout is conducted by comparing the integrated fluorescence of the spin state to be read out with calibration measurements of the integrated fluorescence corresponding to the spin being prepared in the $m_s = 0$ and $\pm1$ spin states \cite{steiner10}.

Differing from NV$^-$, NV$^0$ does not have detectable magnetic resonances associated with its spin doublet ground ($^2E$) and excited ($^2A$) states. The absence of a NV$^0$ ground state magnetic resonance is believed to be due to the presence of the dynamic Jahn-Teller effect \cite{felton08}. It was recently discovered that NV$^0$ does, however, exhibit an EPR signal under continuous optical illumination \cite{felton08}. This EPR signal has been attributed to a long lived spin quartet excited state ($^4A_2$) that is populated by non-radiative decay from the optically excited $^2A$ \cite{felton08}. No ODMR or optical readout of the NV$^0$ quartet spin state have been demonstrated to date. Yet, the EPR signal strength has been observed to increase with optical excitation intensity, indicating an optically dependent spin-polarisation of $^4A_2$ \cite{felton08}, albeit via a very different process to that which occurs in NV$^-$.

\begin{figure}[hbtp]
\begin{center}
\mbox{
\subfigure[]{\includegraphics[width=0.3\columnwidth] {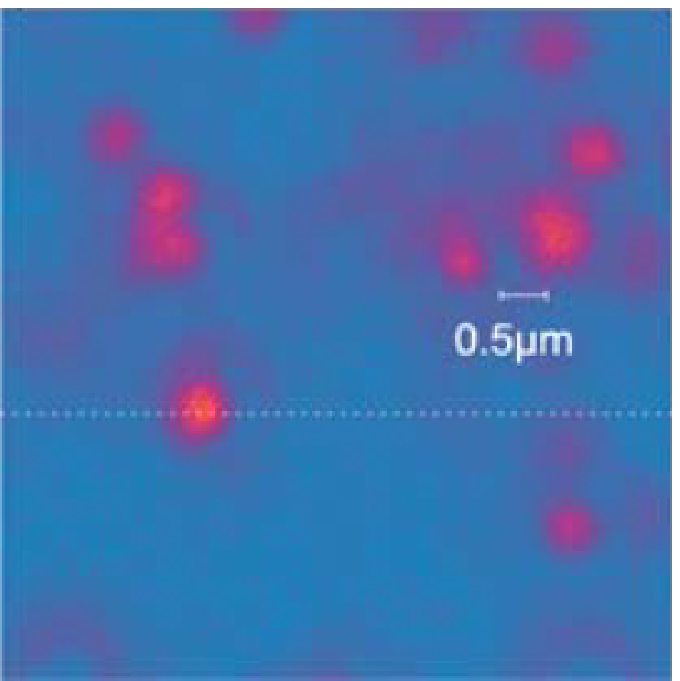}}
\subfigure[]{\includegraphics[width=0.5\columnwidth] {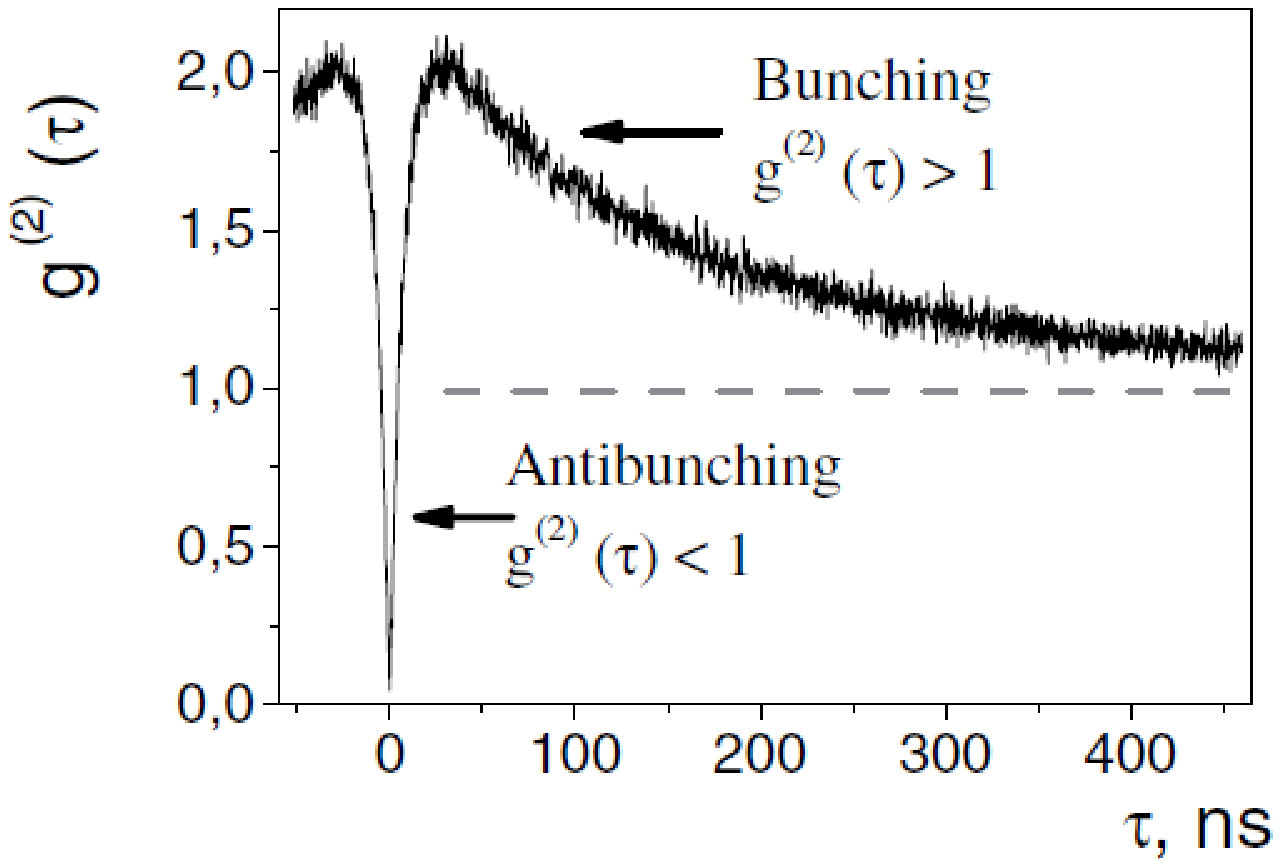}}}
\caption[Example confocal image of individual NV$^-$ centres and example photon autocorrelation function of a single NV$^-$ centre]{(a) Example room temperature confocal image \cite{jelezko06} of single NV$^-$ centres. (b) Example photon autocorrelation function \cite{jelezko06} obtained from a single NV$^-$ centre at room temperature. The antibunching dip below 0.5 at $\tau=0$ is indicative of a single photon source and confirms the presence of a single NV$^-$ centre \cite{jelezko06}.}
\label{fig:introsinglecentredetection}
\end{center}
\end{figure}

Regardless of the competing non-radiative processes in the optical cycles of NV$^-$ and NV$^0$, both charge states have sufficiently strong fluorescence to be detected as single centres using scanning confocal microscopy \cite{gruber97} (see figure \ref{fig:introsinglecentredetection} for an example). Indeed, both centres have been identified as single photon sources \cite{drab99,brouri00,kurt00,treussart06}, as demonstrated by the observed photon statistics in figure \ref{fig:introsinglecentredetection}. The detection of single centres implies certain characteristics of the NV$^-$ and NV$^0$ optical cycles: large optical absorption cross-sections, short excited state lifetimes, high quantum yields, and no efficient shelving in a long lived dark state \cite{gruber97}. Furthermore, both charge states have been observed to be extremely photostable under off-resonance excitation (typically 2.32 eV (532 nm)) \cite{gruber97,brouri00,kurt00,treussart06}, with no evidence of photobleaching except in extreme conditions \cite{bradac10}. However, photoconversion between the charge states and photoblinking under resonant optical excitation has been regularly observed (refer to figure \ref{fig:introtoNV}) \cite{drab99,jelezko02}, but it is not currently understood. The photoconversion efficiency has been observed to be dependent on optical excitation energy and intensity \cite{manson06,manson05,robledo10,iakou00,dumeige04,wee07,acosta09,han10,waldherr11} as well as material properties and the method of NV formation \cite{kennedy03,waldermann07,rabeau07,shen08,fu10}. The spectral diffusion of the NV$^-$optical ZPL (refer to figure \ref{fig:depolarisationdiffusionintro}) is also believed to be related to photoconversion and the ionisation of electron donors in the vicinity of the NV$^-$ centre during optical excitation \cite{fu09,robledo10,jelezko02,shen08,fu10,jelezko03,tamarat06}.

The focus on individual NV$^-$ centres stems from the large inhomogeneous strain broadening of the  optical ZPL and significant spin cross-relaxation effects present in high density ensembles. These ensemble effects prevent the resolution of the NV$^-$ low temperature optical ZPL fine structure and broaden ODMR lines, thereby reducing the visibility of many of the centre's properties.  The evolution towards the study of individual NV centres has been enabled by advances in single molecule spectroscopy \cite{jelezko04c,jelezko01}, NV formation techniques and diamond synthesis. The advent of CVD synthesis techniques provided the means to control the size and quality of the diamond crystal \cite{balasubramanian09}. The employment of nitrogen ion implantation in combination with an annealing step has facilitated the (non-deterministic) creation of single NV centres with nanoscale accuracy \cite{toyli10}. Apart from bulk diamond, single NV centres have been fabricated in nanodiamonds as small as $\sim$5 nm in diameter \cite{bradac10}. Consequently, there exists a diverse range of crystal environments in which the NV centre can be formed and their effects on the centre investigated. Indeed, the advances in spectroscopy and material science have provided a suite of tools to study the NV centre and develop its applications.

\subsection{Overview of the NV in the quantum technologies context}
\label{section:introductionmotivations}

\begin{figure}[hbtp]
\begin{center}
\mbox{
\subfigure[]{\includegraphics[width=0.3\columnwidth] {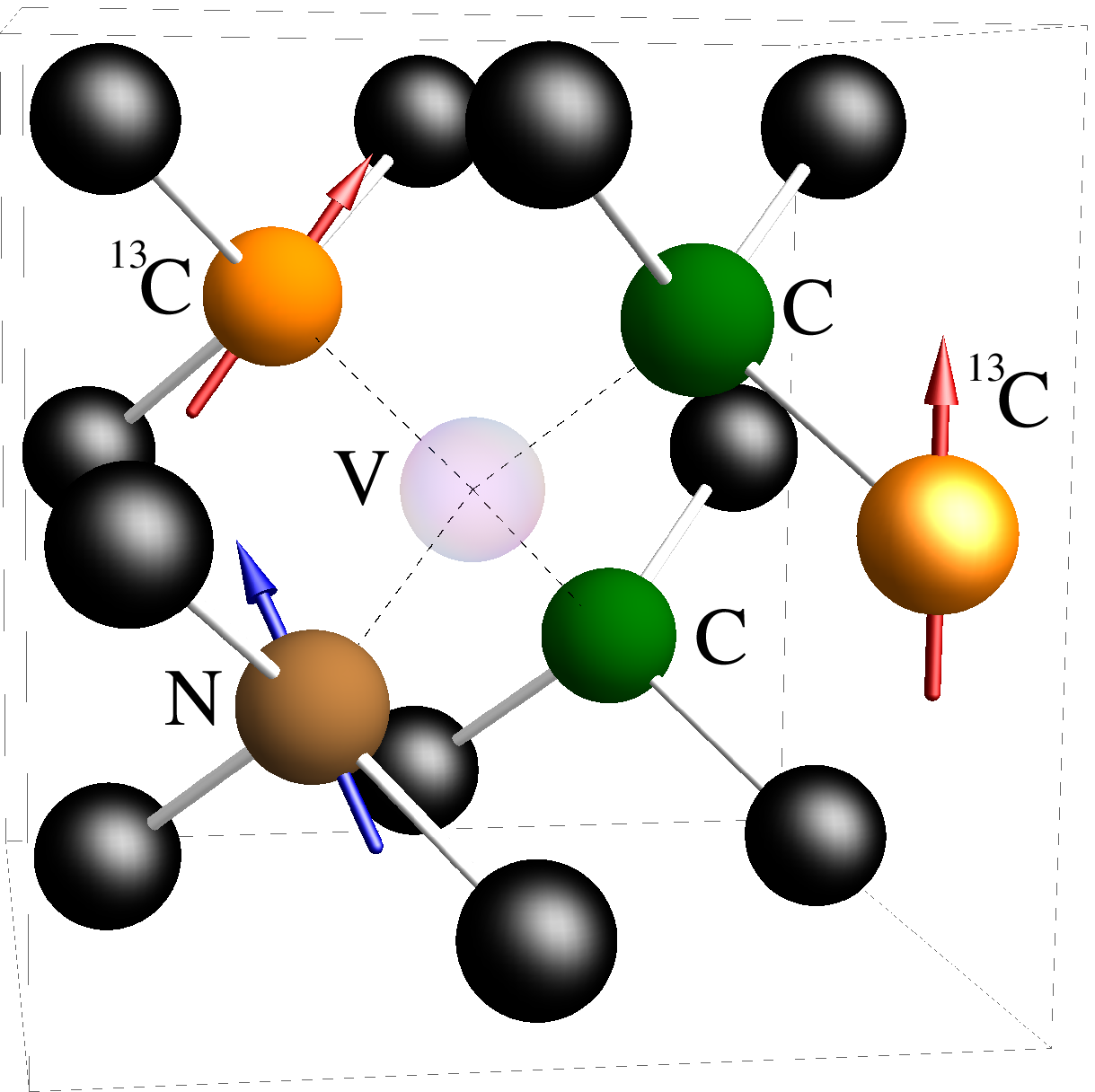}}
\subfigure[]{\includegraphics[width=0.35\columnwidth] {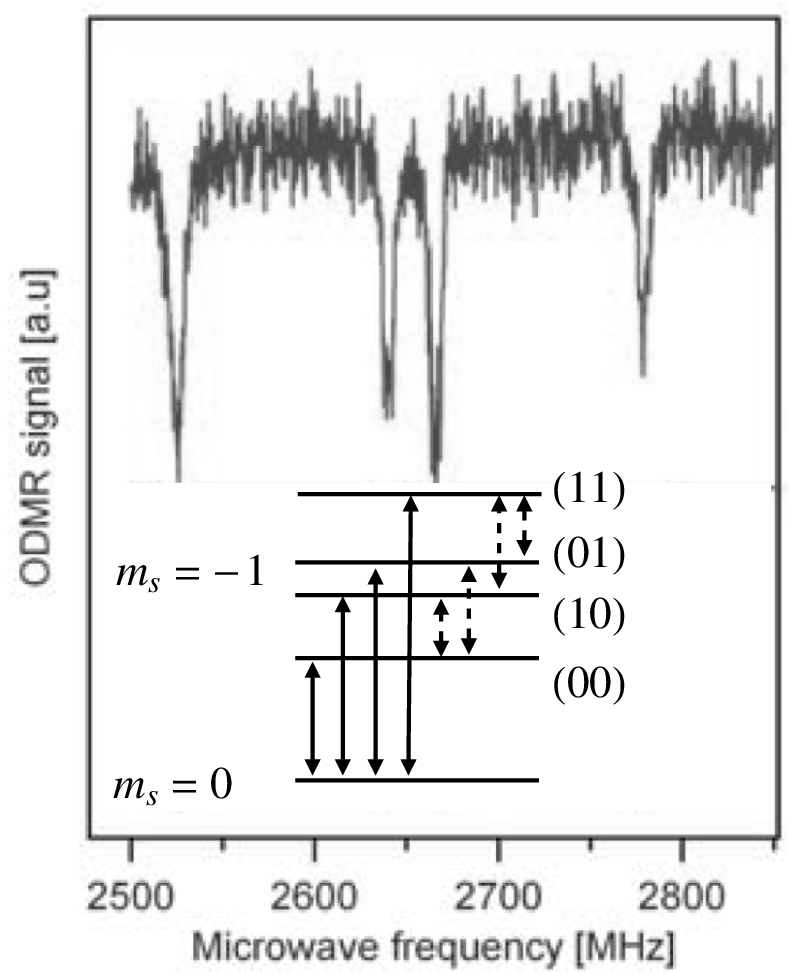}}
\subfigure[]{\includegraphics[width=0.3\columnwidth] {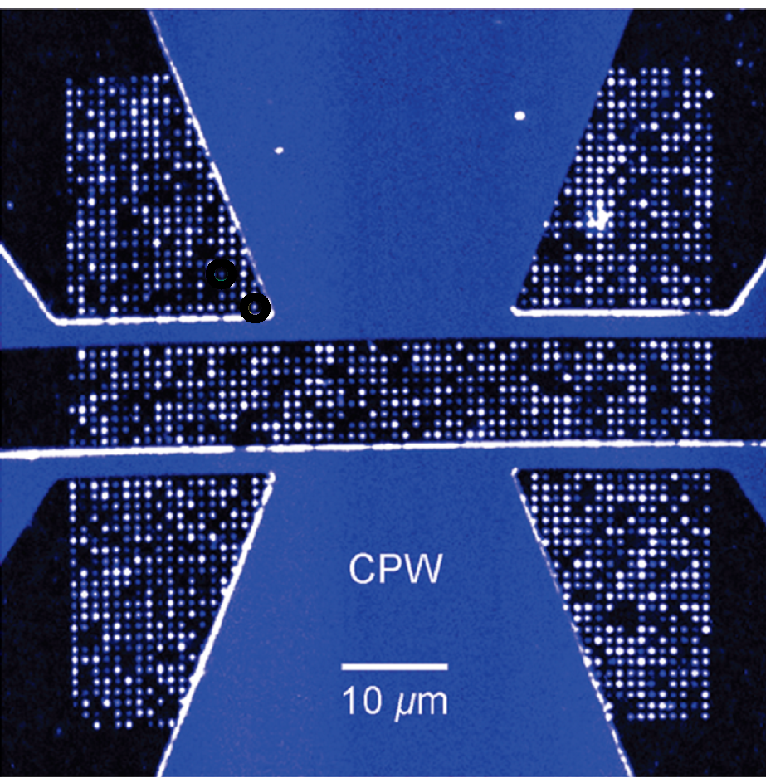}}}
\caption[Example spectra demonstrating the coupling of the NV$^-$ spin with $^{13}$C nuclear spins and an example of an ion implanted NV structure]{(a) Schematic of the proximal nuclear spins (arrows) to the NV centre: the native nitrogen nuclear spin ($^{14}$N or $^{15}$N) and the randomly distributed $^{13}$C nuclear spins that are naturally 1.1\% abundant in diamond. (b) Example ODMR spectra \cite{neumann08} demonstrating the hyperfine structure due to interactions with two $^{13}$C nuclear spins (no coupling to the N nuclear spin observable). Below is the corresponding hyperfine structure with the logical bits (0,1) representing nuclear spin projections ($m_I=-\frac{1}{2},+\frac{1}{2}$) \cite{neumann08}. Electronic spin transitions observable in the ODMR spectra are denoted by solid arrows and the nuclear spin transitions that can be used for control are denoted by dashed arrows \cite{neumann08}. (c) Example spatial photoluminescence image of a co-planar waveguide (CPW) and regular structure of individual NV$^-$ centres (white dots) created by N$^+$ ion implantation \cite{toyli10}.}
\label{fig:spincouplingintro}
\end{center}
\end{figure}

The applications of the NV$^-$ centre each utilise one or a combination of the following five key properties of the centre:
\begin{enumerate}
\item A bright photostable optical transition that is suitable for the detection of individual centres and single photon generation

\item An optical ZPL fine structure that is dependent on electric, magnetic and strain fields at low temperature, but is approximately electric and strain field independent at room temperature

\item A magnetically resonant and controllable ground state electronic spin that exhibits long coherence times and coupling to proximal electronic and nuclear spins

\item Optical spin-polarisation and readout of the ground state spin

\item Flexibility and robustness in fabrication
\end{enumerate}
Some aspects of the above properties have already been discussed. Expanding on the second property in the context of the centre's applications, the strain and electric field dependence of the low temperature optical ZPL fine structure enables the transition energy of the ZPL to be tuned and spin-flip/conserving transitions to be selectively excited. The approximate strain and electric field independence of the room temperature fine structure alternatively offers some degree of uniformity in the fine structure between centres and simplifies the microwave control of the excited state spin of the centre. Whilst the magnetic resonance of the ground state spin has been discussed already, its coherence has not been addressed. Indeed, the ground state spin has the longest room temperature single spin coherence time (T$_2$) of any electronic spin in a solid, being greater than 1.8 ms in highly engineered samples \cite{balasubramanian09}. The long coherence time of the ground state spin enables its couplings with proximal electronic and nuclear spins in the lattice to be resolved and manipulated (see figure \ref{fig:spincouplingintro} for an example) \cite{jelezko04,dutt07,neumann08,neumann10,popa04,gaebel06,hanson06,childress06}.  The final property refers to the advances in material science that have developed a wide range of means to deliberately fabricate both individual centres and ensembles of NV$^-$ centres in bulk and nanodiamond (refer to figure \ref{fig:spincouplingintro} for an example), as well as the observed robustness of the first four properties to the fabrication techniques employed \cite{meijer05,treussart06,bradac10,acosta09,waldermann07,rabeau07,rabeau05,rabeau06,sonne08,nayed10,pezz10,vlasov10}.

\begin{figure}[hbtp]
\begin{center}
\includegraphics[width=0.9\columnwidth] {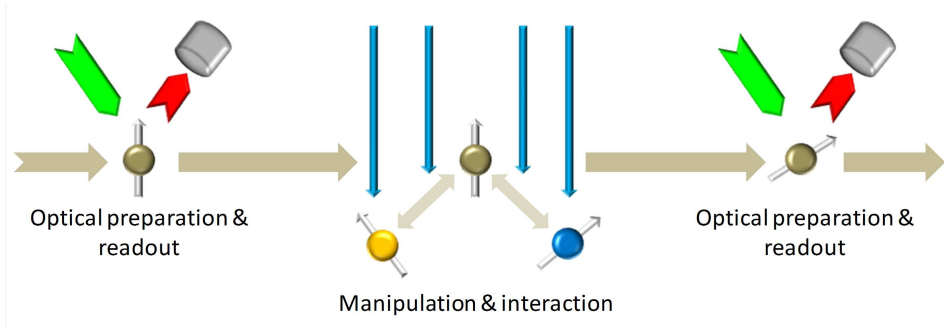}
\caption[The mode of operation of the NV$^-$ spin qubit]{Schematic of the prepare-manipulate/interact-readout mode of operation of the NV$^-$ spin qubit (light blue circle with arrow). An initial off-resonance (green) optical pulse is used to prepare the spin via optical spin-polarisation. Control fields such as microwaves and static electric, magnetic and strain fields are applied to manipulate the spin and the other spins (other circles with arrows) that it is interacting with. Finally, an off-resonance (green) optical pulse is used to readout the spin via the integrated red shifted fluorescence. The readout pulse also prepares the spin for the next manipulate/interact step. }
\label{fig:modeofoperationintro}
\end{center}
\end{figure}

Each of the room temperature implementations of the NV$^-$ centre as a spin qubit employ the same prepare-manipulate/interact-readout mode of operation (see figure \ref{fig:modeofoperationintro}). An initial off-resonance optical pulse is used to prepare the ground state spin via optical spin-polarisation. This is followed by the application of static and microwave fields to manipulate the spin (and potentially radio frequency fields to manipulate coupled nuclear spins) and a period to allow the NV to interact with other spins and evolve. Finally, an off-resonance optical pulse is used to readout the ground state spin by measuring the integrated Stokes shifted fluorescence. It is clear that this mode of operation relies upon the performance of the optical spin-polarisation and readout mechanisms, which are currently only crudely understood. Indeed, the degree of ground state optical spin-polarisation is not consistently reported in the literature, with many different values ranging from 42-96$\%$ reported \cite{jelezko04,jelezko04b,togan10,neumann10,howard06,harrison04,manson92,waldherr11,childress06,harrison06,felton09}. Since the ground state spin-polarisation represents the preparation fidelity of the qubit, it is vital to the operation of the qubit that it is well characterised. However, to date there has been no systematic study of spin-polarisation and its variation with applied electromagnetic fields, strain, and temperature. A second significant problem is the readout contrast, which due to low collection efficiencies of current apparatus, is limited to a difference of a fraction of a photon between the $m_s=0$ and $\pm1$ spin-projections \cite{steiner10,jiang09}. Consequently, a single qubit operation must be performed many times before sufficient shot-noise statistics are acquired to distinguish the spin-projections \cite{steiner10,jiang09}. Recently, there have been reports of readout enhancement techniques using hyperfine coupling with N \cite{steiner10} and $^{13}$C \cite{jiang09} nuclear spins, however it is quite possible that a fully developed understanding of the optical readout mechanism will provide alternate means to enhance readout more significantly.

\begin{figure}[hbtp]
\begin{center}
\mbox{
\subfigure[]{\includegraphics[width=0.4\columnwidth] {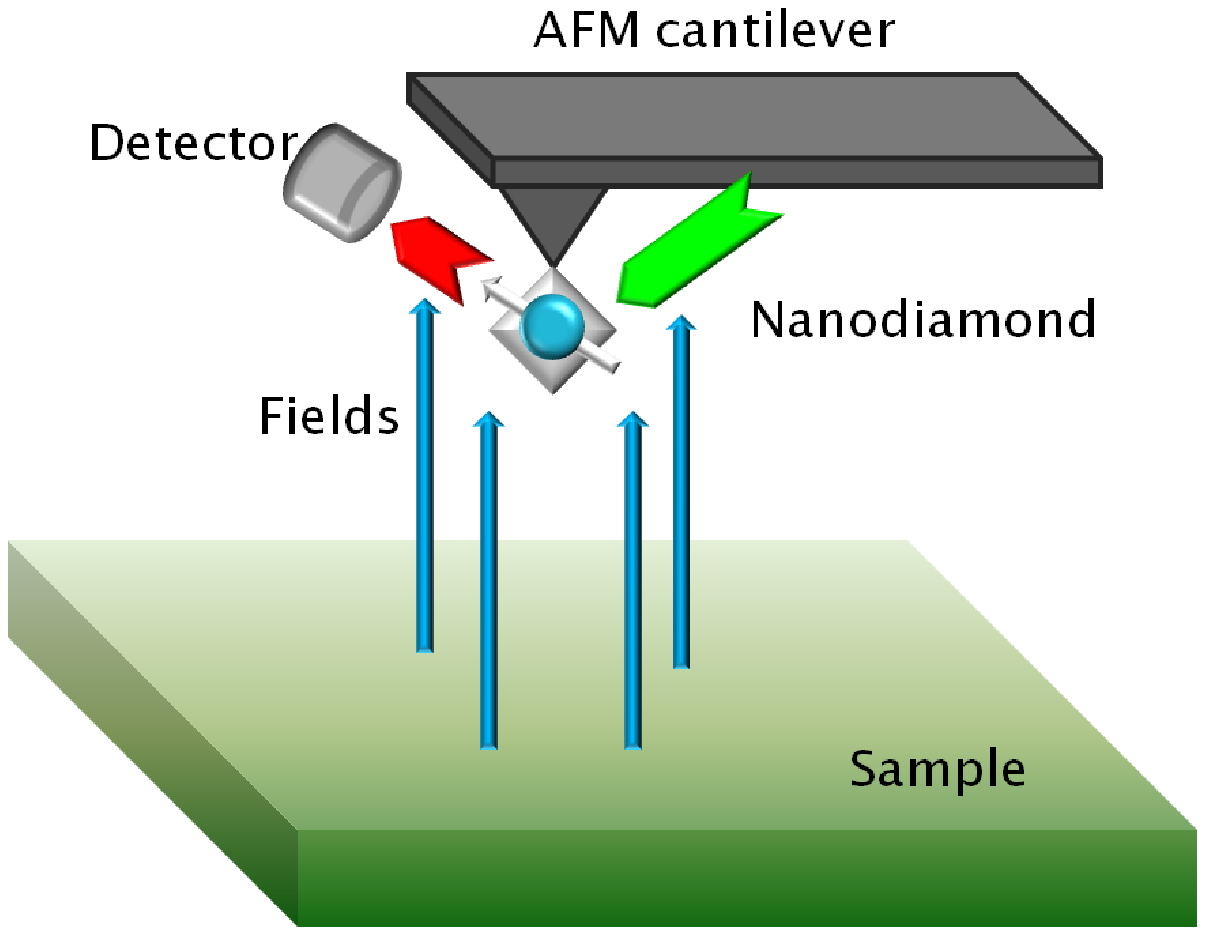}}
\subfigure[]{\includegraphics[width=0.6\columnwidth] {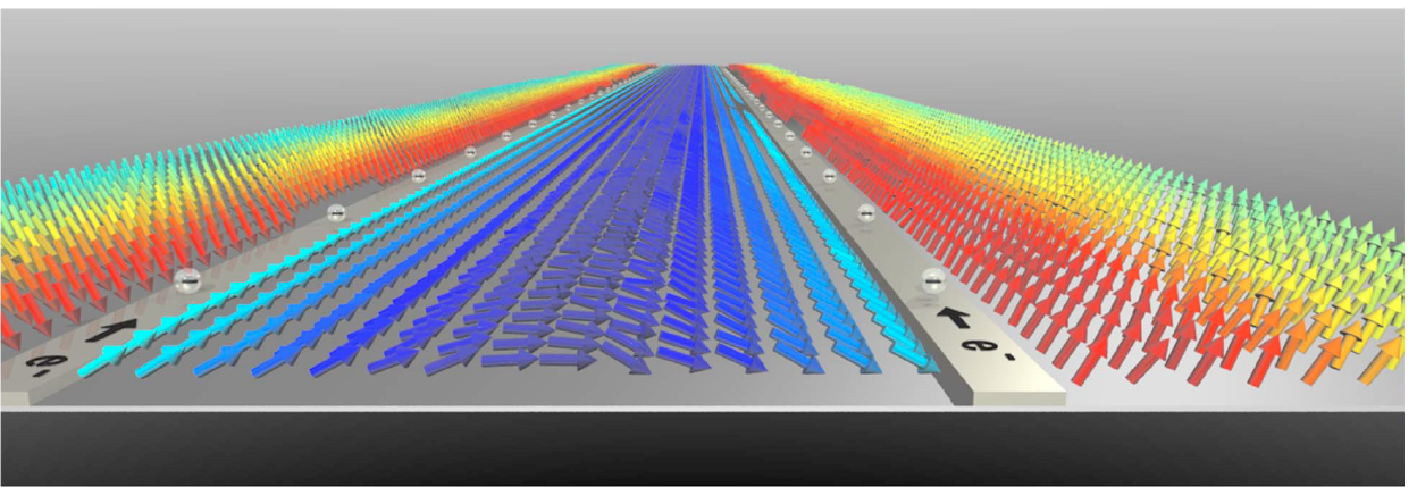}}}
\caption[Scanning and ensemble architectures for the NV$^-$ field sensing applications]{(a) Schematic of one of the NV$^-$ field sensing architecture employing an NV$^-$ centre in a nanodiamond attached to an AFM cantilever to scan a sample. As for the centre's qubit implementations, the field sensing implementations use optical preparation and readout and static and microwave fields for control of the NV$^-$ ground state spin. (b) Example vector magnetic field measurement of the magnetic field induced by parallel current carrying wires using an ensemble of NV$^-$ centres in a diamond substrate \cite{steinert10}.}
\label{fig:magnetometryintro}
\end{center}
\end{figure}

The use of the NV as a nanoscale sensor has been proposed for DC, AC and fluctuating magnetic fields \cite{degen08,taylor08,cole09,hall09,chernobrod05}. Each of the sensing applications utilise one or more of three possible architectures (refer to figure \ref{fig:magnetometryintro}): a single NV$^-$ centre in a nanodiamond attached to a scanning atomic force microscopy (AFM) cantilever, an array of individual NV$^-$ centres in a bulk crystal substrate or in separate nanodiamonds, and an ensemble of NV$^-$ centres in a bulk crystal substrate. The spatial resolution and sensitivity of these architectures are predominately governed by the proximity of the NV$^-$ centre to the sample (as determined by the radius of nanodiamond or depth of the NV$^-$ centre from the surface of the bulk crystal), the coherence time of the NV$^-$ ground state spin and the number of NV$^-$ centres in the sensing volume \cite{degen08,taylor08}. Consequently, given the inclusion of NV$^-$ centres in nanodiamonds, the ability to fabricate high density ensembles or arrays of centres with nanoscale tolerances in bulk diamond, and sub-diffraction limit imaging and magnetic field gradient techniques, the NV$^-$ centre offers nanoscale spatial resolution in each of the three architectures \cite{degen08,taylor08}. Isotopically pure diamond \cite{balasubramanian08} and dynamic decoupling techniques \cite{hall10b,delange10,naydenov11,ryan10} both yield effective spin coherence times of $T_2 \geq 1.8$ ms and magnetic field sensitivities of single centres as high as $\sim$4 nT/$\sqrt{\mathrm{Hz}}$ \cite{balasubramanian09}.  Each of the three architectures employ the same mode of operation as the implementations of the NV$^-$ centre as a spin qubit and are thus subject to the same polarisation and readout problems. However, additional problems arise in the centre's sensing applications because of the ground state spin's sensitivity to not only magnetic fields, but to also electric and strain fields \cite{dolde11,doherty11b}. Being a multi-field sensor, the NV$^-$ centre must be optimised for each sensing mode, otherwise the sensitivity to one field will be inhibited by unwanted sensitivity to the other two fields \cite{dolde11,doherty11b}. A further added complexity is the temperature dependence of the zero field magnetic resonance of the ground state spin \cite{acosta10}, which is currently not well understood and must be a consideration in the optimisation of the sensor. Due to the potential for significant temperature fluctuations when operating at room temperature, the temperature dependence of the NV$^-$ spin will particularly limit the field sensitivity of the spin when operating at room temperature if it is not correctly optimised \cite{acosta10}. Alternatively, the NV$^-$ spin may be optimised to sense the fluctuations in temperature and acts as a nanoscale high sensitivity thermometer \cite{acosta10,toyli12}.

\subsection{Scope of this review}
\label{subsection:scopeofreview}

This review is the first time that the key empirical and \textit{ab initio} results have been extracted from the extensive NV literature and assembled into one consistent picture of the current understanding of the centre. Given this objective, the scope of this review has been restricted to the intrinsic properties of the NV centre and rather than the centre's various applications. There are a number of reviews in existence that deal specifically with the centre's applications \cite{weber10,jelezko04c,jelezko06,wrachtrup06,aharonovich11}. Motivated by the numerous applications of the negatively charged system NV$^-$, it will be the primary subject of this review. However, as it is clear that NV$^0$ offers valuable insights into the properties of the NV$^-$, NV$^0$ will not be ignored. The rapid advance of NV literature also necessitates the restriction of this review to works published prior to 2012.

The structure of this review is as follows. The first section examines the identification of the physical structure of the NV centre and its various charge states by the initial NV studies. The following electronic structure section details the key studies that have provided the current electronic structures of NV$^-$ and NV$^0$, including the variations of the structures due to electromagnetic and strain fields. This section also discusses the key \textit{ab initio} calculations of the electronic structures. The vibronic structure section discusses the optical and infrared vibronic bands of NV$^-$, including the asymmetry of the optical bands, the recent work on the Jahn-Teller effect within the $^3E$ and $^1E$ electronic levels and the temperature variations of the centre's properties. The final optical and spin dynamics section discusses the extrinsic (i.e. the photoconversion mechanisms) and intrinsic (i.e. the spin-polarisation mechanism) aspects of the NV$^-$ optical dynamics as well as the intrinsic coherence properties of the NV$^-$ ground state spin. After the comprehensive review offered in the preceding sections, the ongoing issues in the understanding of the NV centre are identified and the possible avenues for their resolution are examined in the conclusion.

\section{Defect structure and charge states}
\label{section:reviewdefectstructureandchargestates}

In 1965, du Preez observed the formation of the NV$^-$ optical band in type Ib diamond following radiation damage and annealing \cite{duPreez}. Remarkably, he speculated that the band was most likely generated by a substitutional nitrogen-vacancy pair, since type Ib diamond contains a significant concentration of single substitutional nitrogen (N$_s$) defects and the process of radiation damage and annealing creates mobile lattice vacancies. Du Preez supported his argument by an observation of the correlated decay of the intensity of the optical band of the lattice vacancy (GR1) centre with the growth of the intensity of the NV$^-$ optical band. The trigonal symmetry of the centre and the $A$-$E$ electronic nature of the optical transition were established by the photoluminescence polarisation study conducted by Clark and Norris \cite{clark71} and the subsequent uniaxial stress study conducted by Davies and Hamer \cite{davies76}. The establishment of the centre's symmetry supported du Preez's speculation of a nitrogen-vacancy pair formed along the [111] direction. Davies and Hamer took du Preez's speculation a step further and concluded that this must indeed be the structure of the centre, using their analysis of the optical absorption and luminescence bandshapes.

\begin{figure}[hbtp]
\begin{center}
\includegraphics[width=0.8\columnwidth] {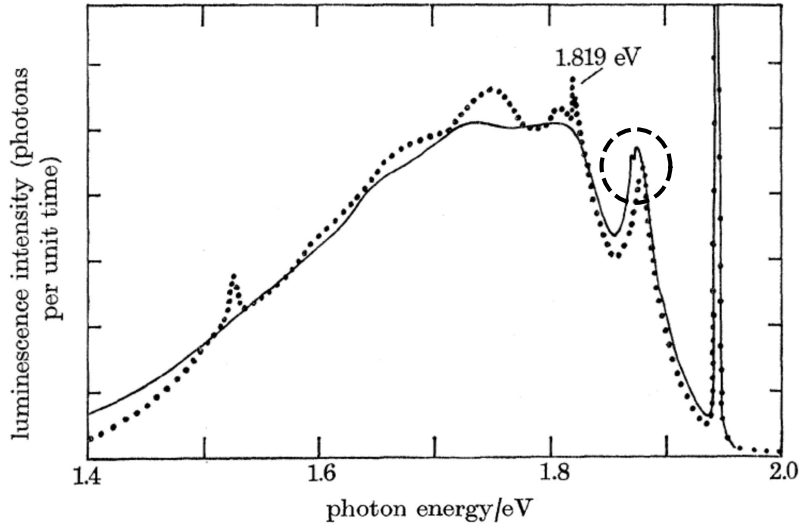}
\caption[Comparison of the absorption and luminescence bands of NV$^-$]{Comparison of the absorption (solid line) and luminescence (dotted line) bands of NV$^-$ \cite{davies76}. Note that the absorption band has been reflected in energy about the ZPL at 1.945 eV in order to directly compare with the luminescence band. The ZPL present at 1.819 eV is due to another defect and not the NV centre \cite{davies76}.}
\label{fig:doublebumpreview}
\end{center}
\end{figure}

Davies and Hamer noted a slight asymmetry in the two bandshapes, which is most pronounced in a small splitting of the one-phonon peak in the absorption band that does not occur in the luminescence band (refer to figure \ref{fig:doublebumpreview} for a comparison of the absorption and luminescence bands). Such a vibronic splitting is typically the signature of the Jahn-Teller effect \cite{davies81}, however, Davies and Hamer observed that after eliminating random lattice strains in the NV$^-$ ensemble, the remaining depolarisation of the one-phonon peak of the luminescence band (typically taken to be secondary evidence of the Jahn-Teller effect \cite{davies81}) was exceedingly small and they concluded that there was no significant Jahn-Teller effect. Instead, Davies and Hamer adopted an explanation of the vibronic band asymmetry based upon the elegant notion that the nitrogen nucleus tunnels between its substitutional site and the vacancy site. The tunneling model is analogous to, and no doubt drew inspiration from, the inversion tunneling of nitrogen in ammonia. Whilst the tunneling model was attractive in its elegance, it appears to have led to a fortuitous identification of the NV centre, given the recent empirical \cite{fu09,Hizhnyakov03,kaiser09} and \textit{ab initio} \cite{mainwood94,zyubin08,gali11,zhang11} evidence in support of the Jahn-Teller explanation. The failure of the tunneling model and the readoption of the Jahn-Teller explanation will be discussed later in this review.

\begin{figure}[hbtp]
\begin{center}
\includegraphics[width=0.8\columnwidth] {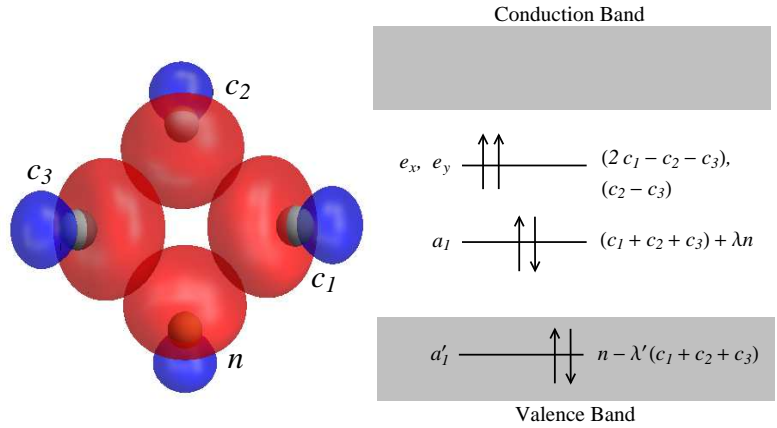}
\caption[Loubser and van Wyk's simple molecular model]{The simple molecular model used by Loubser and van Wyk to interpret their EPR measurements of NV$^-$. Loubser and van Wyk \cite{loubser77, loubser78} explained the existence of an $^3A$ spin triplet state by the depicted occupation (arrows indicating electron spin-up/down) of the defect levels ($a_1^\prime$, $a_1$, $e$) by six electrons. They also provided simple expressions (to right of defect levels) for the defect orbitals in terms of the sp$^3$ orbitals ($c_1$, $c_2$, $c_3$, $n$) of the nearest neighbour nitrogen and carbon atoms to the vacancy (depicted below the defect levels) by applying the linear combinations of atomic orbitals technique.}
\label{fig:loubsersmodelreview}
\end{center}
\end{figure}

Confirmation of the nitrogen-vacancy structure and the proposal of the negative charge state were provided by Loubser and van Wyk \cite{loubser77, loubser78} through their observation and interpretation of an EPR signal correlated with the NV$^-$ optical band. The EPR spectra corresponded to that of a $^3A$ triplet state, whose presence Loubser and van Wyk explained by six electrons occupying two $A_1$ ($a_1$, $a_1^\prime$) and degenerate $E$ ($e_x$, $e_y$) molecular orbitals (MOs), formed from the dangling bonds of the nitrogen and carbon atoms surrounding the vacancy, such that the two unpaired electron spins of the triplet state occupied the $E$ MOs (see figure \ref{fig:loubsersmodelreview} for Loubser's and van Wyk's model). This most simple version of the molecular model was typical of the models used to explain EPR spectra of defects in semiconductors at the time \cite{stoneham75}. The adoption of a molecular model of the NV centre has been supported by both EPR \cite{felton08,felton09,he93} and \textit{ab initio} studies of NV$^-$ and NV$^0$, which have confirmed that the NV MOs ($a_1$, $e_x$, $e_y$) within the diamond band gap are highly localised ($>72\%$) to the nearest neighbours of the vacancy. Note that the majority of \textit{ab initio} studies place the lowest energy MO $a_1^\prime$ predicted by Loubser and van Wyk within the diamond valence band.

In a chemical bonding picture, five of the six electrons were attributed to the electrons occupying the dangling bonds of the nitrogen and carbon atoms surrounding the vacancy \cite{loubser77, loubser78}. The sixth electron that constituted the formation of negative charge state was assumed to have been accepted from a nearby donor (presumed N$_s$) \cite{loubser77, loubser78}. The six-electron/negative-charge model has been generally accepted since its proposal \cite{goss97}, with one notable exception being Lenef and Rand \cite{lenef96}, who tentatively proposed a two-electron/neutral model, but later rejected it based upon Mita's assignment of NV$^0$ \cite{mita96}. Direct evidence in support of the negative-charge model has been obtained through the observation of increases in positron trapping that are correlated with the formation of the NV$^-$ optical band \cite{uedono99}. Having established the number of electrons associated with the NV charge states, the MO $a_1^\prime$ that occurs within the diamond band gap will be ignored in the remainder of this review, since it is filled for all observable electronic levels of NV$^-$ and NV$^0$. As will be discussed in the next section, the observable electronic structures of the charge states are completely described by the various occupations of the MOs ($a_1$, $e_x$, $e_y$) within the diamond band gap by the four/ three (NV$^-$/ NV$^0$) electrons that are in excess of the filled valence band.

Loubser and van Wyk did, however, make the incorrect conclusion that the $^3A$ triplet state was an excited metastable state that existed between optically active $^1A$ and $^1E$ singlets. This conclusion was understandably based upon their observation that the EPR signal existed only under optical illumination and increased when the ZPL and phonon peaks were excited. Indeed, their incorrect conclusion persisted as a source of contention for over a decade and, as will be discussed in the next section, was ultimately resolved by the observation of the EPR signal in the absence of illumination \cite{holliday89,redman91}. The triplet state is now accepted as the ground state of the NV$^-$ centre \cite{reddy87,vanOort88,holliday89,redman91,vanOort91b} and the dependence of the EPR signal strength on optical excitation is now understood as being due to the process of optical spin-polarisation \cite{reddy87}. Notably, Loubser and van Wyk did recognise that the triplet spin sub-levels were being selectively populated under optical excitation as the high field lines in their EPR spectra were in emission.

The optical band of NV$^0$ was initially only observed in cathodoluminescence (note that NV$^-$ is not generally cathodoluminescent) due to its weak photoabsorption signal \cite{davies79}. The NV$^0$ optical band was eventually observed in photoabsorption and photoluminescence by Davies \cite{davies79}, who noted the presence of the Jahn-Teller effect in the optically active states. Although Davies identified the optical band as an $E$-$A$ electronic transition occurring at a trigonal centre, it was not until Mita \cite{mita96} observed correlated changes in the concentrations of NV$^-$ and NV$^0$ under heavy neutron irradiation, that it was asserted that the defect centre responsible for the NV$^0$ optical band was indeed the neutral charge state of the NV centre \cite{mita96}. Mita supported his assertion using the argument that damage induced by neutron irradiation lowers the diamond Fermi level, such that NV$^-$ reduces in concentration relative to NV$^0$. Mita's conclusion was generally correct, although it was subsequently concluded that Fermi level arguments do not provide correct predictions of the relative concentrations of charge states of defects in diamond and that the concentrations are instead governed by the microscopic distributions of donors (such as N$_s$) in the proximity of each individual defect \cite{collins02}. Consequently, in type Ib diamond (substantial N$_s$ concentration), there is a large mean number of proximal N$_s$ donors and their microscopic distributions about the NV centres are reasonably uniform. Thus, NV$^-$ is the most prevalent charge state in type Ib diamond \cite{manson05}. Whereas in type IIa diamond (undetectable N$_s$ concentration), there is a small mean number of proximal N$_s$ donors and their microscopic distributions about the NV centres vary significantly. Therefore, there is no clearly dominant charge state and there is significant variation in the relative concentrations of the NV charge states in type IIa diamond \cite{collins02}. The dependence of the charge state concentrations on the microscopic distributions of donors also implies that for centres formed using N$^+$ ion implantation in type IIa diamond, the charge state concentrations depend sensitively on ion fluence, straggling and the other implantation and annealing parameters that determine the nitrogen distribution \cite{toyli10,nayed10,pezz10}. The microscopic distributions of acceptors (such as the lattice vacancy and surface traps of nanodiamonds) have also been observed to affect the charge state of the NV centre \cite{treussart06,bradac10,waldermann07,fu10,rondin10}.

Mita's identification of NV$^0$ was eventually confirmed by observations of photoconversion between the two charge states via the dependence of the relative emission intensities of the charge state optical bands on the excitation intensity \cite{manson05,dumeige04} and energy \cite{manson05,iakou00}. As previously mentioned, it is now believed that photoconversion leads to an equilibrium of the charge state concentrations under optical excitation and that the equilibrium concentrations are dependent on the excitation energy and intensity \cite{gaebel06}. In spite of the observations of photoconversion and the dependence of the centre's charge state on the various formation factors discussed, there does not presently exist a model that accounts for both the microscopic origins of the NV centre charge state and of the photoconversion process. This outstanding issue will be discussed further at a later stage in the review.

Before preceding to a review of the known aspects of the electronic structures of NV$^-$ and NV$^0$, it is worth mentioning that a third optical ZPL in the range 4.325-4.328 eV has been attributed to the NV centre \cite{deSa77,walker79}. This third optical ZPL has been positively correlated with the growth of the NV$^-$ optical ZPL following radiation damage and annealing in type Ib diamond \cite{deSa77,walker79}. This third optical ZPL has not been studied in any detail, however it has been speculated that it corresponds to the transition of an electron between the valence band and the partially filled $E$ defect levels ($e_x$, $e_y$) \cite{kupriyanov00}. If this speculation is correct, then the third ZPL will be another manifestation of the microscopic origins of the NV charge state and may be involved in the photoconversion process.

\section{Electronic structure}
\label{section:reviewelectronicstructure}

\begin{figure}[hbtp]
\begin{center}
\mbox{
\subfigure[]{\includegraphics[width=0.33\columnwidth] {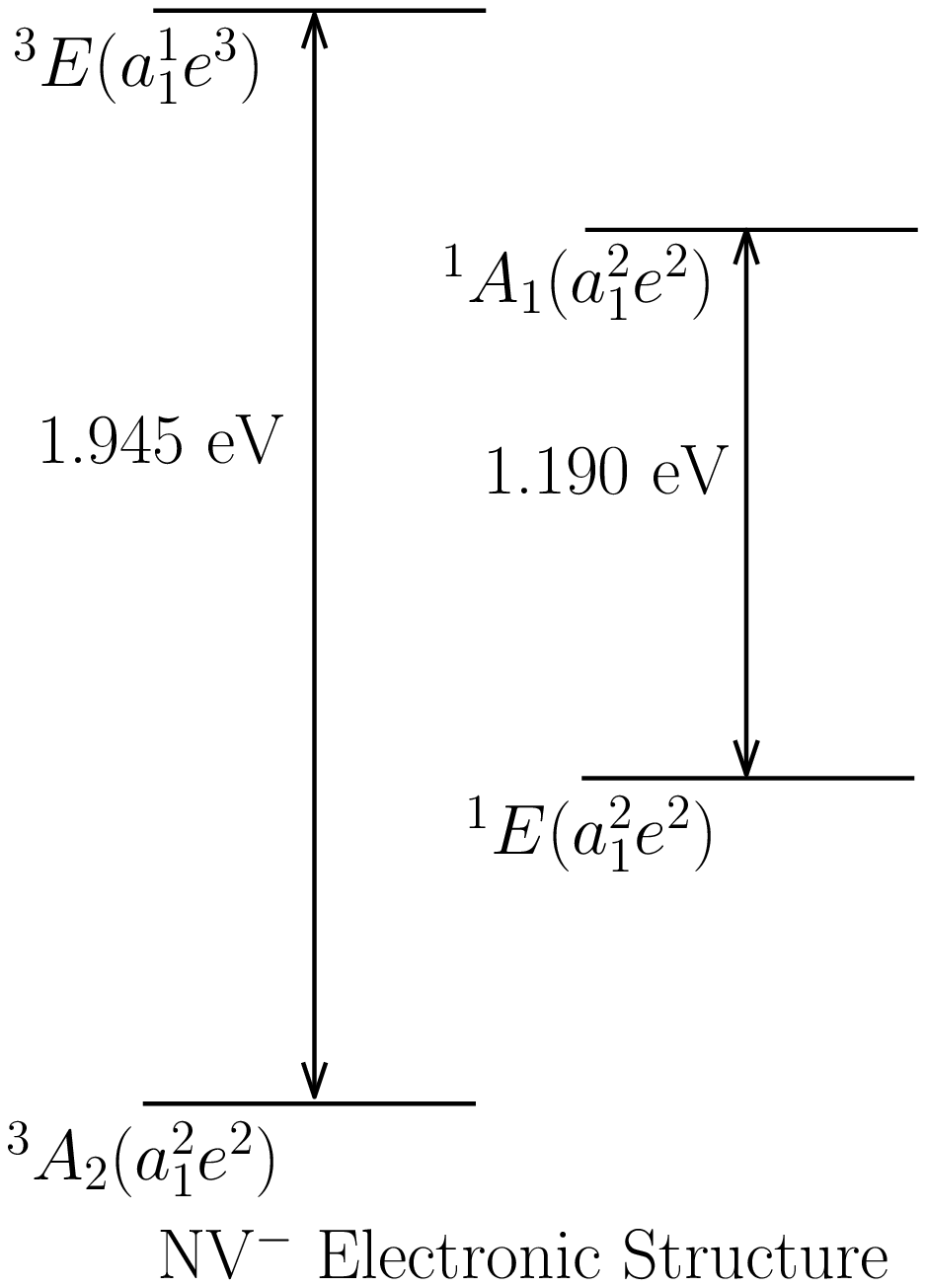}}
\subfigure[]{\includegraphics[width=0.33\columnwidth] {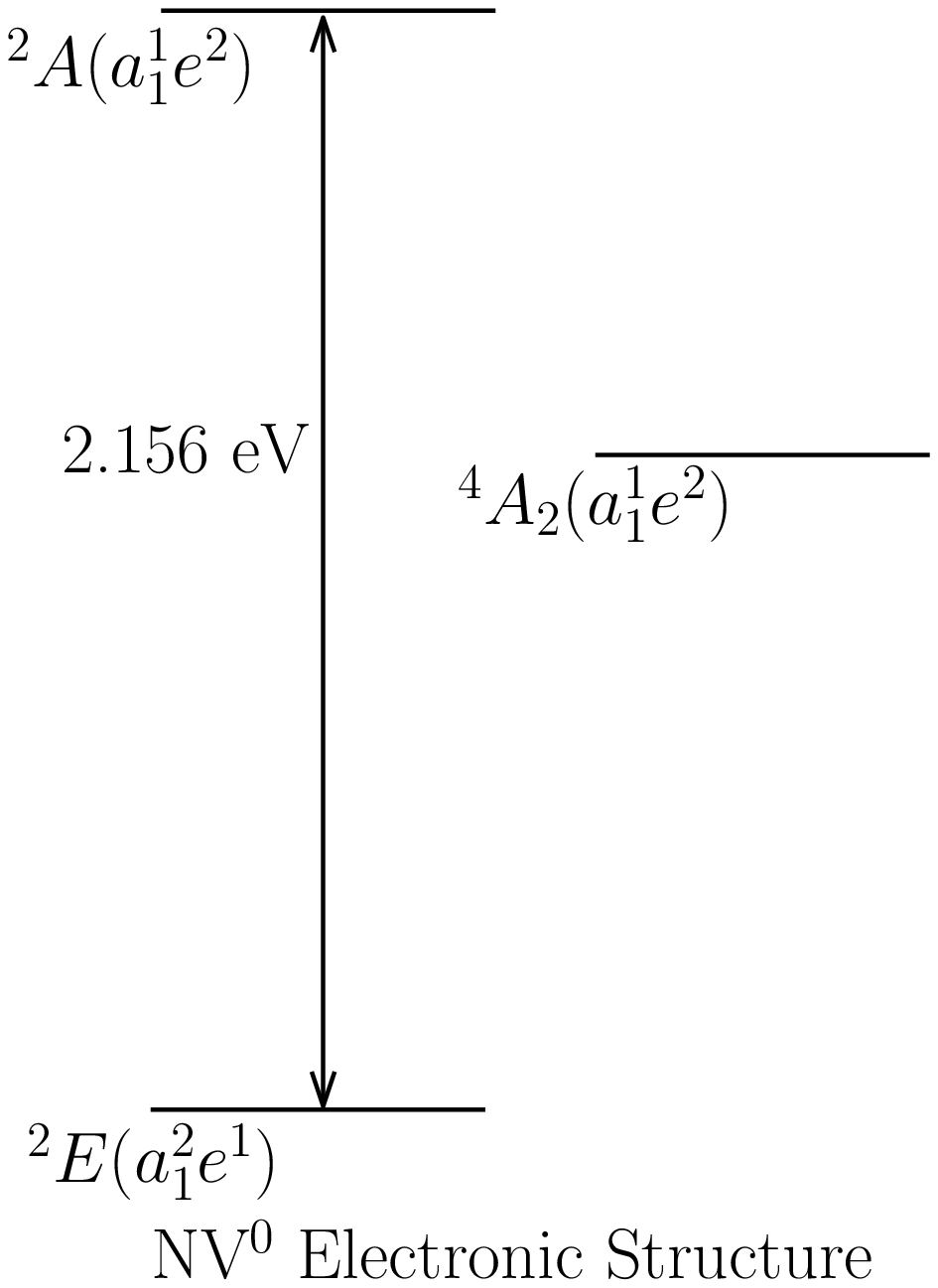}}
}
\caption[Known aspects of the electronic structures of NV$^0$ and NV$^-$]{Schematics of the known aspects of the (a) NV$^-$ and (b) NV$^0$ electronic structures \cite{doherty11,felton08}. The energies of the observed ZPL transitions between the electronic levels are as indicated. The electron occupations of the NV MOs associated with each electronic structure level are denoted in brackets.}
\label{fig:reviewelectronicstructures}
\end{center}
\end{figure}

The electronic structures of NV$^-$ and NV$^0$ have been under constant debate since their first reports in the 1970s \cite{davies76,davies79} and several aspects are still under contention. The currently accepted electronic structures are depicted in figure \ref{fig:reviewelectronicstructures}. As discussed in the previous section, the orbital symmetries of the electronic states involved in the optical transitions of NV$^-$ and NV$^0$ were established very early by photoluminescence polarisation and uniaxial stress studies \cite{davies76,davies79,clark71}. However, since the NV$^-$ EPR signal was initially only detected under optical illumination \cite{loubser77, loubser78}, it took some time for the spin multiplicities of the NV$^-$ ground and optically excited states to be correctly identified. The first evidence of the spin triplet nature of the NV$^-$ ground state was obtained by Reddy et al \cite{reddy87} in their optical hole burning and magnetic circular dichroism study. Further evidence of a triplet ground state was obtained using ODMR \cite{vanOort88}, spin-locking \cite{vanOort88} and spin cross-relaxation \cite{vanOort91b} studies. Confirmation was ultimately provided by observations of the EPR signal in the dark \cite{holliday89,redman91}. Similar spin identification problems to NV$^-$ existed in establishing the spin multiplicities of the NV$^0$ ground and optically excited states. The long absence of any EPR signal that was correlated to the NV$^0$ optical band, facilitated only speculations that the ground state was a spin doublet whose EPR signal was undetectable due the Jahn-Teller effect \cite{felton08}, which had been observed very early in studies of the optical transition \cite{davies79}. The collection of sound evidence for the doublet nature of the NV$^0$ ground and optically excited states required the correct identification of the charge state \cite{mita96} and the subsequent observation of a metastable spin quartet EPR signal that was consistent with the neutral charge state identification \cite{felton08}. Notably, the exact symmetry (possibly $^2A_1$ or $^2A_2$) of the optically excited doublet has not yet been determined, which is particularly important to the establishment of the NV$^0$ electronic structure, since the molecular model predicts three doublet states ($^2A_1$, $^2A_2$, $^2E$) associated with the $a_1e^2$ configuration.

\subsection{The fine structure of the NV$^-$ $^3A_2$ level}

The $\sim$2.88 GHz fine structure splitting of the NV$^-$ $^3A_2$ ground state was measured in the first EPR study \cite{loubser77, loubser78}. However, it was not possible to energetically order the $m_s=0$, $\pm1$ spin sub-levels until Raman-Heterodyne electron nuclear double resonance (ENDOR) and double nuclear magnetic resonance (double NMR) measurements were performed \cite{manson90,he93a}. There was initially some debate over whether the fine structure splitting was due to electronic spin-spin interaction or spin-orbit interaction or both. Simple symmetry considerations \cite{lenef96} imply that spin-spin interaction induces a first order energy splitting of $^3A_2$, whereas spin-orbit interaction only induces a second order energy splitting. The observations of GHz scale spin-orbit interactions in the $^3E$ fine structure \cite{batalov09,tamarat08}, combined with the absence of any indications of an excited state within a similar energy scale to the ground state, have led to the consensus that second order spin-orbit terms are likely to be too small to be principally responsible for the ground state fine structure splitting. Thus, the splitting has been attributed to first order electronic spin-spin interaction \cite{manson06,lenef96,martin99b}.

\begin{figure}[hbtp]
\begin{center}
\includegraphics[width=0.7\columnwidth] {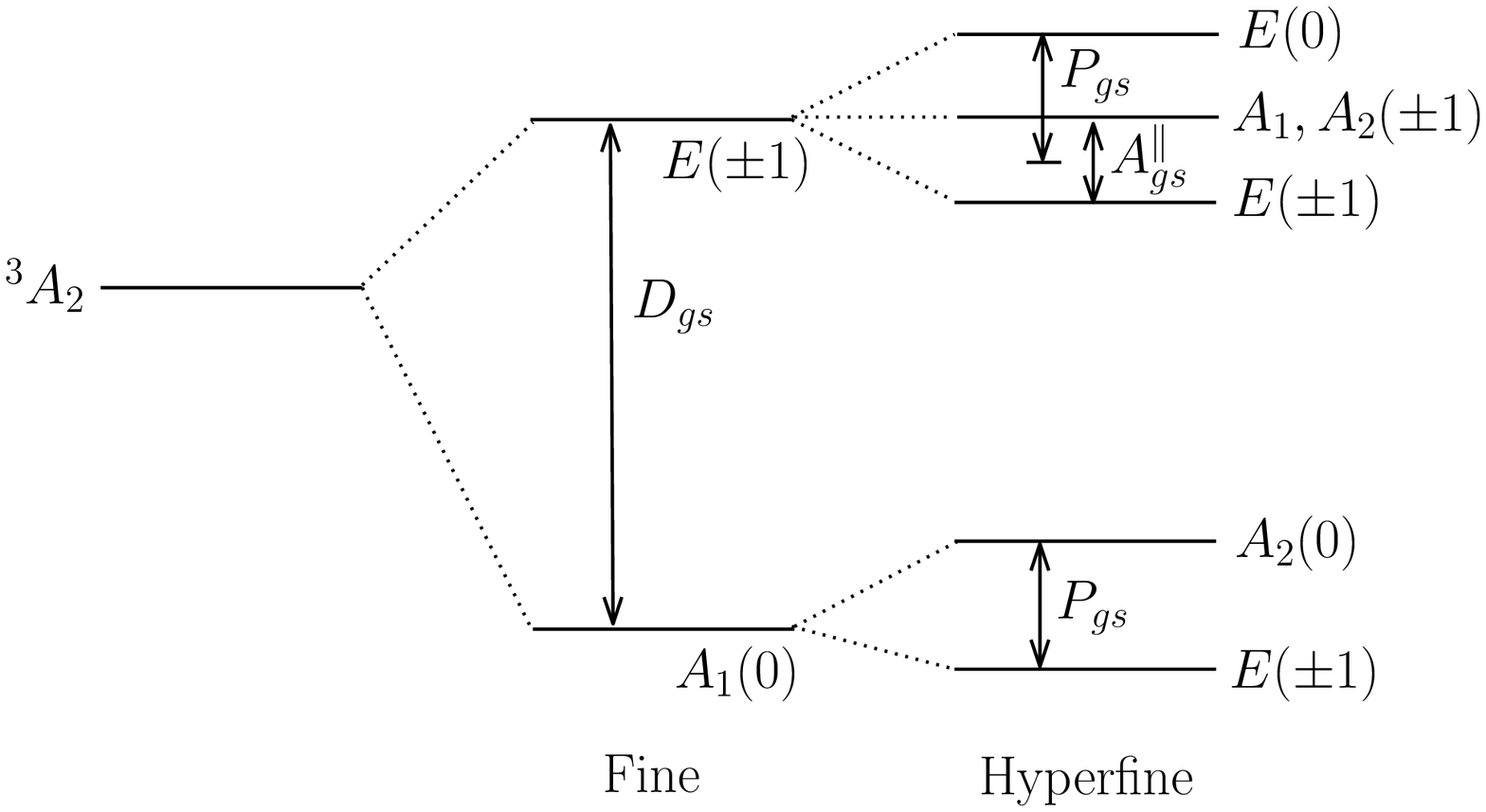}
\caption[Fine and hyperfine structures of the $^{14}$NV$^-$ ground state]{Fine and hyperfine structures of the $^{14}$NV$^-$ ground state \cite{doherty11b}. The fine structure levels are denoted by their spin-orbit symmetry and $m_s=0$, $\pm1$ spin projections and the hyperfine structure levels are denoted by their hyperfine symmetry and $m_I=0$, $\pm1$ spin projections.}
\label{fig:NVgroundstatefinestructure}
\end{center}
\end{figure}

\begin{figure}[hbtp]
\begin{center}
\includegraphics[width=0.65\columnwidth] {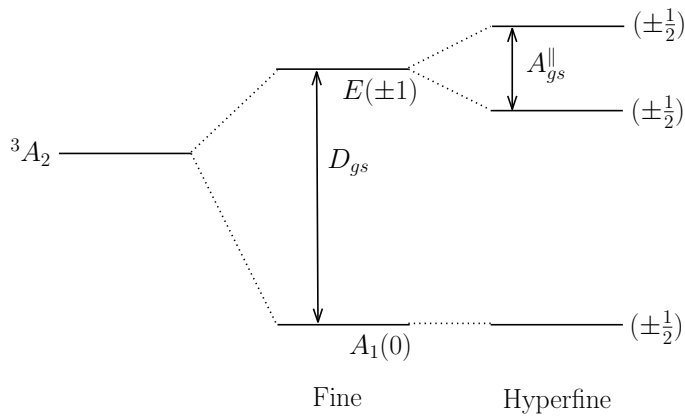}
\caption[Fine and hyperfine structures of the $^{15}$NV$^-$ ground state]{Fine and hyperfine structures of the $^{15}$NV$^-$ ground state \cite{rabeau06}. The fine structure levels are denoted by their spin-orbit symmetry and $m_s=0$, $\pm1$ spin projections and the hyperfine structure levels are denoted by their $m_I=\pm\frac{1}{2}$ spin projections.}
\label{fig:15Ngroundstatehyperfinestructure}
\end{center}
\end{figure}

The first EPR, ENDOR and NMR studies of NV$^-$ also provided measurements of the ground state hyperfine structure due to the $I=1$ nuclear spin of the centre's native $^{14}$N nucleus. The fine and hyperfine structures of the $^{14}$NV$^-$ ground state are depicted in figure \ref{fig:NVgroundstatefinestructure}. The fabrication of NV$^-$ centres using $^{15}$N$^+$  ion implantation has facilitated the observation of the different ground state hyperfine structure arising from the $I = \frac{1}{2}$ nuclear spin of $^{15}$N (refer to figure \ref{fig:15Ngroundstatehyperfinestructure}). The ground state fine and hyperfine structures for both nitrogen isotopes can be described by the canonical spin-Hamiltonian of trigonal defects \cite{stoneham75}
\begin{eqnarray}
\hat{H}_{gs} &=& D_{gs}\left[\hat{S}_z^2-S(S+1)/3\right]+A_{gs}^\parallel \hat{S}_z\hat{I}_z+A_{gs}^\perp\left[\hat{S}_x\hat{I}_x+\hat{S}_y\hat{I}_y\right] \nonumber\\
&& +P_{gs}\left[\hat{I}_z^2-I(I+1)/3\right]
\label{eq:NVgroundstatespinHamiltonian}
\end{eqnarray}
where $D_{gs}\sim2.88$ GHz is the fine structure splitting, $P_{gs}$ is the nuclear electric quadrupole parameter and $A_{gs}^\parallel$ and $A_{gs}^\perp$ are the axial and non-axial magnetic hyperfine parameters. Note that since $^{15}$N is a spin-half nucleus, it does not have an electric quadrupole moment. Whilst the magnitudes and signs of $D_{gs}$ and $P_{gs}$ were accepted from their first measurement \cite{loubser77, loubser78,he93}, the signs of the magnetic hyperfine parameters have been subject to debate. Table \ref{tab:groundstatespinparametersreview} contains the various measured and calculated values of the NV$^-$ ground state hyperfine parameters. Note that at room temperature the $D_{gs}$ reduces to $\sim$2.87 GHz \cite{gruber97}. The temperature dependence of $D_{gs}$ will be discussed later in this review.

\begin{table}
\caption[The hyperfine parameters of the NV$^-$ ground state tabulated by reference and nitrogen isotope]{\label{tab:groundstatespinparametersreview} Experimental measurements and \textit{ab initio} calculations of the hyperfine parameters of the NV$^-$ ground state tabulated by reference and nitrogen isotope. Uncertainty ranges are included if they were provided by the source references. Note that the $^{15}$N nucleus does not have a quadrupole moment.}
\begin{tabular}{lllll}
\noalign{\smallskip}
\toprule
Reference & Iso. & $A_{gs}^\perp$ (MHz) & $A_{gs}^\parallel$ (MHz) & $P_{gs}$ (MHz) \\
\midrule
Loubser$^e$ \cite{loubser77, loubser78} & $^{14}$N &  & $\pm2.32\pm0.01$ &  \\
He$^e$ \cite{he93} & $^{14}$N & $+2.10\pm0.10$ & $+2.30\pm0.02$ & $-5.04\pm0.05$ \\
Felton$^e$ \cite{felton08} & $^{14}$N & $-2.70\pm0.07$ & $-2.14\pm0.07$ & $-5.01\pm0.06$ \\
Steiner$^e$ \cite{steiner10} & $^{14}$N &  & $-2.166\pm0.01$ & $-4.945\pm0.01$ \\
Smeltzer$^e$ \cite{smeltzer09} & $^{14}$N &  & $-2.162\pm0.002$ & $-4.945\pm0.005$ \\
Gali$^a$ \cite{gali08} & $^{14}$N & $-1.7$ & $-1.7$ &  \\
Rabeau$^e$ \cite{rabeau06} & $^{15}$N & $-3.1$ & $-3.1$ & n/a \\
Fuchs$^e$ \cite{fuchs08} & $^{15}$N & 3.01$\pm0.05$ & 3.01$\pm0.05$ & n/a \\
Felton$^e$ \cite{felton09} & $^{15}$N & $3.65\pm0.03$ & $3.03\pm0.03$ & n/a \\
Gali$^a$ \cite{gali09b} & $^{15}$N & 2.7 & 2.3 & n/a \\
\bottomrule\noalign{\smallskip}
\multicolumn{2}{l}{$^e$Experiment, $^a$\textit{ab initio}}
\end{tabular}
\end{table}

The debate over the signs of the magnetic hyperfine parameters has its origins in the unpaired spin density distribution of the NV$^-$ ground state and the relative magnitudes of the Fermi contact and dipolar contributions to the parameters. The magnetic hyperfine parameters are expressed in terms of the Fermi contact ($f_{gs}$) and dipolar ($d_{gs}$) contributions by $A_{gs}^\parallel = f_{gs}+2d_{gs}$ and $A_{gs}^\perp = f_{gs}-d_{gs}$ \cite{stoneham75}. The Fermi contact contribution is directly proportional to the unpaired spin density at the N nucleus and the dipolar contribution is related to the p atomic orbital character of the spin density in the region centred on the N nucleus \cite{loubser77, loubser78}. Using the simple molecular model of Loubser and van Wyk \cite{loubser77, loubser78}, the unpaired spin density is described by the occupation of the $E$ MOs ($e_x$, $e_y$) by two electrons. Since these MOs are not symmetric, their amplitude along the [111] symmetry axis of the NV centre, where the N nucleus resides, is zero. Furthermore, as the $E$ MOs have no contribution from the symmetric N dangling sp$^3$ orbital, the unpaired spin density produced by the $E$ MOs is small in the region of the N nucleus as it is mainly located in the vicinity of the vacancy and its nearest neighbour carbon atoms. Therefore, the simple molecular model of Loubser and van Wyk suggests that the Fermi contact and dipolar contributions should be zero and very small, respectively, in contradiction to observation. Given that the unpaired spin density of the valence electrons does not adequately account for the hyperfine parameters, the parameters must arise from modifications of the orbitals of the N core electrons \cite{gali08}. If the N core orbitals are slightly different for different electron spin-projections (known as core polarisation), then a significant Fermi contact contribution will result due to the significant amplitudes of the core orbitals at the N nucleus. Furthermore, the Fermi contact contribution of the core electrons will be negative, as opposed to the (by definition) positive contribution of the valence electrons \cite{stoneham75}. Since the N core orbitals are expected to be principally s atomic orbital in character, the dipolar contribution of the core electrons is expected to be very small. Thus, the hyperfine parameters are dominated by the negative Fermi contact contribution of the core electrons. This conclusion was overlooked in the first measurements of the hyperfine parameters \cite{loubser77, loubser78,manson90,he93a,rabeau06} and was eventually identified in an \textit{ab initio} study \cite{gali08}. Since the conclusion was overlooked in the first measurements, the signs of the magnetic hyperfine parameters were incorrectly assigned as positive in order to be consistent with their assumed valence spin density origin. Later hyperfine measurements \cite{felton08,fuchs08,steiner10,felton09,smeltzer09} have had the benefit of the identification of core polarisation and have correctly assigned the signs of the hyperfine parameters.

\begin{figure}[hbtp]
\begin{center}
\includegraphics[width=0.45\columnwidth] {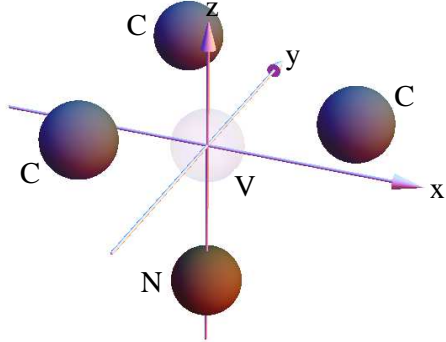}
\caption[Coordinate system of the NV centre]{Schematic of the NV centre depicting the nearest neighbour nitrogen (brown) and carbon (gray) atoms to the vacancy (transparent) as well as the adopted coordinate system (xyz). The coordinate system is defined such that the z axis is coaxial with the trigonal symmetry axis of the centre, the origin resides at the vacancy site (the centre of the defect), and the x axis passes below one of the nearest neighbour carbon atoms.}
\label{fig:fieldcoordsysreview}
\end{center}
\end{figure}

The influence of static electric $\vec{E}$, magnetic $\vec{B}$ and strain $\vec{\delta}$ fields on the NV$^-$ ground state has been observed to be well described by the addition of the following potential to the spin-Hamiltonian (\ref{eq:NVgroundstatespinHamiltonian})
\begin{eqnarray}
\hat{V}_{gs} & = & \mu_Bg_{gs}^\parallel\hat{S}_zB_z+\mu_Bg_{gs}^\perp\left(\hat{S}_xB_x+\hat{S}_yB_y\right)+\mu_Ng_N\vec{I}\cdot\vec{B} \nonumber \\
&& +d_{gs}^\parallel (E_z+\delta_z)\left[\hat{S}_z^2-S(S+1)/3\right]+
d_{gs}^\perp(E_x+\delta_x)\left(\hat{S}_y^2-\hat{S}_x^2\right) \nonumber \\
&& +d_{gs}^\perp (E_y+\delta_y)\left(\hat{S}_x\hat{S}_y+\hat{S}_y\hat{S}_x\right)
\label{eq:NVgroundstateeffectiveHamiltonianfields}
\end{eqnarray}
where $\mu_B$ is the Bohr magneton, $\mu_N$ is the nuclear magneton, $g_{gs}^\parallel$ and $g_{gs}^\perp$ are the components of the ground state electronic g-factor tensor, $g_N$ is the isotropic nuclear g-factor of $^{14}$N or $^{15}$N as required, and $d_{gs}^\parallel$ and $d_{gs}^\perp$ are the components of the ground state electric dipole moment. The cartesian components of the fields are defined with respect to the coordinate system depicted in figure \ref{fig:fieldcoordsysreview}. Note that crystal strain has been expressed as an effective electric field in the above \cite{doherty11} and that all fields are assumed approximately constant over the dimensions of the NV centre. Measured values for the ground state electronic g-factor and electric dipole components are contained in table \ref{tab:groundstateinteractionsreview}. Being of $A_2$ orbital symmetry, the ground state cannot possess a first order orbital magnetic moment or a non-axial permanent electric dipole moment. Consequently, the ground state electronic g-factor components can only differ from the isotropic free electron value $g_e\approx 2.0023$ if higher order terms in the centre's interactions with magnetic fields are of adequate magnitude or if excited electronic states with orbital magnetic moments are significantly coupled (via electronic spin-orbit and spin-spin interactions) with the ground state \cite{stoneham75}. Likewise, the ground state may only possess a permanent electric dipole moment if excited electronic states with electric dipole moments are significantly coupled (also via electronic spin-orbit and spin-spin interactions) with the ground state \cite{dolde11,doherty11b}.

\begin{table}
\caption[The electric and magnetic field parameters of the NV$^-$ ground state tabulated by reference]{\label{tab:groundstateinteractionsreview} Experimental measurements of the electronic g-factor and electric dipole components of the NV$^-$ ground state tabulated by reference.}
\begin{tabular}{lllll}
\noalign{\smallskip}
\toprule
Reference & $g_{gs}^\perp$ & $g_{gs}^\parallel$ & $d_{gs}^\perp$ & $d_{gs}^\parallel$  \\
& & & (Hz cm/V)& (Hz cm/V) \\
\midrule
Loubser \cite{loubser77, loubser78} & 2.0028$\pm0.0003$ & 2.0028$\pm0.0003$ &  &  \\
Felton \cite{felton09} & 2.0031$\pm0.0002$ & 2.0029$\pm0.0002$ &  &  \\
van Oort \cite{vanOort90} &  &  & 17$\pm$3 & 0.35$\pm$0.02 \\
\bottomrule
\end{tabular}
\end{table}

\subsection{The low temperature fine structure of the NV$^-$ $^3E$ level}

\begin{figure}[hbtp]
\begin{center}
\mbox{
\subfigure[]{\includegraphics[width=0.45\columnwidth] {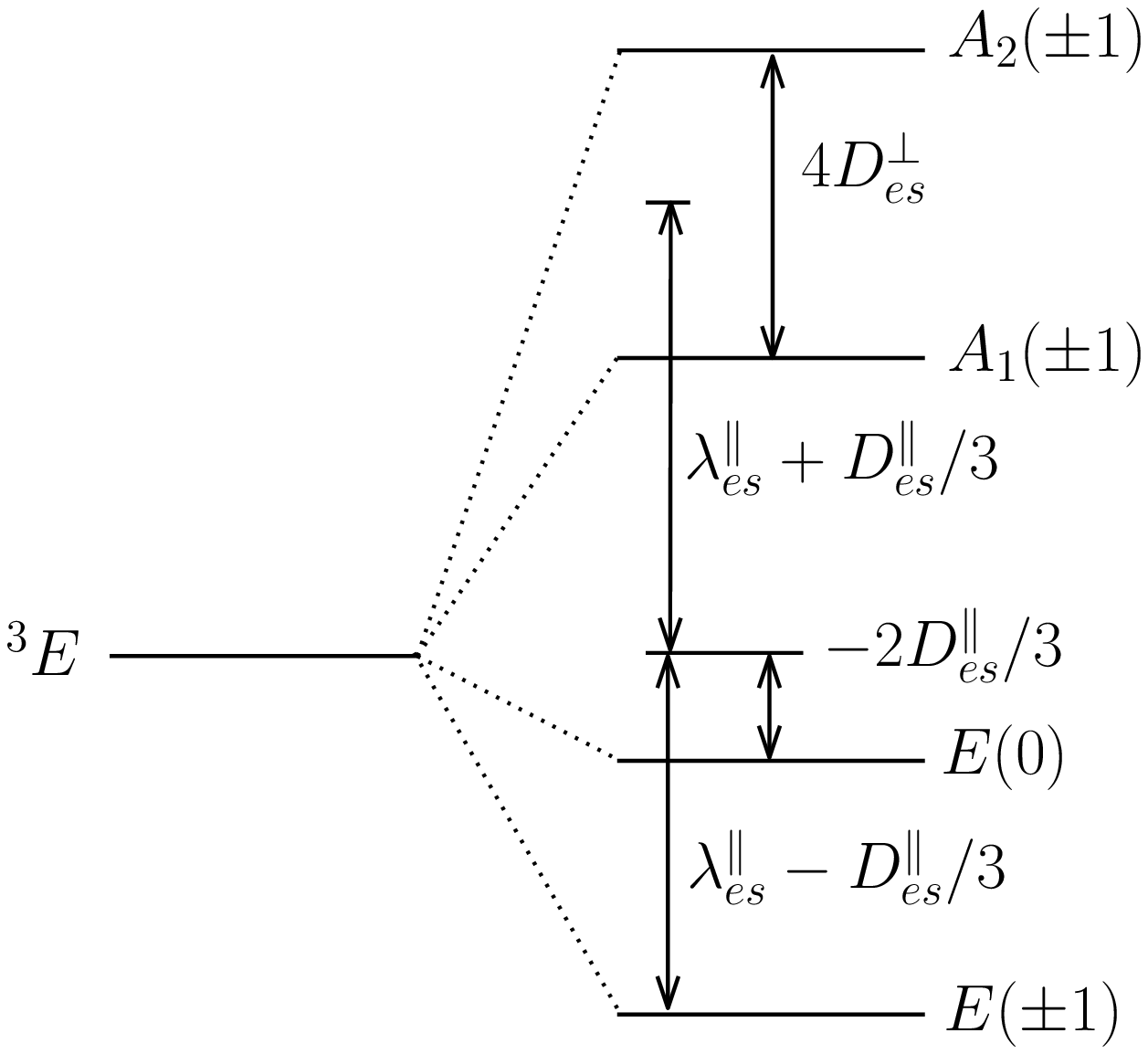}}
\subfigure[]{\includegraphics[width=0.55\columnwidth] {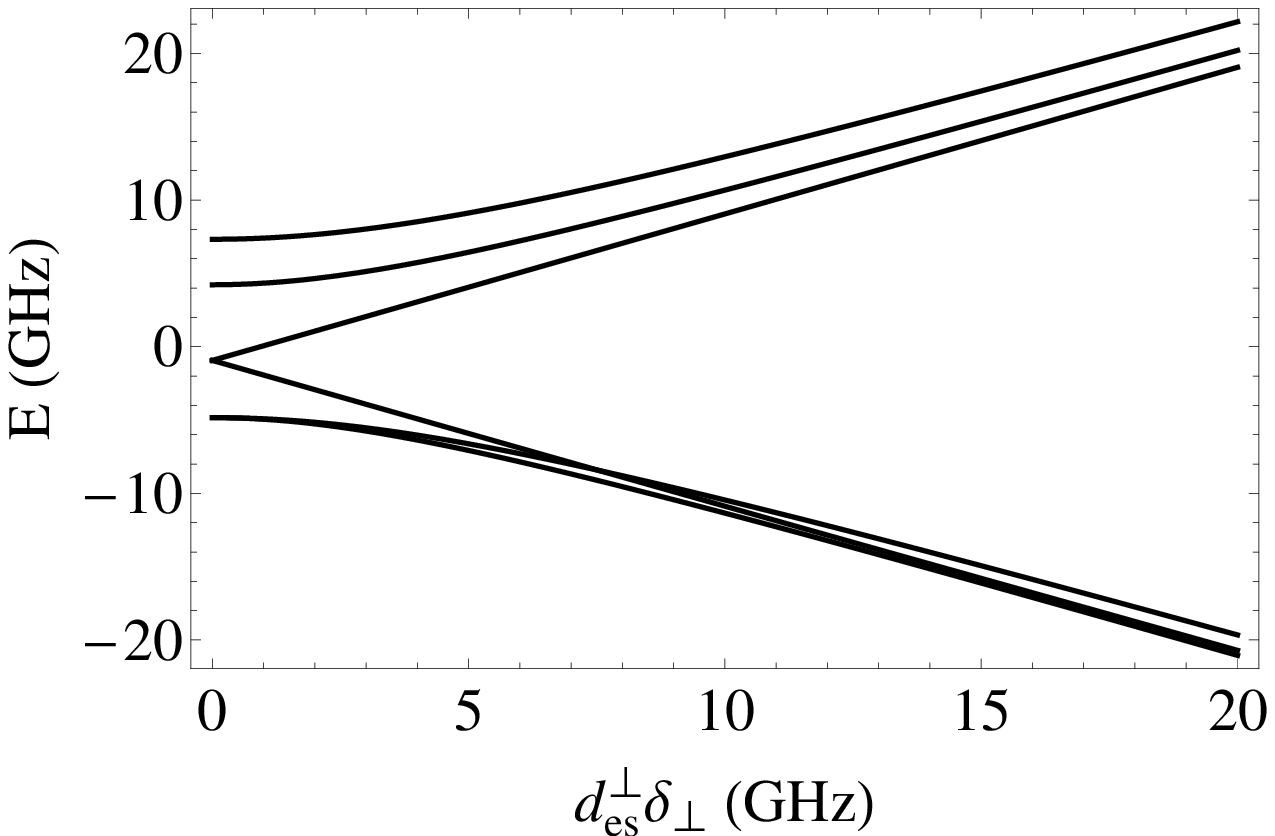}}}
\caption[NV$^-$ $^3E$ low temperature fine structure and its dependence on crystal strain]{(a) Low temperature zero field fine structure of the NV$^-$ $^3E$ excited state \cite{doherty11}. The fine structure splittings are indicated in terms of the parameters of the effective Hamiltonian (\ref{eq:NVexcitedstateeffectiveHamiltonian}). Spin-orbit symmetries and $m_s=0$, $\pm1$ spin-projections of the fine structure states are as labelled on the right hand side. (b) The dependence of the fine structure on the transverse strain energy $d_{es}^\perp\delta_\perp = d_{es}^\perp\sqrt{\delta_x^2+\delta_y^2}$ as modeled by (\ref{eq:NVexcitedstateeffectiveHamiltonian}) and (\ref{eq:NVexcitedstateeffectiveHamiltonianfields}). The model parameters used to produce the plot were those of Batalov et al \cite{batalov09} (refer to table \ref{tab:lowtempexcitedstatespinparametersreview}).}
\label{fig:NVexcitedstatefinestructurereview}
\end{center}
\end{figure}

The low temperature fine structure of the NV$^-$ $^3E$ excited state (refer to figure \ref{fig:NVexcitedstatefinestructurereview}) was only recently observed directly \cite{batalov09,tamarat08}. The direct observations were made using photoexcitation spectroscopy of spectrally stable single centres that exhibited narrow optical linewidths. Direct observation of the fine structure proved elusive in earlier spectroscopic attempts due to the confounding effects of strain broadening in ensembles, spectral diffusion, spin mixing in the strain-split fine structure and optical spin-polarisation. A number of pioneering works employed spectral hole burning \cite{reddy87,santori06,redman92,manson94}, narrow band ODMR \cite{vanOort91} and four-wave mixing techniques \cite{lenef96b,rand94} to probe the strain broadened optical ZPL of NV$^-$ ensembles. Some of these works \cite{santori06,lenef96b,redman92,manson94,rand94} gave aspects of the fine structure that were eventually observed directly. Notably, qualitatively correct models of the $^3E$ fine structure existed very early \cite{reddy87,lenef96}, but without the direct observation of the structure, it was not possible to correctly assign the measured features to the splittings predicted by the models.

The first low temperature spectroscopy studies of single NV$^-$ centres \cite{drab99,jelezko02,jelezko01} were also unable to directly observe the fine structure of $^3E$ because the centres investigated had broad optical ZPLs due to poor spectral stability. The single centres selected for the studies also typically had low strain \cite{jelezko01}, which due to spin mixing in the lower fine structure branch, ensured that the lower branch did not have adequate fluorescence to be visible \cite{tamarat08}. The combination of optical excitation and microwaves tuned to the ground state magnetic resonance, enabled the sufficient increase of the fluorescence of the lower branch fine structure levels for them to be observed in the later single centre studies \cite{batalov09,tamarat08}. This increase in the fluorescence of the lower branch occurred due to the microwaves maintaining steady state population in all spin-projections. The poor spectral stability of the first single centres to be studied is believed to be due to their formation in type Ib diamond, which contains significant concentrations of electron donors (such as N$_s$). The donors undergo photoionisation during the optical excitation of NV$^-$ and produce fluctuating electric fields in the vicinity of the NV$^-$ centre that broaden the optical ZPL \cite{jelezko02,jelezko01}. The spectrally stable centres used to directly observe the fine structure were consequently formed in type IIa diamond, which has very small concentrations of donors. Thus, it can be seen that the achievement of direct observation of the NV$^-$ optical fine structure required several advances in the understanding of NV$^-$ as well as spectroscopy and fabrication techniques.

Given the assignment of the energy ordering of the ground state spin sub-levels as well as the direct observation of the strain and electric field dependence of the low temperature $^3E$ fine structure \cite{batalov09,tamarat08}, it was straight forward to assign the spin-projections of the $^3E$ fine structure levels using optical spin selection rules (refer to figure \ref{fig:NVexcitedstatefinestructurereview}). The double degeneracy of the lower two levels imply that they are of $E$ spin-orbit symmetry, which is in keeping with the predictions of the molecular model \cite{doherty11,manson06,reddy87,lenef96,manson07,maze11}. Whilst the molecular model predicted the $A_1$ and $A_2$ spin-orbit symmetries of the remaining non-degenerate levels, their energy ordering had to be interpreted from experiment. The ordering of the upper levels has very recently been assigned as depicted in figure \ref{fig:NVexcitedstatefinestructurereview} using lifetime arguments \cite{togan10}. These lifetime arguments involve non-radiative processes and will be discussed further at a later stage in this review.

The low temperature $^3E$ fine structure can be described by the following effective Hamiltonian
\begin{eqnarray}
\hat{H}_{es}^{LT} & = & D_{es}^\parallel\left[\hat{S}_z^2-S(S+1)/3\right]-\lambda_{es}^\parallel\hat{\sigma}_y\otimes \hat{S}_z \nonumber\\
&& +D_{es}^\perp\left[\hat{\sigma}_z\otimes(\hat{S}_y^2-\hat{S}_x^2)-\hat{\sigma}_x\otimes(\hat{S}_y\hat{S}_x+\hat{S}_x\hat{S}_y)\right] \nonumber \\
&&+\lambda_{es}^\perp\left[\hat{\sigma}_z\otimes(\hat{S}_x\hat{S}_z+\hat{S}_z\hat{S}_x)-\hat{\sigma}_x\otimes(\hat{S}_y\hat{S}_z+\hat{S}_z\hat{S}_y)\right]
\label{eq:NVexcitedstateeffectiveHamiltonian}
\end{eqnarray}
where $\hat{\sigma}_x$, $\hat{\sigma}_y$ and $\hat{\sigma}_z$ are the standard Pauli matrices that represent the fine structure orbital operators in the basis of orbital states $\{\ket{^3E_x},\ket{^3E_y}\}$ associated with the $^3E$ level, and $\hat{S}_x$, $\hat{S}_y$ and $\hat{S}_z$ are the $S=1$ spin operators. The values of the fine structure parameters $D_{es}^\parallel$, $D_{es}^\perp$, $\lambda_{es}^\parallel$ and $\lambda_{es}^\perp$ as obtained from the direct observations \cite{batalov09,tamarat08} of the low temperature fine structure are tabulated in table \ref{tab:lowtempexcitedstatespinparametersreview}.

\begin{table}
\caption[The parameters of the low temperature NV$^-$ $^3E$ fine structure tabulated by reference]{\label{tab:lowtempexcitedstatespinparametersreview} Experimental measurements of the parameters of the low temperature NV$^-$ $^3E$ fine structure tabulated by reference. Uncertainty ranges were not provided by the source references.}
\begin{tabular}{lllll}
\noalign{\smallskip}
\toprule
Reference & $D_{es}^\parallel$ (GHz) & $D_{es}^\perp$ (GHz) & $\lambda_{es}^\parallel$ (GHz) & $\lambda_{es}^\perp$ (GHz)  \\
\midrule
Tamarat \cite{tamarat08}  & 1 & 0.5 & 4.4 & 0.2$/\sqrt{2}$  \\
Batalov \cite{batalov09}  & 1.42 & 1.55/2 & 5.3 & 0.2$/\sqrt{2}$  \\
\bottomrule
\end{tabular}
\end{table}

Applications of the molecular model \cite{doherty11,maze11,batalov09} clearly indicate that the physical origins of $D_{es}^\parallel$ and $D_{es}^\perp$ are spin-spin interactions and the origin of $\lambda_{es}^\parallel$ is axial spin-orbit interaction. The origin of $\lambda_{es}^\perp$ has, however, experienced a degree of contention. Manson et al \cite{manson06,manson07} asserted that $\lambda_{es}^\perp$ arises from non-axial spin-orbit interaction, whereas Lenef and Rand asserted that it could not arise from spin-orbit interaction using arguments based upon the Hermitian properties of the spin-orbit operator. It was later shown separately by Doherty et al \cite{doherty11} and Maze et al \cite{maze11} that $\lambda_{es}^\perp$ has contributions from spin-spin interaction. Furthermore, Doherty et al extended Lenef and Rand's argument to definitively prove that spin-orbit interaction does not contribute to $\lambda_{es}^\perp$. However, the contention has not yet been completely resolved because two-electron spin-orbit interactions (spin-own-orbit and spin-other-orbit) have not yet been considered in the molecular model and are likely to also contribute to $\lambda_{es}^\perp$.

The influence of static electric $\vec{E}$, magnetic $\vec{B}$ and strain $\vec{\delta}$ fields on the low temperature $^3E$ fine structure has been observed to be well described by the addition of the following potential to the effective Hamiltonian (\ref{eq:NVexcitedstateeffectiveHamiltonian})
\begin{eqnarray}
\hat{V}_{es}^{LT} & = & d_{es}^\parallel (E_z+\delta_z)+
d_{es}^\perp(E_x+\delta_x)\hat{\sigma}_z-d_{es}^\perp (E_y+\delta_y)\hat{\sigma}_x \nonumber \\
&&+\mu_B(l_{es}^\parallel\hat{\sigma}_y+g_{es}^\parallel\hat{S}_z)B_z+\mu_Bg_{es}^\perp\left(\hat{S}_xB_x+\hat{S}_yB_y\right)
\label{eq:NVexcitedstateeffectiveHamiltonianfields}
\end{eqnarray}
where $l_{es}^\parallel$ is the $^3E$ orbital magnetic moment, $g_{es}^\parallel$ and $g_{es}^\perp$ are the components of the $^3E$ electronic g-factor tensor, and $d_{es}^\parallel$ and $d_{es}^\perp$ are the components of the $^3E$ electronic electric dipole moment. Out of all of the interaction parameters, only $l_{es}^\parallel$ has been measured directly at low temperature. The magnetic circular dichroism measurements of Reddy et al \cite{reddy87} and Rogers et al \cite{rogers09} both obtained $l_{es}^\parallel=0.1\pm0.01$ at $\sim$1.5 K. The low temperature $^3E$ Stark effect measurements of Tamarat et al \cite{tamarat08,tamarat06} indicate that the $^3E$ electric dipole moment components are of the order $\sim$6 kHz m/V. However, as the direction of the electric field was not known in their experiments, the individual components of the $^3E$ electric dipole moment could not be distinguished. The components of the $^3E$ electronic g-factor tensor have only been measured at room temperature and were found to be $g_{es}^\perp\sim g_{es}^\parallel\sim g_{es}^{RT}=2.01\pm0.01$ \cite{fuchs08,neumann09}. Due to the presence of the Jahn-Teller effect in $^3E$, it is likely that the low temperature g-factor tensor and electric dipole components differ from those measured at room temperature \cite{fu09}. A detailed analysis of the influence of the Jahn-Teller effect on the $^3E$ orbital magnetic moment, electric dipole moment and g-factor tensor is yet to be presented in the literature.

\subsection{The room temperature fine structure of the NV$^-$ $^3E$ level}

The first observation of the room temperature $^3E$ fine structure was made by Epstein et al \cite{epstein05} in their ODMR study of single centres and ensembles. Using precise magnetic field alignment, Epstein et al detected a LAC at an axial magnetic field of $\sim$500 G, which they assigned to $^3E$. This assignment was later confirmed by direct ODMR detection of the room temperature fine structure using single centres  \cite{fuchs08,neumann09}. These direct observations were made at similar times to the first direct observations of the low temperature fine structure \cite{batalov09,tamarat08}. However, the observations \cite{neumann08,fuchs08} incorrectly identified the room temperature fine structure as the upper branch of the low temperature fine structure. The misidentification was justifiably made for two reasons. The first, was that there were only three fine structure levels observed at room temperature, which is consistent with one of the low temperature branches. The second reason was that the upper branch was known to undergo bright spin-conserving transitions in contrast to the lower branch. The lower branch was thus concluded to be unobservable at room temperature due to the presence of spin-flip transitions and increased non-radiative decay at room temperature. The room temperature fine structure was correctly identified as being the result of phonon mediated averaging of both low temperature fine structure branches by Rogers et al \cite{rogers09}.

Rogers et al observed the temperature dependence of their CW ODMR detection of the excited state LAC in a NV$^-$ ensemble. For temperatures down to 25 K, there existed a single LAC at $\sim$ 510 G that was consistent with the room temperature fine structure observed in the single centre studies \cite{fuchs08,neumann09}. Below 25 K, Rogers et al observed an additional LAC feature at $\sim$ 250 G and that this feature varied with applied stress. By modeling the strain dependence of the low temperature fine structure, Rogers et al showed that the behaviour of the LAC features at $\sim 250$ G and $\sim$ 510 G in response to applied stress was consistent with the low temperature fine structure. Hence, Rogers et al had demonstrated the transition between the low and room temperature $^3E$ fine structures. In order to explain their observation, Rogers et al proposed that phonon transitions between the low temperature fine structure levels resulted in an average fine structure that varied with temperature, such that at low temperature the averaging effect was minimal and at room temperature the averaging resulted in the observed room temperature fine structure. Specifically, the averaging process is believed to occur via the redistribution of population optically excited to a given $^3E$ low temperature fine structure level to all the other fine structure levels with the same spin-projection through spin-conserving phonon transitions. The degree of population redistribution that occurs within the lifetime of $^3E$ depends on the rates of the phonon transitions, which in turn depend on temperature. At sufficiently high temperature, the phonon transition rates will greatly exceed the $^3E$ decay rate and the population will be equally distributed between the low temperature fine structure levels of the occupied spin-projection. Therefore, at sufficiently high temperatures the energy of the centre's state is the average of the low temperature fine structure levels of the occupied spin-projection. The equivalence of the average of the low temperature fine structure and the room temperature fine structure was confirmed by the single centre experiments of Batalov et al \cite{batalov09}. A detailed model of the averaging process is yet to be presented and a thorough analysis of the transition between the low and room temperature fine structures remains an outstanding issue to be resolved.

The observed room temperature $^3E$ fine and hyperfine structures can be described by a spin-Hamiltonian that is equivalent to that used to describe the $^3A_2$ ground state
\begin{eqnarray}
\hat{H}_{es}^{RT} & = & D_{es}^\parallel\left[\hat{S}_z^2-S(S+1)/3\right]+A_{es}^\parallel \hat{S}_z\hat{I}_z+A_{es}^\perp\left[\hat{S}_x\hat{I}_x+\hat{S}_y\hat{I}_y\right] \nonumber\\
&& +P_{es}\left[\hat{I}_z^2-I(I+1)/3\right]
\label{eq:NVexcitedstateeffectiveHamiltonianroomtemp}
\end{eqnarray}
where the fine and hyperfine parameters are tabulated in table \ref{tab:roomtemperatureexcitedstatespinparametersreview}. Note that the hyperfine structure has only been observed at room temperature and no model has yet been developed for the low temperature hyperfine structure. Comparing the above with $\hat{H}_{es}^{LT}$ (\ref{eq:NVexcitedstateeffectiveHamiltonian}), it is evident that the averaging processes can be described by a partial trace over the orbital states $\hat{H}_{es}^{RT}=\frac{1}{2}\mathrm{tr}_{\hat{\sigma}}\{\hat{H}_{es}^{LT}\}$, which is conceptually consistent.

The influence of static electric $\vec{E}$, magnetic $\vec{B}$ and strain $\vec{\delta}$ fields on the room temperature $^3E$ fine structure has been observed to be well described by the addition of the following potential to the above spin-Hamiltonian (\ref{eq:NVexcitedstateeffectiveHamiltonianroomtemp})
\begin{eqnarray}
\hat{V}_{es}^{RT} & = & d_{es}^\parallel (E_z+\delta_z)+g_{es}^{RT}\vec{S}\cdot\vec{B}+\xi(\hat{S}_y^2-\hat{S}_x^2)
\label{eq:NVexcitedstateeffectiveHamiltonianfieldsroomtemp}
\end{eqnarray}
where $\xi$ is a parameter that has been related to strain \cite{fuchs08,neumann09,batalov09}, and as previously discussed, $g_{es}^{RT}=2.01\pm01$ \cite{fuchs08,neumann09} is the measured isotropic room temperature electronic g-factor of $^3E$. Comparing the above with the low temperature potential (\ref{eq:NVexcitedstateeffectiveHamiltonianfields}) and extending the notion of the room temperature averaging being consistent with a partial trace over the orbital states, it is not clear where the strain related `$\xi$' term originates from, since all of the non-axial strain related terms in (\ref{eq:NVexcitedstateeffectiveHamiltonianfields}) are traceless. Fuchs et al \cite{fuchs08} measured $\xi\sim70\pm30$ MHz for one of the centres they investigated and related the parameter to strain because the strain was very large in that particular centre (as indicated by a ground state strain splitting of $\sim$6 MHz). Neumann et al \cite{neumann09} only observed $\xi>0$ for $<20\%$  of the centres they surveyed in bulk diamond. However, Neumann et al did observe a splitting consistent with the `$\xi$' term in a single centre within a nanodiamond. As internal strains are typically large in nanodiamonds, this observation is consistent with Fuchs et al's conclusion that $\xi$ is related to strain. Consequently, it appears that the averaging process is not complete for centres with very large strains, leading to an additional splitting being observed in their room temperature fine structures that is otherwise absent in centres with smaller strains. Hence, the objective of any attempt at a detailed model of the averaging process should be the description of the conditions required to observe the strain related splitting of the room temperature $^3E$ fine structure.

\begin{table}
\caption[The parameters of the room temperature NV$^-$ $^3E$ fine and hyperfine structure tabulated by reference and nitrogen isotope]{\label{tab:roomtemperatureexcitedstatespinparametersreview} Experimental measurements and \textit{ab initio} calculations of the room temperature NV$^-$ $^3E$ fine and hyperfine structure parameters tabulated by reference and nitrogen isotope. Measurement uncertainty ranges are indicated if provided by source reference. Note that the $^{15}$N nucleus does not have a quadrupole moment.}
\begin{tabular}{llllll}
\noalign{\smallskip}
\toprule
Reference & Iso. & $D_{es}^\parallel$ (MHz) & $A_{es}^\perp$ (MHz) & $A_{es}^\parallel$ (MHz) & $P_{es}$\\
\midrule
Steiner$^e$ \cite{steiner10} & $^{14}$N & 1420 & $(\pm)$40 & $(\pm)$40 &   \\
Fuchs$^e$ \cite{fuchs08} & $^{15}$N & 1425$\pm3$ & $(\pm)$61$\pm6$ & $(\pm)$61$\pm6$ & n/a  \\
Neumann$^e$ \cite{neumann09} & $^{15}$N & 1423$\pm10$ &  &  & n/a  \\
Gali$^a$ \cite{gali09b} & $^{15}$N &  & -39.2 & -57.8 & n/a \\
\bottomrule\noalign{\smallskip}
\multicolumn{2}{l}{$^e$Experiment, $^a$\textit{ab initio}}
\end{tabular}
\end{table}

Comparing the room temperature $^3E$ magnetic hyperfine parameters to those of the $^3A_2$ ground state (refer to table \ref{tab:groundstatespinparametersreview}), it is evident that the magnetic hyperfine interaction is much larger in $^3E$. This difference can be explained by once again returning to the simple molecular model of Loubser and van Wyk \cite{loubser77, loubser78}. The unpaired spin density of $^3E$ arises from the unpaired electrons occupying both the $A_1$ ($a_1$) and $E$ ($e_x$, $e_y$) MOs in the first excited configuration $a_1e^3$. Unlike the $E$ MOs, the $A_1$ MO has a contribution from the dangling N sp$^3$ atomic orbital and thus contributes to both the Fermi contact and dipolar components of the magnetic hyperfine interaction \cite{fuchs08}. Furthermore, since these contributions arise from unpaired valence electrons and not core polarisation, the contributions will be opposite in sign to the core polarisation contributions in the ground state \cite{gali09b}. Thus, the magnetic hyperfine parameters are expected to change sign, as supported by the \textit{ab initio} calculation of Gali \cite{gali09b}.

\subsection{The fine structure of the NV$^0$ $^4A_2$ level}

At this stage, it will be instructive to compare the fine and hyperfine parameters of NV$^0$ with those of NV$^-$, since they are expected to have common origins. By noting the change to $S=\frac{3}{2}$ operators, the $^4A_2$ metastable excited state of NV$^0$ can be described by an equivalent spin-Hamiltonian to that of the $^3A_2$ ground state (and the $^3E$ excited state at room temperature) of NV$^-$ (\ref{eq:NVgroundstatespinHamiltonian}). The NV$^0$ parameters as measured by Felton et al \cite{felton08} and calculated by Gali \cite{gali09} are contained in table \ref{tab:NV0spinHamiltonianparameters}. It is clear that the NV$^0$ $^4A_2$ hyperfine parameters are similar to the room temperature NV$^-$ $^3E$ parameters. This can be explained by the fact that the $A_1$ MO is unpaired for both $^4A_2$ $(a_1e^2)$ and $^3E$ $(a_1e^3)$, which implies that similar unpaired spin density will occur at the N nucleus for both $^4A_2$ and $^3E$ due to the $A_1$ MO being non-zero at the N nucleus in contrast to the $E$ MOs. There also appears to be a strong relationship between the electronic g-factor components of NV$^0$ $^4A_2$ (measured to be $g_{NV^0}^\perp = 2.0035(2)$ and $g_{NV^0}^\parallel = 2.0029(2)$ \cite{felton08}) and NV$^-$ $^3A_2$ (refer to table \ref{tab:groundstateinteractionsreview}). This is likely due to the common orbital symmetry of the two states leading to similar origins of the shifts of the components away from the free electron value, but no analysis of the shifts has been performed to confirm.

\begin{table}
\caption[The parameters of the NV$^0$ $^4A_2$ fine and hyperfine structures tabulated by reference and nitrogen isotope]{\label{tab:NV0spinHamiltonianparameters} Experimental measurements and \textit{ab initio} calculations of the NV$^0$ $^4A_2$ fine and hyperfine structure parameters tabulated by reference and nitrogen isotope. Note that the $^{15}$N nucleus does not have a quadrupole moment.}
\begin{tabular}{llllll}
\noalign{\smallskip}
\toprule
Reference & Iso. & $D_{NV^0}^\parallel$ (MHz) & $A_{NV^0}^\perp$ (MHz) & $A_{NV^0}^\parallel$ (MHz) & $P_{NV^0}$\\
\midrule
Felton$^e$ \cite{felton08} & $^{15}$N & 1685$\pm5$ & -23.8$\pm0.3$ & -35.7$\pm0.3$ & n/a  \\
Gali$^a$ \cite{gali09} & $^{15}$N &  & -23.4 & -39.0 & n/a \\
\bottomrule\noalign{\smallskip}
\multicolumn{3}{l}{$^e$Experiment, $^a$\textit{ab initio}}
\end{tabular}
\end{table}

\subsection{The NV$^-$ $^1E$ and $^1A_1$ levels}

Returning to NV$^-$, the existence and energy ordering of the intermediate singlet states ($^1A_1$ and $^1E$) that are predicted by Loubser's and van Wyk's simple molecular model \cite{loubser77, loubser78} has been a constant source of contention throughout the study of NV$^-$. The early contention was fueled by the interpretation of very different storage phenomena in the centre's optical dynamics as evidence of a metastable state. In their four wave mixing study, Redman et al \cite{redman91} attributed a very long storage of $\sim0.3$ s to a metastable state in the optical cycle that they speculated to be the $^1A_1$ singlet (ignoring the $^1E$ singlet). However, their observation is inconsistent with more recent measurements of the optical cycle and is likely to be related to spin-lattice relaxation and not a metastable state \cite{glasbeek92}. Another interpretation of a metastable state in the centre's optical dynamics was based upon observations of fluorescence intensity fluctuations in the first single centre study \cite{gruber97}. The presence of a metastable state was also suggested by the observations of `bunching' in the fluorescence autocorrelation functions of single centres (refer to figure \ref{fig:introsinglecentredetection}) in the second single centre study \cite{drab99}.

The models of the centre's optical dynamics that were employed in the early single centre studies only contained three levels (corresponding to the  $^3A_2$ and $^3E$ states and a metastable state assumed to be $^1A_1$) \cite{drab99,brouri00,kurt00,jelezko01,beveratos00}. In order to explain observations of biexponential radiative decay and temperature dependent fluorescence saturation, the three level models required the metastable state to repopulate $^3E$ with a rate that has an Arrhenius temperature dependence and a 37 meV activation energy \cite{drab99}. The energy of the metastable state (assumed $^1A_1$) was consequently concluded to be within 37 meV of $^3E$ \cite{drab99}. The subsequent use of five \cite{manson06,nizovtsev03} and seven \cite{nizovtsev03b} level models nullified these conclusions regarding the metastable state because these more sophisticated models could explain the observations (and a number of other aspects of the centre's fluorescence) by correctly accounting for spin-dependent fluorescence and temperature dependent decay from the metastable state to the ground state.

In order to describe the intensity dependence of the observed fluorescence autocorrelation functions, the three level models also required the metastable state to optically repopulate the $^3E$ with a rate that is linear in optical intensity \cite{beveratos00}. This is now understood as another demonstration of the inadequacy of the three level model to describe the centre's optical dynamics and not a property of an intermediate singlet state of NV$^-$. The intensity dependence of the fluorescence autocorrelation function is intead related to the photoionisation process to NV$^0$ \cite{manson06,manson07}. Indeed, this is an example of how the assignment of the metastable state to one of the singlet states was confounded by the presence of two storage processes in the centre's dynamics \cite{manson06,manson07}: photoionisation to NV$^0$ that results in long lived storage, and non-radiative decay via the intermediate singlet states that results in relatively short lived storage. Consequently, there have been a great variety of lifetimes associated with the metastable state under different optical excitation conditions that are seemingly at odds with the identification of one of the singlet states as the metastable state. The models of the centre's optical dynamics and the deconvolution of the two storage processes will be discussed further at a later stage of the review.

\begin{figure}[hbtp]
\begin{center}
\includegraphics[width=0.7\columnwidth] {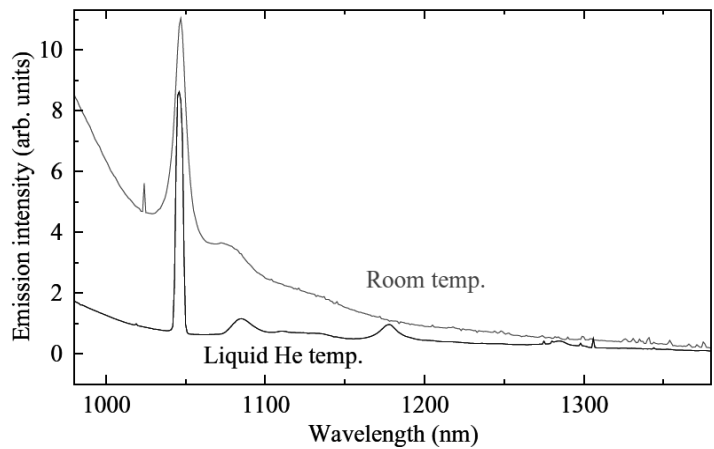}
\caption[The NV$^-$ infrared luminescence band and the change of its ZPL under applied magnetic fields]{The NV$^-$ infrared luminescence band at two different temperatures, including its ZPL at 1.190 eV (1043 nm) and vibronic band extending towards lower energy, as measured using a high density NV$^-$ ensemble \cite{rogers08}. Note that the liquid He trace has been divided by a factor of five in order to compare directly to the room temperature trace. The tail of the NV$^-$ optical luminescence band is evident underneath the infrared band.}
\label{fig:infraredbandreview}
\end{center}
\end{figure}

The existence of the intermediate singlet states ($^1A_1$ and $^1E$) was unambiguously confirmed for the first time by Rogers et al \cite{rogers08} through their observation of the 1.190 eV infrared band under optical excitation (see figure \ref{fig:infraredbandreview}). Rogers et al used uniaxial stress to demonstrate that the infrared transition occurred between electronic states of $E$ and $A_1$ orbital symmetry. Evidence that the transition occurred between spin singlets was obtained through the absence of any Zeeman splitting or broadening of the infrared ZPL. However, in the presence of a static magnetic field, the infrared fluorescence increased in correlation with a decrease in the optical fluorescence. These changes in fluorescence are consistent with the increased non-radiative population of the upper intermediate singlet state due to the mixing of the triplet spin-projections, thereby providing evidence that the non-radiative decay suspected in the centre's optical dynamics was indeed occurring via the intermediate singlet states. Rogers et al established that the upper singlet state had a very short lifetime (estimated $\ll30$ ns) through interpreting the transient response of the infrared fluorescence under optical excitation. Rogers et al thus concluded that the lower singlet state was the longer lived storage state in the non-radiative decay path. Using a model of the centre's optical dynamics, Rogers et al further concluded that the infrared fluorescence was less than expected and speculated that the infrared emission had significant competition from a non-radiative decay path between the singlets. The measurements performed by Rogers et al were later extended by Acosta et al \cite{acosta09,acosta10b}, who successfully measured the lifetime of the upper singlet to be 0.9(5) ns and the lifetime of the lower singlet, which was much longer and exhibited a significant temperature dependence, ranging from 462 ns at 4.4 K to 142 ns at 450 K.

In spite of the comprehensive studies of the infrared band performed by Rogers et al and Acosta et al, the energy ordering of the singlet states could not be ascertained because the band was only observed in emission. Rogers et al speculated that $^1A_1$ was lower in energy than $^1E$ based upon the assumption that spin-polarisation occurred via the emission of only symmetric ($A_1$) phonons. This spin-polarisation mechanism requires the $^1A_1$ to be lowest so that the $m_s=0$ spin sub-level of the $^3A_2$ ground state is preferentially populated due to its $A_1$ spin-orbit symmetry \cite{manson06}. Indeed, this is still the generally assumed spin-polarisation mechanism, even in light of the recent conclusive theoretical and experimental evidence that the ordering of the singlets is the opposite to that speculated by Rogers et al. The theoretical evidence for the now accepted energy ordering of the $^1E$ singlet lower than the $^1A_1$ singlet, was provided by a combination of improved \textit{ab initio} techniques \cite{delaney10,ma10} and a detailed application of the molecular model and electrostatic identities \cite{doherty11}. The experimental evidence was provided in a follow up study performed by Manson et al \cite{manson10}, who observed shifts of the infrared ZPL emission under uniaxial stress that were indicative of a LAC between vibronic levels of $^1E$, and were only consistent with $^1E$ being the lower energy singlet. Note that a recent \textit{ab initio} calculation by Ma et al \cite{ma10} added an additional element to the contention surrounding the intermediate singlet states by predicting the presence of a third singlet ($^1E^\prime$) between the triplet states and that the additional $E$ symmetric singlet was actually the upper singlet in the infrared transition (i.e. predicted a $^1E^\prime\rightarrow$$^1A_1$ infrared transition). The prediction of this calculation is, however, in direct conflict with the conclusive theoretical and experimental evidence just discussed.

The observation of a vibronic LAC by Manson et al is also direct evidence of the Jahn-Teller effect in $^1E$ \cite{manson10}. The Jahn-Teller effect in $^1E$ and the vibronic bandshape of the infrared transition is yet to be analysed in any detail and is not well understood. Aspects of the response of the infrared transition to magnetic and strain fields are also not well understood. The aspects are the origins of the $^1E$ strain splitting and the absence of a magnetic splitting arising from the expected orbital magnetic moment of $^1E$. The simplest version of the molecular model predicts that there should be no strain splitting of $^1E$. Doherty et al \cite{doherty11}, however, applied the molecular model to a higher degree of complexity and showed that the observed strain splitting could arise from the exchange coupling of $^1E$ and the higher energy singlet $^1E^\prime$. The relationships between the strain splittings of $^1E$ and $^3E$ (at low temperature) and their connections with the optical and infrared vibronic bands have not been fully explored \cite{manson10}. Continuing with the convention employed by Davies and Hamer \cite{davies76} to describe the stress response of the optical ZPL, the stress parameters $A1$, $A2$, $B$ and $C$ of the optical and infrared ZPLs are contained in table \ref{tab:stressparametersreview}. The stress parameters denote the shift/splitting of the respective ZPLs due to different components of applied stress. The molecular model also predicts that $^1E$ should possess an orbital magnetic moment equivalent to that of $^3E$ that splits the infrared ZPL in the presence of a magnetic field \cite{doherty11}. However, no comparable splitting has been observed. Given the evidence of the Jahn-Teller effect in $^1E$ via the vibronic LAC \cite{manson10}, the most likely explanation for the absent orbital magnetic moment is the Jahn-Teller quenching of orbital interactions \cite{fu09}. Clearly, there is a great deal of analysis to be performed on the infrared vibronic band and its interactions with applied fields. The connections between the optical and infrared vibronic bands and the Jahn-Teller effect will be addressed more thoroughly in the next section.

\begin{table}
\caption[The stress response parameters of the optical and infrared ZPLs of NV$^-$]{\label{tab:stressparametersreview} The stress response parameters of the optical and infrared ZPLs of NV$^-$ \cite{manson10}. The $A1$ and $A2$ parameters correspond to shifts of the ZPLs due to two different symmetric ($A_1$) distortions of the lattice under stress. The $B$ and $C$ parameters correspond to splittings of the ZPLs due to two different non-symmetric ($E$) distortions of the lattice under stress. Each parameter is in units of meV/ GPa.}
\begin{tabular}{lllll}
\noalign{\smallskip}
\toprule
ZPL &  $A1$  & $A2$ & $B$  & $C$ \\
\midrule
Optical 1.945 eV  & 1.53 & -3.91 & 0.987 & 1.77 \\
Infrared 1.190 eV & 0.484 & -0.378 & 1.22 & 0.693 \\
\bottomrule
\end{tabular}
\end{table}

The observation of the NV$^-$ infrared band established the energy separation of the intermediate singlet states, however, it did not provide the energies of the singlet states relative to the triplet states. As it currently stands, the singlet-triplet energy separations are experimentally unknown. Aspects of the centre's optical dynamics provide qualitative indications of the singlet energies. The rapid non-radiative depopulation of the $m_s=\pm1$ sub-levels of $^3E$, which is of the same order of magnitude as the optical spontaneous decay rate \cite{manson06}, suggests that $^1A_1$ is energetically close to $^3E$ in order for such rapid non-radiative transitions to occur. The comparable energy of $^1A_1$ to $^3E$ implies that $^1E$ is roughly $\sim$0.7 eV above the ground state, which is itself consistent with the observed relatively long lifetime of $^1E$. For more quantitative estimates of the intermediate singlet energies, one must turn to \textit{ab initio} calculations (refer to table \ref{tab:abinitioelectronicenergies}). However, as clearly evident in table \ref{tab:abinitioelectronicenergies}, the \textit{ab initio} calculations differ greatly in their estimates of the singlet energies. Apart from the differences in calculation method, geometry and size, the inconsistencies of the calculations have their origins in the limitations of current \textit{ab initio} methods.

\begin{table}
\caption[Summary of \textit{ab initio} calculations of the NV$^-$ electronic structure]{\label{tab:abinitioelectronicenergies} Summary of \textit{ab initio} calculations of the NV$^-$ electronic structure. The Method column details the calculation method (DFT: Density Functional Theory, TD-DFT: Time Dependent DFT, CI: Configuration Interaction, CASSCF: Complete Active Space Self Consistent Field, DFT-MBPT: Combined DFT and Many Body Perturbation Theory) and exchange-correlation functional for DFT methods (LSDA: Linear Spin Density Approximation, BP: Becke-Perdew, PBE: Perdew-Burke-Ernzerhof, GGA: Generalised Gradient Approximation, HSE - Hartree-Fock hybrid). The Size column details the number of carbon atoms (clusters) or total number of atoms (supercells) in the geometries employed. All energies (except for those denoted) were calculated in the ground state nuclear equilibrium geometry.}
\begin{tabular}{llllll}
\noalign{\smallskip}
\toprule
State & Reference & Method & Geometry & Size & Energy (eV) \\
\midrule
$^1E$ & Goss \cite{goss96} & DFT-LSDA & Cluster & 36 & 0.44 \\
& Gali \cite{gali08} & DFT-LSDA & Supercell & 512 & $\sim$0.9 \\
& Delaney \cite{delaney10} & DFT-BP & Cluster & 42,284 & 0.422,0.482 \\
& & CI & Cluster & 42 & 0.629-0.644$^c$ \\
& Ma \cite{ma10} & DFT-MBPT & Supercell & 256 & 0.40 \\
$^1A_1$ & Goss \cite{goss96} & DFT-LSDA & Cluster & 36 & 1.67 \\
& Gali \cite{gali08} & DFT-LSDA & Supercell & 512 & $\sim$0.0 \\
& Delaney \cite{delaney10} & DFT-BP & Cluster & 42,284 & 2.096,2.028 \\
& & CI & Cluster & 42 & 2.060 \\
& Ma \cite{ma10} & DFT-MBPT & Supercell & 256 & 0.99 \\
$^3E$ & Goss \cite{goss96} & DFT-LSDA & Cluster & 36 & 1.77$^a$ \\
& Larrson \cite{larrson08} & DFT-BP & Cluster & 163,284 & 1.847,1.867 \\
& Gali \cite{gali08} & DFT-LSDA & Supercell & 512 & 1.91 \\
& Hossain \cite{hossain08} & DFT-GGA & Supercell & 256 & 1.912 \\
& Lin \cite{lin08} & TD-DFT & Cluster & 85,89,104 & 2.17$^b$ \\
&  & CI & Cluster & 85,89,104 & 2.66$^b$ \\
&  & CASSCF & Cluster & 85,89,104 & 2.53$^b$ \\
& Gali \cite{gali09c} & DFT-PBE & Supercell & 512 & 1.910 \\
&  & DFT-HSE & Supercell & 512 & 2.213 \\
& Delaney \cite{delaney10} & DFT-BP & Cluster & 42,284 & 1.270,1.898 \\
& & CI & Cluster & 42 & 1.932-1.958$^c$ \\
& Ma \cite{ma10} & DFT-MBPT & Supercell & 256 & 2.32 \\
$^1E^\prime$ & Larrson \cite{larrson08} & DFT-BP & Cluster & 163,284 & 2.277,2.296 \\
& Gali \cite{gali08} & DFT-LSDA & Supercell & 512 & $\sim$2.41 \\
& Ma \cite{ma10} & DFT-MBPT & Supercell & 256 & 2.25 \\
\bottomrule\noalign{\smallskip}
\multicolumn{6}{l}{$^a$ZPL energy; $^b$average over cluster sizes; $^c$split level due to $C_s$ symmetry}
\end{tabular}
\end{table}

\subsection{\textit{Ab initio} calculations of the NV$^-$ electronic structure}

The vast majority of calculations have been performed using density functional theory (DFT) or a variant. The key idea of DFT is to avoid the computational complexity of directly solving for the many-body electronic wavefunction and instead solve for the simpler ground state electronic charge density $\rho(\vec{r})$. This idea is supported by the fact that many quantities can be expressed as functionals of $\rho(\vec{r})$. For instance, the ground state electronic energy of a system can be described by the functional
\begin{equation}
E[\rho] = T_s[\rho]+\int V(\vec{r})\rho(\vec{r})d\vec{r}+\frac{1}{8\pi\epsilon_0}\int\frac{\rho(\vec{r})\rho(\vec{r}^\prime)}{|\vec{r}-\vec{r}^\prime|}d\vec{r}d\vec{r}^\prime+E_{xc}^\mathrm{exact}[\rho]
\end{equation}
where $T_s[\rho]$ is the (calculable) kinetic energy of the non-interacting system of electrons with the same density, $V(\vec{r})$ is the electrostatic electron-nucleus attraction energy, the third term is the electrostatic electron-electron repulsion energy and $E_{xc}^\mathrm{exact}[\rho]$ is the exact, albeit unknown, exchange-correlation functional that captures all of the remaining contributions to the total electronic energy. The non-interacting electronic Kohn-Sham system \cite{kohn65} with the same density is useful as its wavefunction is simple, being typically given by one or a few Slater determinants constructed from a set of Kohn-Sham orbitals $\phi_i(\vec{r})$, and as its kinetic energy $T_s[\rho]$ captures a large part of the unknown kinetic energy of the interacting system. The central approximation that is made in the application of DFT is the selection of an approximate exchange-correlation functional $E_{xc}^\mathrm{approx}[\rho]$, such as the Local Spin Density Approximation functional or the Becke-Perdew functional. Then the application of DFT proceeds by the minimisation of $E[\rho]$ with respect to $\rho$ by varying the orthonormal $\phi_i(\vec{r})$, thereby resulting in the approximate total energy and charge density of the electronic ground state.

Time-dependent DFT aims to go beyond ground state properties, obtaining information on excited states through studying the response to a time-dependent potential $V(\vec{r},t)$. However the analogue to $E_{xc}^\mathrm{approx}$ is more complex and, being a more recent development, has been less studied.
Alternatively, Gunnarsson and Lundqvist \cite{gun76} have shown that DFT can be extended beyond the overall ground state to the lowest state of each distinct spin and spatial symmetry, e.g.\ allowing estimates of the $^1E$, $^1A_1$ and $^3E$ states in addition to $^3A_2$ in the NV centre. However though $E_{xc}[ \rho ]$ should then depend on the symmetry no prescription was given for this dependence, and codes typically implement symmetry-independent $E_{xc} [\rho]$ which  can lead to nonsensical results (e.g.\ for atomic carbon \cite{barth79}). Fortunately, as explained by von Barth \cite{barth79}, symmetry-independent $E_{xc}$ give reasonable results for states which reduce in the non-interacting system to single Slater determinants of Kohn-Sham orbitals. Now in the NV centre, the $^3A_2\, (m_s=\pm1)$ and $^3E$ states can be approximately written as single Slater determinants, enabling e.g.\ Goss's 1.77 eV estimate of the NV$^-$ $^3A_2\rightarrow$$^3E$ vertical excitation energy \cite{goss97,goss96}.
However each of the non-interacting $^3A_2 \, (m_s=0)$, $^1E$ and $^1A_1$ wavefunctions needs two determinants.
This is a common issue whenever two electrons are to be distributed between two orthogonal spatial orbitals, here $e_x$ and $e_y$. von Barth also proposed to apply DFT to calculate the energy expectation values of non-eigenstate single Slater determinants formed from well-defined mixtures of symmetries, such as $\frac{1}{\sqrt{2}}($$^1E_y+$$^3A_2(m_s=0))$ or $\frac{1}{\sqrt{2}}($$^1A_1-$$^1E_x)$. The energies of these orthogonal mixtures will give an average of the desired energies of the components, and given the $^3A_2$ energy those of $^1E$ and $^1A_1$ follow. von Barth's procedure has been applied a number of times in calculations of the NV$^-$ centre's electronic structure and, if the calculation performed by Gali et al \cite{gali08} is ignored, von Barth's procedure has enabled the intermediate singlet states to be correctly energetically ordered by DFT calculations \cite{goss96,delaney10}.

The Configuration Interaction (CI) method is a more rigorous means of calculating the energies of excited states than DFT. CI attempts to directly solve for the many-body wavefunction of each electronic state by expanding the state in terms of the many Slater determinants formed from an expansive basis of single electron orbitals. The method therefore avoids the inherent restrictions of DFT and can in theory calculate the energy of any excited state at the cost of a significant increase in the required computational resources. Consequently, CI calculations are currently limited to small cluster models where the confinement of the defect wavefunction potentially perturbs the calculated energies. Indeed, the effects of confinement are believed responsible for $^1A_1$ being predicted to have higher energy than $^3E$ in the CI calculation performed by Delaney et al \cite{delaney10} because the $^3E$ energy was found to be dependent on cluster size. The CI calculation performed by Delaney et al obtained the correct ordering of the intermediate singlet levels, thereby resolving the disagreement in the orderings estimated by applications of von Barth's DFT procedure. The correct ordering of the intermediate singlets states has also been obtained using various quantum chemical methods employed by Zyubin et al \cite{zyubin08} and post-DFT methods employed by Ma et al \cite{ma10}.

The complementary nature of the different theoretical methods and experiment is clear in how the results of the current calculations support the ordering the centre's electronic states inferred from the molecular model \cite{doherty11} and experiment \cite{manson10}. Furthermore, \textit{ab initio} calculations have provided a direct means to confirm the structure of the NV centre, its charge states, and aspects of the molecular model, such as the number and symmetry of MOs associated with the centre within the diamond bandgap.

It is important to note in interpreting the calculated energies of table \ref{tab:abinitioelectronicenergies} that they correspond to the energies of the electronic states in the ground state nuclear equilibrium coordinates (i.e. vertical excitation energies) and are not directly related to the observed optical and infrared ZPLs. To reproduce the ZPL energies, the Stokes and anti-Stokes shifts associated with the correlated relaxations of nuclear coordinates with electronic transitions must be considered. These shifts, alongside the other observed vibronic effects in the NV centre, will be discussed in the next section.

\section{Vibronic structure}
\label{section:reviewvibronicstructure}

\subsection{The Huang-Rhys model of the NV$^-$ optical vibronic band}

\begin{figure}[hbtp]
\begin{center}
\includegraphics[width=0.7\columnwidth] {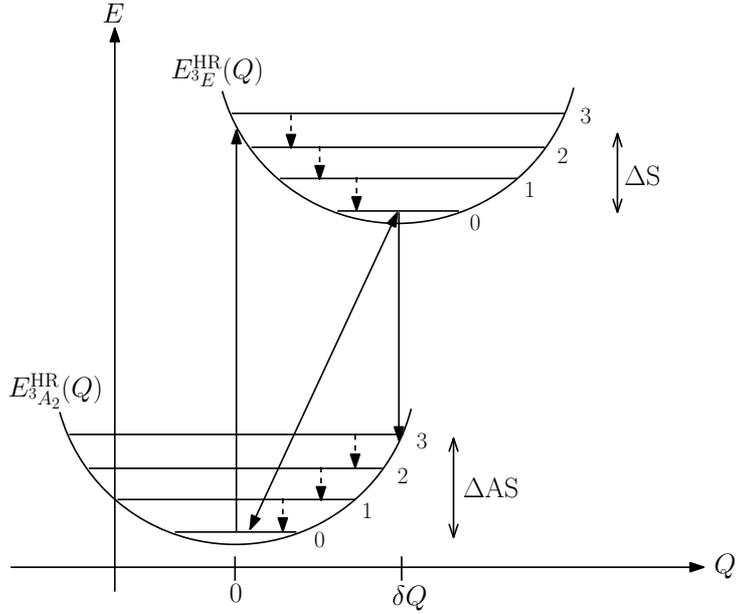}
\caption[Configuration coordinate diagram depicting vibronic transitions in the Frank-Condon approximation]{Configuration coordinate diagram depicting the vibronic transitions of the Huang-Rhys model in the Frank-Condon picture. $E_{^3A_2}^\mathrm{HR}(Q)$ and $E_{^3E}^\mathrm{HR}(Q)$ are the electronic energies of the NV$^-$ ground $^3A_2$ and optically excited $^3E$ electronic states as harmonic functions (curved lines) of the effective nuclear displacement coordinate $Q$. The horizontal lines represent the vibrational levels of each electronic state. $\Delta$S and $\Delta$AS are the Stokes and anti-Stokes shifted energies of the vertical optical transitions (solid arrows), respectively. The diagonal transition (solid arrow) is a zero-vibration transition and contributes to the optical ZPL. The non-radiative transitions within the vibronic structures of the electronic states that follow the optical transitions are depicted as dashed arrows. Note that only the vertical transitions originating from the ground vibronic levels and the zero-vibration transition between the ground vibronic levels have been depicted for reasons of clarity and that transitions between all of the vibronic levels are indeed allowed.}
\label{fig:frankcondonreview}
\end{center}
\end{figure}

The vibronic structure of the NV centre is generated by the interactions of its electronic and vibrational degrees of freedom. The observable aspects of vibronic structure are most evident in the coupled electronic and vibrational changes that occur in the process of an electronic transition, such as the optical transition of the NV$^-$ centre. The simplest approximate description of these changes for a particular electronic transition is provided by the Huang-Rhys model \cite{stoneham75}, within which, the quasi-continuum of vibrational modes of the diamond lattice are replaced by a single effective vibrational mode whose normal coordinate corresponds to the displacement of the nuclei between their equilibrium coordinates associated with the electronic states of the transition. As depicted in figure \ref{fig:frankcondonreview}, the use of an adiabatic approximation (such as the Born-Oppenheimer approximation) results in the vibrational motion associated with a given electronic state being described by the potential formed by the state's electronic energy and its dependence on the nuclear coordinates \cite{bransden}. The harmonic approximation facilitates a further simplification through the approximation of the potential by a simple quadratic function of the nuclear coordinates \cite{bransden}, which for the $^3A_2$ and $^3E$ electronic states of NV$^-$ take the form
\begin{eqnarray}
E_{^3A_2}^\mathrm{HR}(Q) & = & \frac{1}{2}\omega^2Q^2 \nonumber \\
E_{^3E}^\mathrm{HR}(Q) & = & E_{^3E}+aQ+\frac{1}{2}(\omega^2+b)Q^2 \nonumber \\
& = & E_{^3E}-E_R+\frac{1}{2}(\omega^2+b)(Q-\delta Q)^2
\label{eq:huangrhys}
\end{eqnarray}
where $E_{^3E}$ is the electronic energy of the excited state at the ground state nuclear equilibrium coordinate ($Q=0$), $a$ and $b$ are the linear and quadratic electron-vibration coupling parameters of the $^3E$ state, respectively, $\omega$ and $\sqrt{\omega^2+b}$ are the frequencies of the mode $Q$ in the $^3A_2$ and $^3E$ states, respectively, $\delta Q=-a/(\omega^2+b)$ is the equilibrium displacement of the $^3E$ state, $E_R=a^2/2(\omega^2+b)=S\hbar\sqrt{\omega^2+b}$ is the relaxation energy of the $^3E$ state and $S$ is the Huang-Rhys factor of the optical transition. The physical origins of the linear and quadratic electron-vibration coupling terms in $E_{^3E}^\mathrm{HR}(Q)$ can be understood as being the nuclear displacement and vibrational frequency shift that result from the redistribution of electronic charge in the excited electronic state \cite{bransden}. Given the harmonic potentials, the vibrational states associated with the $^3A_2$ and $^3E$ electronic states will be simple harmonic oscillator states with energies $\hbar\omega(\nu+\frac{1}{2})$ and $\hbar\sqrt{\omega^2+b}(\nu+\frac{1}{2})$, respectively, where $\nu$ is the vibrational occupation \cite{davies81}.

Applying the Frank-Condon principle \cite{bransden}, at low temperature the electronic transition from the $^3A_2$ ground state to the optically excited $^3E$ state originates from the ground vibronic level of $^3A_2$, it approximately occurs within the $^3A_2$ nuclear equilibrium coordinates and is followed by a non-radiative relaxation to the ground vibronic level of $^3E$ that effectively displaces the nuclei to their $^3E$ equilibrium coordinates. The electronic transition from $^3E$ back to $^3A_2$ proceeds in the opposite sense. Due to the difference in the nuclear equilibrium coordinates (and to a lesser degree the difference in the vibrational frequencies) associated with the two electronic states, the overlap of the vibrational wavefunctions of the two electronic states are non-zero for vibrational states with different vibrational occupancies \cite{stoneham75}. Consequently, transitions are allowed between all of the vibronic levels of the two electronic states with rates that are proportional to the vibrational Frank-Condon factors, thereby producing the vibronic band of the electronic transition that is Stokes shifted in excitation and anti-Stokes shifted in decay \cite{stoneham75} (see figure \ref{fig:frankcondonreview}). The transitions between vibronic levels with the same occupation of the effective mode contribute to the ZPL of the electronic transition.

\begin{figure}[hbtp]
\begin{center}
\includegraphics[width=0.5\columnwidth] {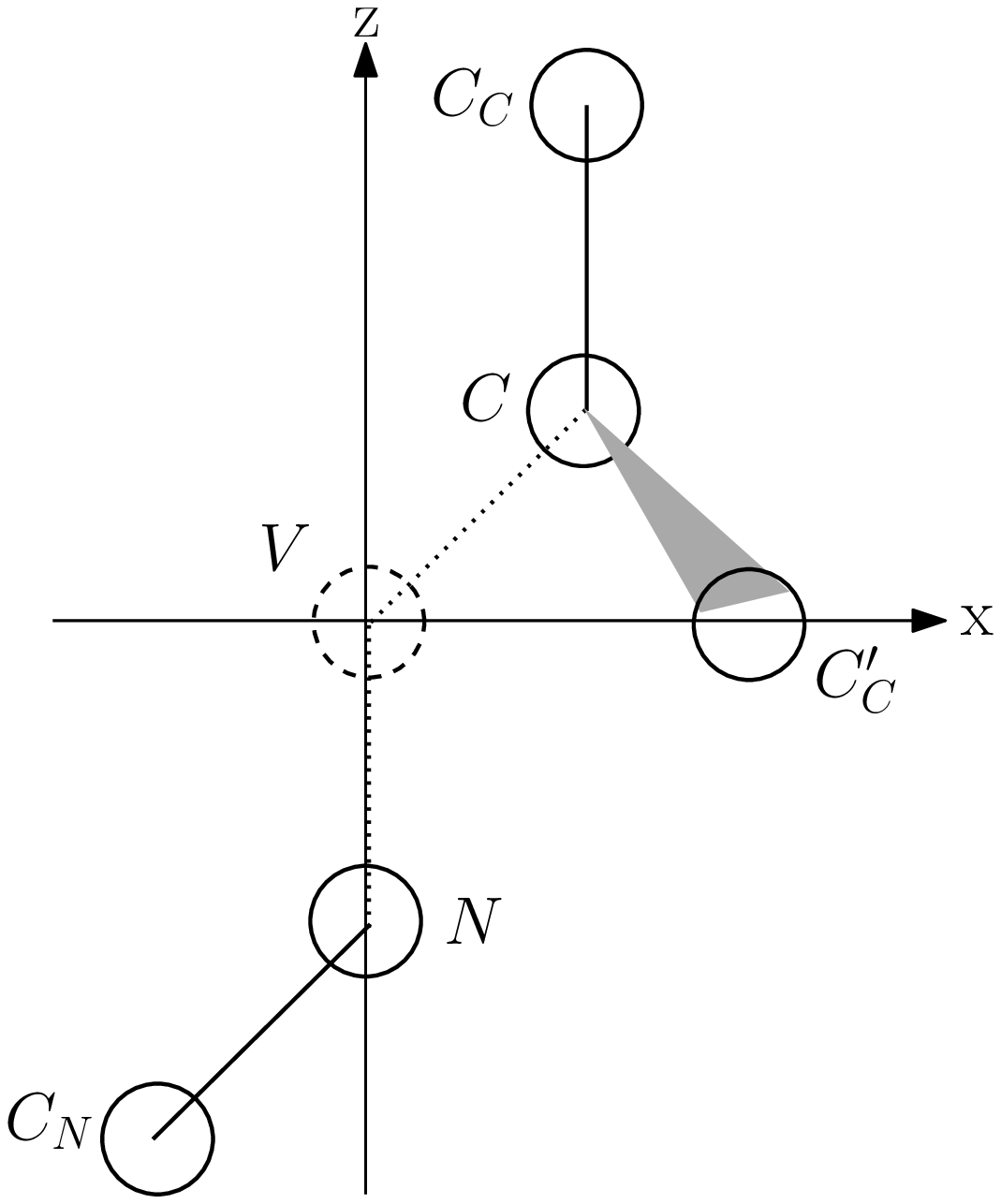}
\caption[Schematic of the geometry of the NV centre]{Schematic of the geometry of the NV centre depicting the nuclei that exists within the $xz$-plane of the coordinate system defined in figure \ref{fig:fieldcoordsysreview}, and the carbon nuclei $C_C^\prime$ that are bonded to the carbon nearest neighbour of the vacancy and exist out of the $xz$-plane and within the $xy$-plane. Given the $C_{3v}$ symmetry of the centre, the depicted elements completely describe the geometry of the centre.}
\label{fig:nuclearequilibriumcoordinatesreview}
\end{center}
\end{figure}

\textit{Ab initio} studies have calculated the nuclear equilibrium coordinates of the ground and excited electronic states of the optical transitions of NV$^-$ \cite{goss96,luszczek04,larrson08,gali08,zyubin08,delaney10,ma10,zhang11} and NV$^0$ \cite{gali09,zyubin08}. Referring to figure \ref{fig:nuclearequilibriumcoordinatesreview}, the positions of the nearest neighbour nuclei to the vacancy can be specified relative to the positions of the next-to-nearest neighbours by the parameters contained in table \ref{tab:nuclearequilibriumcoordinatesreview}. Note that the \textit{ab initio} calculations consistently indicate that the next-to-nearest neighbour nuclei and beyond undergo negligible relaxation from their positions in a defect-free crystal lattice. As determined by the \textit{ab initio} studies performed by Gali et al \cite{gali08}, the ground state equilibrium positions of NV$^-$ correspond to a symmetry preserving outwards (away from the vacancy) relaxation of the nearest neighbour nitrogen and carbon nuclei from their positions in a defect-free crystal lattice. In the ground state relaxation, the nitrogen relaxes further than the carbon nuclei. The $^3E$ excited state equilibrium positions of NV$^-$ correspond to a counter relaxation of the carbon and nitrogen nuclei from their ground state positions, where the nitrogen moves further inwards than the carbon nuclei move outwards \cite{gali08}. In a separate study, Gali et al \cite{gali09} calculated that the ground state equilibrium positions of NV$^0$ corresponded to a lowering of the centre's symmetry from $C_{3v}$ to $C_{1h}$, which is indicative of the Jahn-Teller effect in the degenerate $^2E$ ground state. \textit{Ab initio} calculations \cite{gali08,gali09c,ma10} have also obtained the energies of the NV$^-$ electronic states in each nuclear configuration, which can be used to estimate the optical ZPL energy and associated Stokes and anti-Stokes shifts (refer to table \ref{tab:abinitiostokesshiftsreview}). The calculated Stokes shifts may be used to relate the electronic energies in table \ref{tab:abinitioelectronicenergies} to the observed optical ZPL energy of NV$^-$.

\begin{table}
\caption[Summary of \textit{ab initio} calculations of the NV$^-$ geometry tabulated by reference and electronic state]{\label{tab:nuclearequilibriumcoordinatesreview} Summary of \textit{ab initio} calculations of the NV$^-$ geometry tabulated by reference and electronic state. The unrelaxed geometry of the NV centre as calculated using the empirical diamond bond length $1.54$ {\AA} is provided in the first line for comparison. All distances are provided in units of {\AA} and the angle $\angle C_NN$ between the $C_N$-$N$ bond and the centre's symmetry axis is provided in units of degrees.}
\begin{tabular}{lllllllll}
\noalign{\smallskip}
\toprule
State & Reference & \multicolumn{2}{l}{Distance ({\AA})} &  &  &  &  & Angle ($^\circ{}$) \\ \cline{3-4}\cline{9-9} \noalign{\smallskip}
 & & $\overline{NC_N}$ & $\overline{VN}$ & $\overline{NC}$ & $\overline{VC}$ & $\overline{CC_C}$ & $\overline{CC_C^\prime}$ & $\angle C_NN$ \\
\midrule
\multicolumn{2}{l}{Unrelaxed} & 1.54 & 1.54 & 2.51 & 1.54 & 1.54 & 1.54 & 109.5 \\
$^3A_2$ & Goss \cite{goss96} & 1.44 &  &  &  & 1.45 & 1.45$^a$ &  \\
& Luszczek \cite{luszczek04} & 1.50 &  & 2.72 &  & 1.48 & 1.51 &  \\
& Larrson \cite{larrson08} & 1.48 &  & 2.77 &  & 1.51 & 1.51 & 105.1 \\
& Gali \cite{gali08} & 1.46 & 1.68 & & 1.62 & 1.50 & 1.50$^a$ &  \\
& Zyubin \cite{zyubin08} & 1.49 & 1.68 &  & 1.61 & 1.52 & 1.54 &  \\
& Delaney \cite{delaney10} & 1.48$^c$ &  & 2.92$^c$ &  & 1.50$^c$ & 1.50$^c$ & 105.5$^c$ \\
& Zhang \cite{zhang11} &  & 1.70 &  & 1.65 &  &  &  \\
$^3E$ & Gali \cite{gali08} &  & 1.61 &  & 1.67 &  &  &  \\
& Ma \cite{ma10} &  & -0.05$^b$ &  & +0.06$^b$ &  &  &  \\
& Zhang \cite{zhang11} &  & 1.63&  & 1.68 &  &  &  \\
\bottomrule\noalign{\smallskip}
\multicolumn{9}{l}{$^a$did not differentiate $\overline{CC_C}$ and $\overline{CC_C^\prime}$; $^b$displacement from $^3A_2$ value;} \\
\multicolumn{9}{l}{$^c$small cluster.}
\end{tabular}
\end{table}

\begin{table}
\caption[Experimental measurement and \textit{ab initio} calculations of the NV$^-$ optical ZPL, Stokes and anti-Stokes shifted energies tabulated by reference]{\label{tab:abinitiostokesshiftsreview} Experimental measurements and \textit{ab initio} calculations of the NV$^-$ optical ZPL, Stokes and anti-Stokes shifted energies tabulated by reference. Note that the difference in the vibrational zero point energies of the $^3A_2$ and $^3E$ electronic states has been neglected in the calculated optical ZPL energies. This is equivalent to the neglect of quadratic electron-vibration coupling.}
\begin{tabular}{llll}
\noalign{\smallskip}
\toprule
Reference & Stokes (eV) & ZPL (eV) & Anti-Stokes (eV) \\
\midrule
Davies$^a$ \cite{davies76} & 2.180 & 1.945 & 1.760 \\
Gali$^b$ \cite{gali08} & 1.910 & 1.706 & 1.534 \\
Gali$^b$ \cite{gali09c} &  2.213 & 1.955 & 1.738 \\
Ma$^b$ \cite{ma10} & 2.32 & 2.09 & 1.95 \\
\bottomrule\noalign{\smallskip}
\multicolumn{4}{l}{$^a$Experiment; $^b$\textit{ab initio}}
\end{tabular}
\end{table}

\subsection{The linear symmetric mode model of the NV$^-$ optical vibronic band}

To go beyond the overly simple Huang-Rhys model and attempt to quantitatively describe the vibronic bands of NV$^-$, one must consider the quasi-continuum of crystal vibrational modes \cite{stoneham75}. Generalising the expressions (\ref{eq:huangrhys}) of the Huang-Rhys model to the many mode case, the harmonic nuclear potentials of the $^3A_2$ ground and $^3E$ excited electronic states of NV$^-$ are \cite{davies81}
\begin{eqnarray}
E_{^3A_2}(\vec{Q}) & = & \frac{1}{2}\sum_i\omega_i^2Q_i^2 \nonumber \\
E_{^3E}(\vec{Q}) & = & E_{^3E}+\sum_ia_iQ_i+\frac{1}{2}(\omega_i^2+b_{ii})Q_i^2+\frac{1}{2}\sum_{i\neq j}b_{ij}Q_iQ_j
\end{eqnarray}
where $Q_i$, $a_i$, $b_{ii}$ and $\omega_i$ are the normal coordinates, $^3E$ linear and quadratic electron-vibration interaction parameters and frequency of the $i^{th}$ vibrational mode, respectively, and $b_{ij}$ $i\neq j$ is the quadratic electron-vibration interaction parameter that couples the $i^{th}$ and $j^{th}$ modes.  Note that $\vec{Q}$ represents the collection of normal coordinates of all modes. The eigenmode displacement coordinates of a diamond crystal containing a NV centre will have definite $C_{3v}$ symmetry due to the reduction of the cubic and translational symmetries of the perfect lattice by the presence of the NV defect. Simple symmetry arguments imply that $^3E$ state may have linear and quadratic interactions with both symmetric and non-symmetric modes. The electron-vibration interaction of non-symmetric vibrations and electronic states is the root of the Jahn-Teller effect \cite{stoneham75}. Due to the complexity of the many-mode Jahn-Teller effect, linear and quadratic interactions with non-symmetric modes are typically ignored until it is apparent that symmetric modes cannot adequately  describe a vibronic band \cite{davies81}. A further complexity typically ignored is the quadratic coupling of $A_1$ modes \cite{davies81}. Hence, the simplest theory that attempts to account for the quasi-continuum of crystal vibrational modes, considers just the linear interactions with $A_1$ modes. The general theory of such linear symmetric mode models was developed by Maradudin \cite{maradudin66} and applied extensively by Davies \cite{davies74} to describe the optical band of NV$^-$ and other colour centres in diamond. Davies and Hamer \cite{davies76} later extended the model to include their nitrogen-tunneling explanation of the slight asymmetry between the NV$^-$ optical absorption and emission bandshapes (refer to section \ref{section:reviewdefectstructureandchargestates}). The simple linear symmetric mode model nevertheless provides a reasonable description of the optical band and also provides a wealth of information on the electron-vibration interactions in the $^3A_2$ and $^3E$ electronic states.

\begin{figure}[hbtp]
\begin{center}
\mbox{
\subfigure[]{\includegraphics[width=0.4\columnwidth] {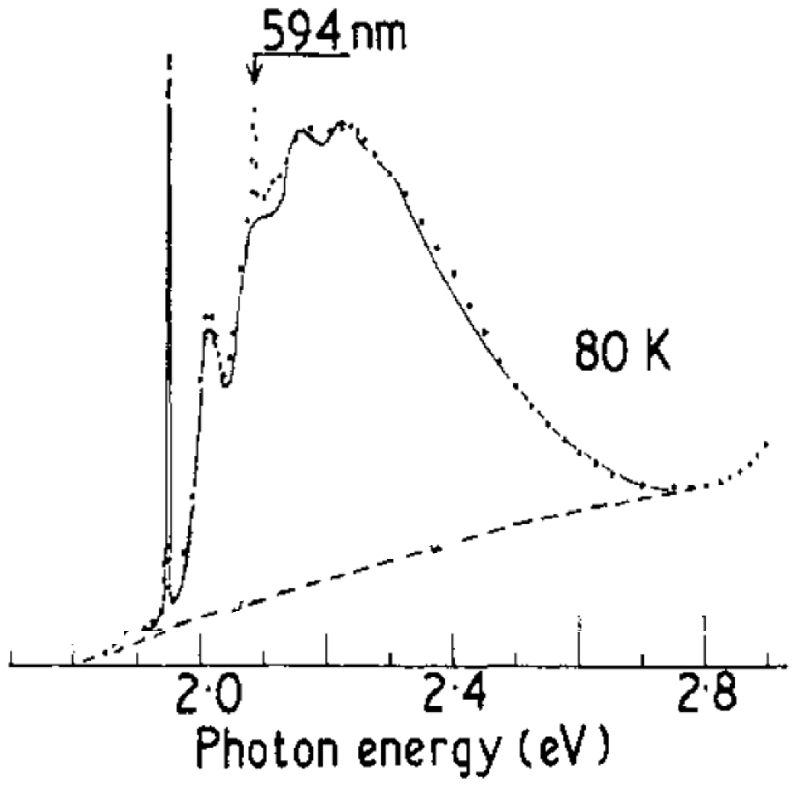}}
\subfigure[]{\includegraphics[width=0.4\columnwidth] {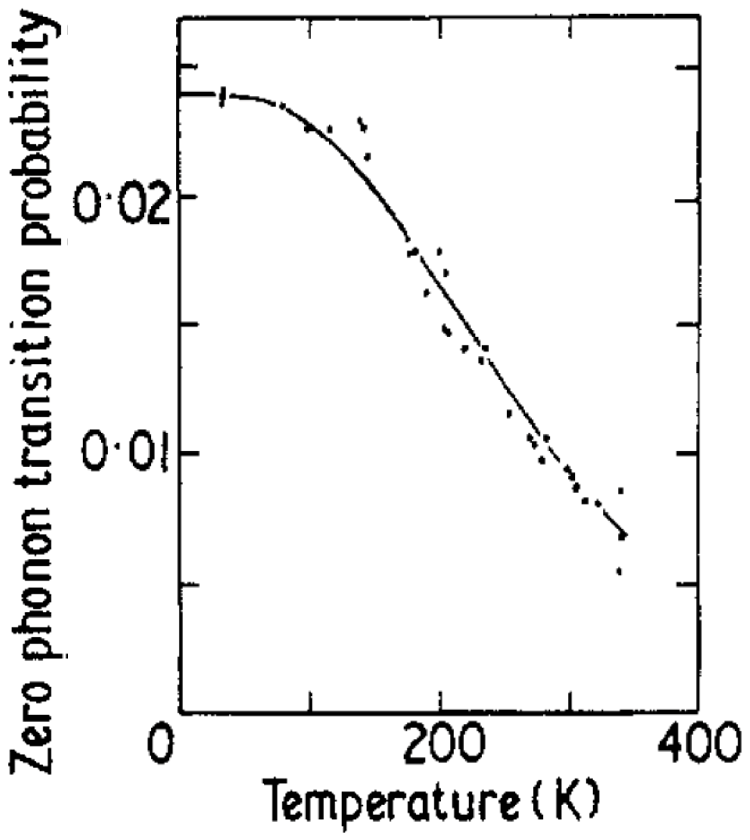}}
}
\mbox{
\subfigure[]{\includegraphics[width=0.7\columnwidth] {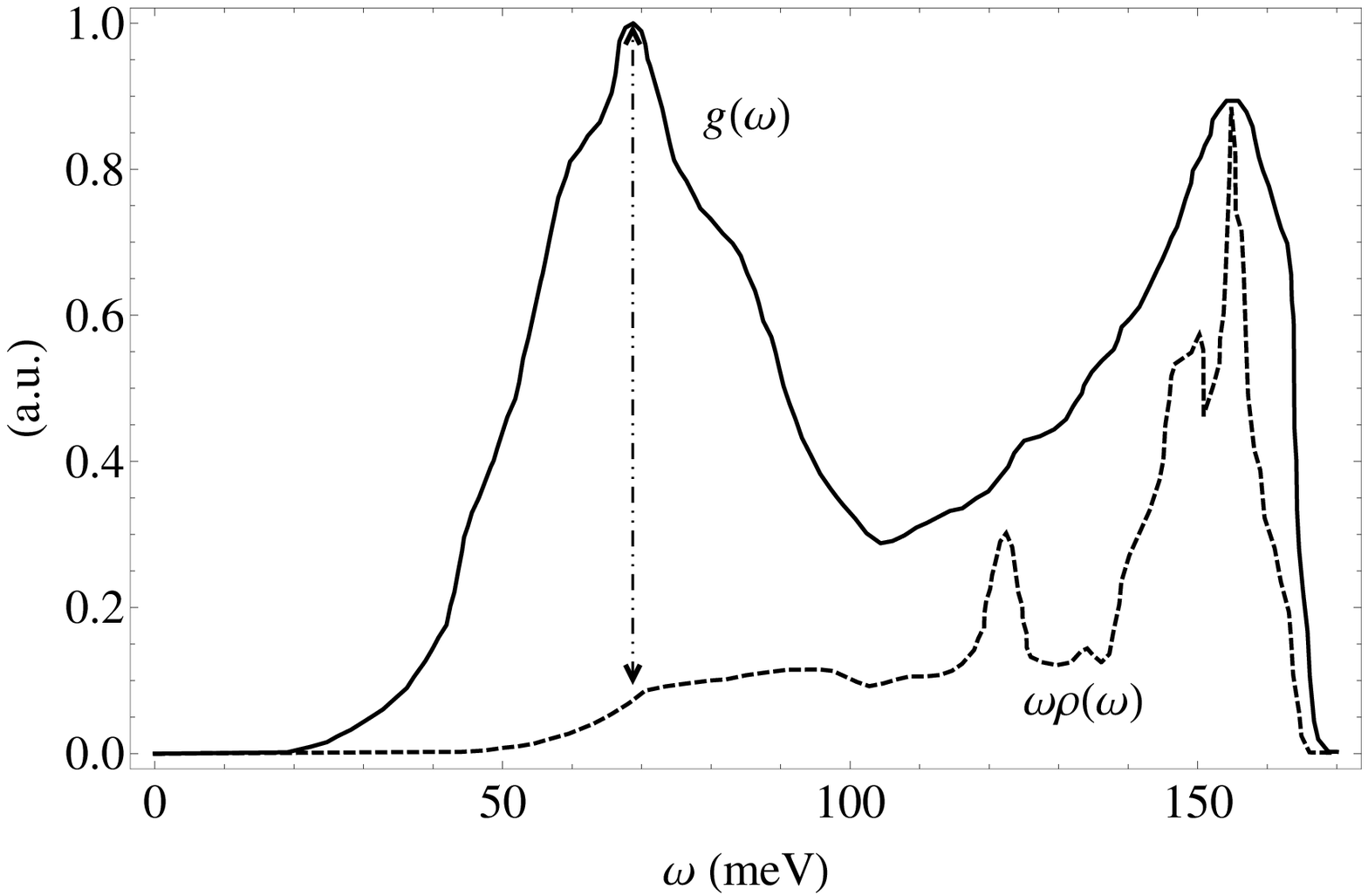}}
}
\caption[The symmetric mode model of the NV$^-$ optical absorption band and ZPL intensity]{The symmetric mode model of the NV$^-$ optical (a) absorption band and (b) normalised ZPL intensity change with temperature \cite{davies74}. The dashed ($--$) curve in (a) indicates the absorption baseline. Note the identified 594 nm feature is the ZPL of another defect and is unrelated to the NV centre. (c) The linear electron-vibration interaction strength with $A_1$ modes $g(\omega)$ (solid curve) as obtained from the absorption band using the linear symmetric mode model \cite{davies74} and the diamond density of phonon modes $\omega\rho(\omega)$ (dashed curve) \cite{wehner67} in normalised units. The vertical dotted arrow marks the peak in $g(\omega)$ at $\sim67$ meV and the correlated feature marking the change from the Debye functional form $\propto \omega^2$ of the density of modes.}
\label{fig:davieslinearfitreview}
\end{center}
\end{figure}

\begin{figure}[hbtp]
\begin{center}
\includegraphics[width=0.7\columnwidth] {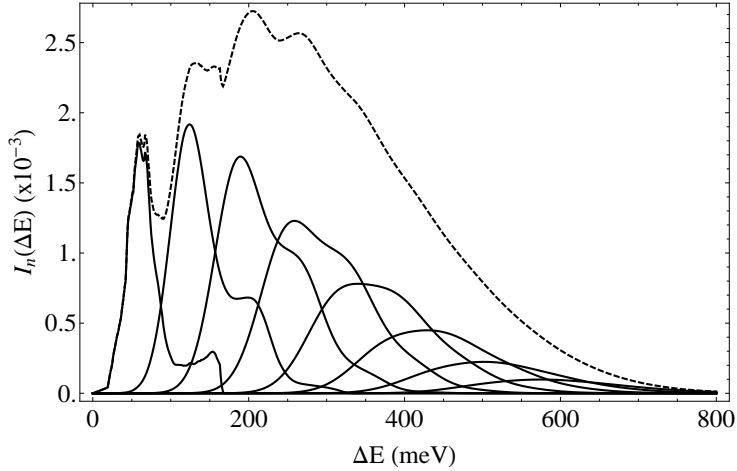}
\caption[The symmetric mode model of the NV$^-$ optical absorption band and ZPL intensity]{The normalized intensities $I_n(\Delta E)$ of the $n=1,2\ldots8$ phonon sidebands (solid curves) and total absorption vibronic band $I(\Delta E)=\sum_nI_n(\Delta E)$ (dashed curve) of NV$^-$ as calculated using the function $g(\omega)$ obtained by Davies \cite{davies74} using the symmetric model model (refer to figure \ref{fig:davieslinearfitreview}). $\Delta E$ denotes the energy difference from the optical ZPL of NV$^-$ and the sidebands begin with the $n=1$ sideband on the left hand side and increase sequentially towards the right hand side.The integrated intensity or the $n^{th}$ phonon sideband is normalized to $e^{-S}S^n/n!$, such that the integrated intensity of the total vibronic band is one.  For further details, refer to Ref. \cite{davies74}.}
\label{fig:daviessideband}
\end{center}
\end{figure}

Using ensemble measurements of the absorption band, Davies obtained an expression for the function $g(\omega)=\hbar^2\omega^2S(\omega)\rho_{A_1}(\omega)$ (where $S(\omega)$ is the average Huang-Rhys factor and $\rho_{A_1}(\omega)$ is the density of symmetric modes of frequency $\omega$) by employing a trial and repeat method, in which he initially defined a trial function, convolved it with the liquid-nitrogen temperature zero-phonon lineshape, compared the calculated band with the observed band and repeated the process until a good fit was obtained. The function that Davies obtained is contained in figure \ref{fig:davieslinearfitreview} and the individual components of the absorption sideband obtained using the function are depicted in figure \ref{fig:daviessideband}. The function also provided an excellent fit for the temperature dependence of the normalised ZPL strength via the simple expression derived in \cite{davies74}.

By noting the definition of $S(\omega)=a^2(\omega)/2\hbar\omega^3$ (where $a^2(\omega)$ is the average of $a_i^2$ over all $A_1$ modes with frequency $\omega$), it can be concluded that the function obtained by Davies $g(\omega)=\hbar^2\omega^2S(\omega)\rho_{A_1}(\omega)=\hbar a^2(\omega)\rho_{A_1}(\omega)/2\omega$ is the average linear interaction strength of the $^3E$ electronic state with $A_1$ modes of frequency $\omega$. Consequently, $g(\omega)$ provides direct insight into the linear interactions of $^3E$ with $A_1$ modes. The peak of the function at $\sim$67 meV indicates that $A_1$ modes of these frequencies interact particularly strongly with $^3E$. Given that the electronic states of the NV centre are highly localised to the nitrogen and nearest neighbour carbon nuclei to the vacancy, the displacements of the $A_1$ modes with energies close to $\sim67$ meV must also be localised to some degree to these nuclei, and the modes may be identified as pseudo-local modes. Note that true local modes that involve just the displacements of the nearest nuclei to the centre must have frequencies greater than the maximum frequency of the defect-free lattice. As there are no such local mode features present in $g(\omega)$, Davies concluded that the NV centre has no true local modes \cite{davies74}. As indicated in figure \ref{fig:davieslinearfitreview}, the peak of $g(\omega)$ at $\sim155$ meV is consistent with the peak in the diamond density of phonon modes at a similar energy. This peak in the density of modes corresponds to the optical modes of the lattice. Thus, the $\sim155$ meV peak can be attributed to the increase in the density of $A_1$ modes and is not necessarily indicative of any significant localisation of the optical modes.

A simple physical explanation for the locations of the peak interactions in $g(\omega)$ can be obtained by considering the phonon modes of the defect-free lattice. The nearest neighbour nitrogen and carbon nuclei to the vacancy occupy equivalent lattice sites in the diamond bravais lattice, each being connected by a primitive lattice vector. Consequently, the nearest neighbours only undergo relative displacements to each other for phonons with reciprocal lattice vectors ($k$-vectors) away from the Brillouin zone centre (the $\Gamma$ point). The nearest neighbours achieve maximum relative displacement for phonons at the boundaries of the Brillouin zone. Indeed, phonons with $k$-vectors at certain high symmetry points on the zone boundary correspond to vibrations where pairs of the nearest neighbours vibrate 180$^\circ$ out of phase. Hence, the highly localised electronic states of the NV centre are expected to interact most strongly with phonons close to the Brillouin zone boundary. As indicated in figure \ref{fig:davieslinearfitreview}, the lowest energy phonons in the vicinity of the zone boundary have energies close to $\sim67$ meV, which is indicated by the change in the functional form of the density of modes from the simple Debye form $\propto\omega^2$ \cite{pavone93}. The modes that reach the zone boundary close to $\sim67$ meV are the transverse acoustic phonon modes \cite{pavone93}, and are thus the modes that mix to form the identifiable pseudo-local modes in $g(\omega)$. The longitudinal acoustic and optical phonon branches approach the zone boundary at various energies between $\sim67$ meV and $\sim160$ meV \cite{pavone93}. Consequently, these modes do not produce a distinct peak in $g(\omega)$, but instead transform the sharp features present in the density of modes into the smooth curve of $g(\omega)$ between $\sim67$ meV and $\sim 155$ meV. The transverse optical phonon modes reach the zone boundary with energies $\sim155$ meV \cite{pavone93}, however given the smaller peak amplitude at $\sim155$ meV compared to $\sim67$ meV in $g(\omega)$, it appears that in spite of the corresponding peak in the density of modes, the transverse optical modes must not interact as strongly as the transverse acoustic modes with the electrons of the NV centre. This qualitative explanation for the form of $g(\omega)$ has been supported by an \textit{ab initio} calculation \cite{zhang11} that obtained the density of phonon modes for a NV$^-$ supercell model and used the displacement coordinates of the phonon modes to produce a projected density of modes that was weighted by the amplitudes of the displacements of the vacancy nearest neighbours.

In addition to his observations of the NV$^-$ optical absorption bandshape and the temperature dependence of the ZPL intensity, Davies also observed a temperature dependent homogeneous broadening and energy shift of the ZPL. These effects cannot be explained by the linear symmetric mode model and require the inclusion of at least quadratic interactions with symmetric modes. In order to avoid the complexity of a rigorous treatment of quadratic electron-vibration interaction, Davies (following Maradudin \cite{maradudin66}) assumed that the quadratic interaction parameters for $A_1$ modes were small ($b_{ij}/\omega_i\omega_j\ll1$) and that they could be expressed as the product of linear interaction parameters $b_{ij}\propto a_ia_j$. The first assumption is supported by recent \textit{ab initio} studies \cite{gali11,zhang11} that have calculated only small differences ($<2\%$) in the vibrational energies between the $^3A_2$ and $^3E$ electronic states. This second assumption can only be justified by the phenomenological evidence that it has provided good descriptions of the vibronic bands of other defects, since there is no fundamental reason for the quadratic and linear interaction parameters to be related in such a way \cite{maradudin66}. Using these assumptions, Davies derived expressions for the temperature dependent homogeneous broadening and energy shift of the ZPL \cite{davies74}. Davies was able to obtain a good fit of the temperature dependence of the ZPL energy. However, due to the significant inhomogeneous broadening present in his ensemble sample, Davies could not directly apply his expression for the homogeneous width, but instead had to fit a convolution of a homogeneous lineshape (with a temperature dependent width given by the above) and an inhomogeneous lineshape (with a width with an unknown temperature dependence) to the observed zero-phonon lineshape at each temperature. Although Davies obtained a good fit using this method, this cannot be taken as evidence that the symmetric mode model correctly described the broadening of the ZPL, because the temperature dependence of the inhomogeneous width could have compensated for any inaccuracy in the symmetric mode model during the fitting procedure.  This point is particularly pertinent since the recent direct measurements of the homogeneous linewidth and its temperature dependence by Fu et al \cite{fu09} conflict with those of Davies \cite{davies74} and can only be explained by the presence of the Jahn-Teller effect. This and other evidence of the Jahn-Teller effect will be discussed further shortly.

\subsection{The asymmetry of the NV$^-$ optical absorption and emission bandshapes}

Although Davies' symmetric mode model provided good descriptions of the NV$^-$ optical absorption bandshape and the temperature dependence of the ZPL intensity and energy, it could not describe the asymmetry of the absorption and luminescence bandshapes observed later by Davies and Hamer \cite{davies76}. The slight asymmetry observed by Davies and Hamer consisted of a small ($<10$ meV) splitting of the one-phonon peak in absorption that does not occur in luminescence as well as broadening of the two- and three-phonon features in absorption compared to luminescence (refer to figure \ref{fig:doublebumpreview}). Given the approximations and assumptions made in the derivation of the symmetric mode model, the failure of the model to describe the band asymmetry could arise from one of the following: (1) neglect of interactions with non-symmetric modes; (2) neglect of quadratic interactions with symmetric modes in the model of the bandshapes; or, (3) the failure of the harmonic approximation. Davies and Hamer considered both the first and third explanations for the failure of the model. Given that recent \textit{ab initio} calculations \cite{gali11,zhang11} as well as Davies' models of the ZPL homogeneous broadening and energy shift with temperature \cite{davies74} all indicate that quadratic interactions with symmetric modes are weak, the second explanation is the least probable out of all three. Davies and Hamer ultimately adopted the third explanation.

\begin{figure}[hbtp]
\begin{center}
\includegraphics[width=0.65\columnwidth] {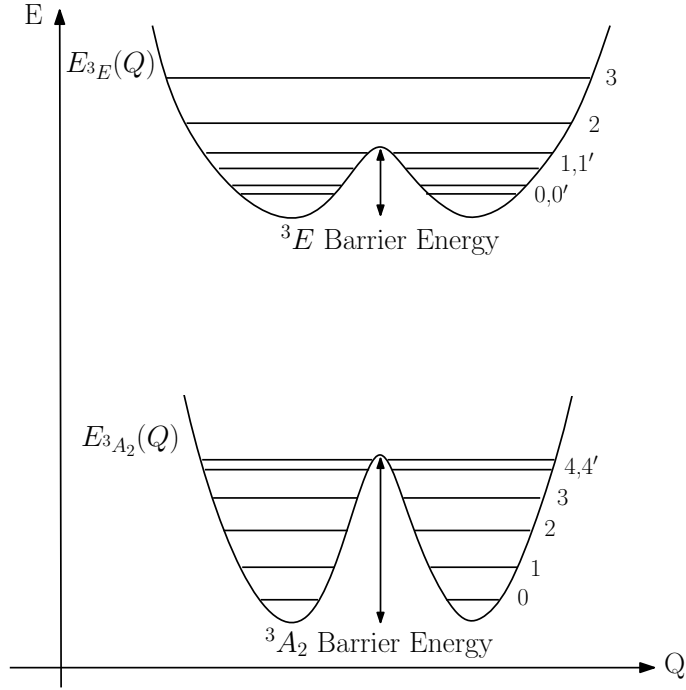}
\caption[Schematic of the nuclear potentials of the nitrogen tunneling model of NV$^-$]{Schematic of the nuclear potentials (solid curves) of the nitrogen tunneling model of NV$^-$ corresponding to the ground $^3A_2$ and optically excited $^3E$ electronic states. The horizontal lines indicate vibronic levels of each electronic state and are labelled by their vibrational number. Vibronic levels marked with a prime are levels that have been split due to inversion doubling arising from the overlap of the vibrational wavefunctions of the left and right potential wells near the top of the respective barrier energies. Davies and Hamer \cite{davies76} proposed that the barrier energy of $^3E$ is much smaller than $^3A_2$, resulting in a vibronic splitting of the one-phonon peak (and ZPL \cite{lin07}) in absorption and no observable splitting in luminescence.}
\label{fig:doublepotentialwellmodelreview}
\end{center}
\end{figure}

As previously discussed, Davies and Hamer attempted to explain the asymmetry in the absorption and luminescence bands by the notion that the nitrogen nucleus tunnels between its substitutional lattice site and the vacancy site whilst preserving the centre's symmetry, which is analogous to the inversion tunneling of nitrogen in ammonia. The notion of the tunneling nitrogen nucleus is too complicated to theoretically treat if the many modes of the lattice are considered. Instead, Davies and Hamer had to consider just one symmetric mode (similar to the Huang-Rhys model detailed at the beginning of this section) with an effective vibrational energy of $\sim67$ meV that describes the tunneling behaviour. Although Davies and Hamer did not explicitly state this at the time, the tunneling model represents the failure of the harmonic approximation, since the nuclear potential is no longer considered to be a simple quadratic function of the nuclear displacement coordinate of the one symmetric mode, but a double-well composed of a barrier between two quadratic functions. As depicted in figure \ref{fig:doublepotentialwellmodelreview}, the double-well potential directly yields the splitting of vibronic levels with energies that are close to the top of the central barrier due to the overlap of the corresponding vibrational wavefunctions of each well (i.e. inversion doubling). Consequently, it is straight forward to conclude that if the barrier energy in $^3E$ was small enough such that the one-phonon levels at $\sim67$ meV were split due to inversion doubling, then the splitting of the one-phonon absorption peak could be simply explained \cite{kilin00}. Furthermore, the absence of any obvious splitting of peaks in the luminescence band would require the barrier energy in $^3A_2$ to be higher than vibronic levels occupied by at least six phonons with energies of $\sim67$ meV \cite{kilin00}. Clearly, the tunneling model was elegant in its explanation of the vibronic splitting in the absorption band, however it required very specific values for the barrier energies to be quantitatively correct and generated implications for other aspects of the optical transition that have since been unobserved or demonstrated to be incorrect.

The tunneling model proposed by Davies and Hamer was developed further by the much later works of Kilin et al \cite{kilin00} and Lin et al \cite{lin07}. Kilin et al fitted the vibronic bands of single centres using a double-well potential and found that the parameters of their model varied significantly between the individual centres, making it improbable that any features attributable to the model could still be visible in the ensemble measurements of Davies and Hamer \cite{manson07}. Lin et al refined the approach by Kilin et al and found that in order to match the vibronic splitting of the one-phonon peak in the absorption band, the ZPL also exhibited a small vibronic splitting of $\sim0.6$ meV. The studies of single centres with ZPL widths much smaller than $\sim0.6$ meV following the work of Lin et al have clearly showed that no such splitting exists \cite{manson07}. Supporting the empirical evidence against the tunneling model, the early \textit{ab inito} calculation of Mainwood \cite{mainwood94} predicted tunneling barrier energies much greater than allowed by the model. Mainwood's calculation has been supported by later and more sophisticated calculations \cite{zyubin08,gali11} that have estimated the barrier energies associated with both $^3A_2$ and $^3E$ to be in excess of 4 eV, which is two orders of magnitude greater than allowed by the model. Hence, a body of conclusive evidence against the tunneling model has been collected and the model is no longer accepted as an explanation of the band asymmetry. In light of recent observations of the Jahn-Teller effect in $^3E$ \cite{fu09,Hizhnyakov03,kaiser09}, the attention has now been turned to a possible Jahn-Teller explanation for the differences in the absorption and luminescence bands.

\subsection{The Jahn-Teller effect and the temperature dependence of the NV$^-$ optical ZPL}

\begin{figure}[hbtp]
\begin{center}
\includegraphics[width=0.9\columnwidth] {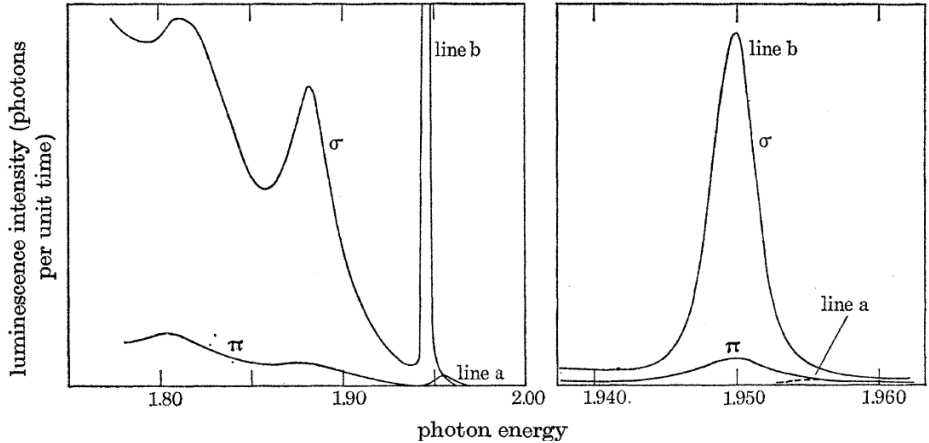}
\caption[Polarised luminescence spectra of NV$^-$ as taken by Davies and Hamer]{Polarised luminescence spectra of NV$^-$ as taken by Davies and Hamer \cite{davies76} at 17 K using a high density ensemble and an applied stress of 1.4 GPa along the $[001]$ direction. The left panel shows that the ZPL has been split into upper (line a) and lower (line b) branches, which are expected to be purely $\pi$-polarised (parallel to stress axis) and $\sigma$-polarised (perpendicular to stress axis), respectively. The right panel shows that the polarisation of the lower branch (solid curves) is a mixture of both polarisations, which Davies and Hamer explained by the presence of random crystal strain. The left panel represents the polarisation of the luminescence band once corrected for the depolarising effects of random crystal strain inferred from the lower ZPL branch.}
\label{fig:davieshamerpolarisationspectra}
\end{center}
\end{figure}

Although Davies and Hamer rejected the first possibility of a Jahn-Teller explanation, their investigation of it deserves some discussion. They conducted ensemble measurements of the polarisation of the luminescence band at low temperature (17 K) using broadband optical illumination from a high pressure mercury vapour lamp. In addition, the measurements were conducted in the presence of a large stress applied along the $[001]$ direction, such that the two branches of the optical ZPL were split by $\sim6$ meV (refer to figure \ref{fig:davieshamerpolarisationspectra}). As expected, the lower energy ZPL branch (line b in figure \ref{fig:davieshamerpolarisationspectra}) is predominately $\sigma$-polarised (perpendicular to stress axis), whereas the higher energy ZPL branch (line a in figure \ref{fig:davieshamerpolarisationspectra}) is predominantly $\pi$-polarised (parallel to stress axis). Davies and Hamer showed that the small depolarisation of the lower branch could be explained by the random crystal strain distribution present in their sample. The polarisation of the luminescence band depicted in figure \ref{fig:davieshamerpolarisationspectra} represents the polarisation of the band once corrected for the depolarisation due to random strain inferred from the ZPL. Since the emission from the upper branch is negligible ($<1\%$) compared to the emission from the lower branch, the polarisation of the luminescence band is governed by the emission from the lower branch. Consequently, in the absence of the Jahn-Teller effect, where optical selection rules are degraded by vibronic couplings within the $^3E$ level arising from interactions with non-symmetric modes, the polarisation of the luminescence band should be the same as the lower branch ($\sigma$-polarisation). Davies and Hamer observed that the one-phonon peak was only slightly depolarised ($<10\%$ $\pi$-polarisation) and thus concluded that the Jahn-Teller effect was very weak in $^3E$ and not capable of explaining the differences in the absorption and luminescence bandshapes.

It is an important point to note that even though the ensemble was being excited with broadband illumination in the experiment performed by Davies and Hamer, the luminescence of the upper branch is much weaker than the lower ZPL branch. This implies that any population excited to the upper branch rapidly decays non-radiatively to the lower branch within the radiative lifetime of $^3E$, and that there exists no commensurate non-radiative process in the opposite direction from the lower branch to the upper branch, thereby making the emission from the lower branch much more intense than from the upper branch. The dominance of the non-radiative decay from the upper branch to the lower branch is due to the presence of the large stress splitting and the low temperature conditions of Davies' and Hamer's measurement.  Consequently, it appears that in using such high stress and low temperature, Davies and Hamer observed just the small Jahn-Teller effect that arises from the time-independent mixing of the vibronic levels of $^3E$ due to relatively strong interactions with pseudo-local phonon modes of $E$ symmetry. This is unlike the recent experiments performed by Fu et al \cite{fu09} and Kaiser et al \cite{kaiser09}  using single centres at low temperatures (similar to those employed by Davies and Hamer) with much smaller strain splittings ($<100$ GHz), where rapid two-way non-radiative transitions between the ZPL branches occurred. As will be now discussed, Fu et al and Kaiser et al observed the Jahn-Teller effect that arises from both the time-independent vibronic coupling due to $E$ symmetric pseudo-local modes and the rapid two-way transitions between the ZPL branches that are mediated by the continuum of non-local $E$ symmetric phonon modes.

In their low temperature ($\sim4$ K) single centre study, Kaiser et al \cite{kaiser09} performed two measurements to establish the polarisation selection rules for the absorption and emission of the NV$^-$ optical ZPL. The centres they studied exhibited small strain splittings (3-10 GHz). Their first experiment consisted of the polarised resonant excitation of one of the ZPL branches and the detection of the fluorescence intensity as a function of excitation polarisation. Kaiser et al observed that the ZPL branches were orthogonal in excitation polarisation with $\sim$100$\%$ contrast, implying that the ZPL absorption selection rules were almost perfect. Their second experiment consisted of the resonant excitation of one of the ZPL branches and the detection of the fluorescence polarisation. Kaiser et al observed that the emitted polarisation contrast was $<60\%$ for all of the centres they investigated and suggested that the variance of the emission contrast between centres is likely due to variations in the local environment between centres. When this second result is combined with the observation of Fu et al \cite{fu09} (in their similar low temperature single centre study) that the emission polarisation of each ZPL branch is well polarised with $\sim100\%$ contrast, they imply that the ZPL emission selection rules are perfect and that the emission depolarisation is occurring due to non-radiative population transfer between the ZPL branches within the radiative lifetime of $^3E$. This conclusion is consistent with the preceding discussion regarding the luminescence spectra of Davies and Hamer \cite{davies76} as well as the observation of orbital averaging of the $^3E$ fine structure at room temperature \cite{rogers09} discussed in the previous section.

\begin{figure}[hbtp]
\begin{center}
\includegraphics[width=0.3\columnwidth] {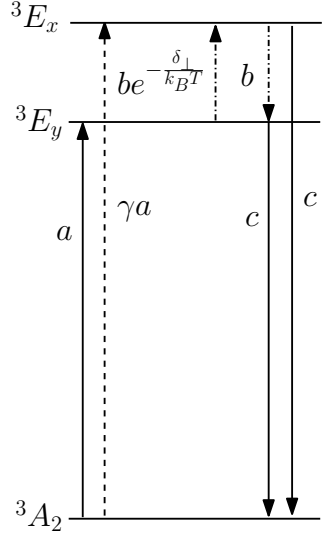}
\caption[The Jahn-Teller model of fluorescence depolarisation and homogeneous ZPL broadening of the NV$^-$ optical transition]{(a) The Jahn-Teller model of fluorescence depolarisation and homogeneous ZPL broadening of the NV$^-$ optical transition provided by Fu et al \cite{fu09}. The three electronic levels denote $^3A_2$ and the two strain split low temperature fine structure branches of $^3E$ that are orthogonally polarised. The solid arrows denote optical absorption and emission transitions with rates $a$ and $c$, respectively. The optical excitation is taken to be $y$-polarised, such that the dashed ($--$) arrow denotes the potential cross transition that populates the $x$-polarised branch due to the modification of off-resonance excitation selection rules by the Jahn-Teller effect. The chain ($-\cdot-$) arrows denote the non-symmetric phonon transitions between the orbital branches with decay rate $b$ and a Boltzmann prefactor $e^{-\delta/k_BT}$ in excitation, where $\delta_\perp$ is the strain splitting between the two branches.}
\label{fig:polarisationrateequationmodelreview}
\end{center}
\end{figure}

Fu et al also measured the temperature dependence of the emission polarisation contrast of the ZPL branches of single centres with various strain splittings (8-81 GHz) under polarised off-resonance (532 nm) excitation in the temperature range 10-50 K. In agreement with the resonant excitation - fluorescence detection measurements of Kaiser et al, Fu et al observed $<60\%$ polarisation contrast at all temperatures. Note the subtle, but complementary, difference between the two investigations is that Fu et al excited into the absorption band and measured just the emission polarisation of the ZPL branches, whereas Kaiser et al excited into the ZPL and measured the polarisation of the entire emission band. Furthermore, Fu et al observed that the emission polarisation contrast depended significantly on temperature, such that by 40 K, the contrast had reduced to $\sim0\%$ and was independent of the excitation polarisation (refer to figure \ref{fig:depolarisationdiffusionintro}). Fu et al noted that the temperature dependence of the emission polarisation contrast differed between individual centres and demonstrated a correlation with strain. In order to explain their observations, both Kaiser et al and Fu et al used a three-level rate equation model to describe the dynamics of the optical transitions between $^3A_2$ and $^3E$ and the non-radiative transitions between the fine structure orbital branches of $^3E$. Fu et al provided a more general model and that model is depicted in figure \ref{fig:polarisationrateequationmodelreview}. The key difference in the models of Fu et al and Kaiser et al was the inclusion of the excitation cross transition with rate $\gamma a$ by Fu et al. Kaiser et al did not include this cross transition because it represents imperfect ZPL absorption selection rules for their case of resonant excitation, which is contradictory to their first observation. Fu et al included the cross transition to account for the possibility of imperfect absorption selection rules when the optical transition is excited off-resonance. For individual centres with strain splittings of $\sim8$ GHz, Fu et al found that the non-radiative rate between the $^3E$ orbital branches depended on temperature as the fifth power of temperature $b/c\propto T^5$, and that the cross transition ratio was $\gamma\sim40\%$. The temperature dependence of the non-radiative rate is indicative of a two-phonon Raman transition between the orbital branches involving phonons with frequencies much greater than the strain splitting of the branches. Given the $T^5$ dependence of the non-radiative transitions between the branches, the cross transition rate for off-resonance excitation governs the temperature dependence of the polarisation contrast at very low temperatures, such that there remains $<60\%$  polarisation contrast in the limit of zero temperature. Fu et al attributed the origins of the cross transition and their consequence of imperfect polarisation contrast at very low temperature to the Jahn-Teller effect, but did not provide a thorough explanation.

Based upon the observations of Davies and Hamer \cite{davies76}, up to $\sim10\%$ of the $>40\%$ depolarisation observed by Fu et al at very low temperature can be attributed to temperature independent cross transitions allowed by the time-independent Jahn-Teller effect arising from the couplings of the vibronic levels of $^3E$ that are induced by $E$ symmetric pseudo-local modes. A small amount of the remaining unexplained depolarisation at low temperature can be attributed to direct one-phonon transitions between the ZPL branches that was neglected in the rate equation model of Fu et al. However, it is evident that a significant portion of low temperature depolarisation remains unexplained and is an important issue in the understanding of the NV$^-$ centre to resolve. One possible avenue of explanation is a strain dependence of the time-independent Jahn-Teller effect, such that in low strain conditions, the depolarisation arising from the time-independent Jahn-Teller effect is much greater than the $\sim10\%$ depolarisation observed by Davies and Hamer in high strain conditions.

The rapid non-radiative transitions between the ZPL branches also contribute to the homogeneous broadening of the ZPL. In their low temperature single centre study, Fu et al \cite{fu09} performed the first direct measurements of the homogeneous ZPL width as a function of temperature. Fu et al's single centre measurements were unlike the ensemble measurements performed by Davies \cite{davies74} and later by Hizhnyakov et al \cite{Hizhnyakov03}, who only obtained the homogeneous ZPL width indirectly via a deconvolution with the ensemble inhomogeneous lineshape.  Fu et al found that the upper ZPL branch homogeneously broadens also as the fifth power of temperature, which correlates with the two-phonon Raman transitions between the orbital branches of $^3E$ that govern the depolarisation of the optical emission. Given that the symmetric mode model of Davies \cite{davies74} yields a significantly different $T^7$ homogeneous broadening of the ZPL, it may be concluded that the fit that Davies obtained via deconvolution with a temperature dependent inhomogeneous lineshape was fortuitous and that the symmetric mode model cannot explain the homogeneous broadening of the ZPL. The observation of a $T^5$ homogeneous broadening is supported by the ultrafast photon echo measurements of ensembles at low temperature performed by Lenef et al \cite{lenef96b}, who observed a $T^5$ dephasing of the upper ZPL branch. However, Lenef et al observed a $T^3$ dephasing of the lower ZPL branch (not measured by Fu et al), which is supported by the indirect ensemble measurements of the inhomogeneously broadened ZPL performed by Hizhnyakov et al \cite{Hizhnyakov03}. Lenef et al speculated that the $T^3$ dephasing arose from the combination of the Jahn-Teller effect and large crystal strain, but did not provide any specific details. Alternatively, Hizhnyakov et al speculated that the $T^3$ broadening is due to a combination of a strong linear and weak quadratic Jahn-Teller effects, but there is no other evidence of a strong linear effect. Consequently, the homogeneous broadening of the lower ZPL branch still requires further experimental investigation, preferably by the direct means employed by Fu et al.

\begin{figure}[hbtp]
\begin{center}
\includegraphics[width=0.5\columnwidth] {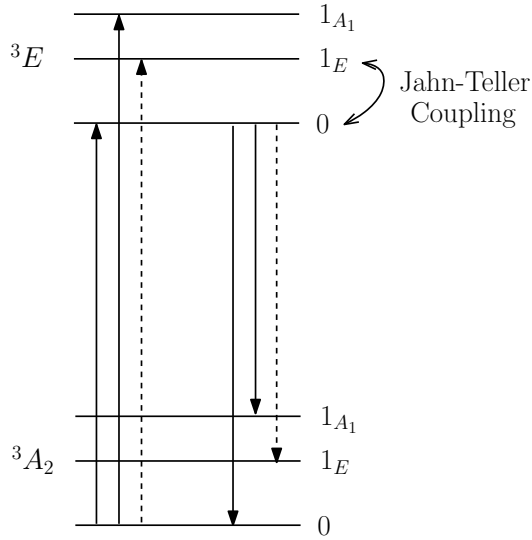}
\caption[The Jahn-Teller model of the NV$^-$ optical transition proposed by Gali et al]{The Jahn-Teller model of the NV$^-$ optical transition proposed by Gali et al \cite{gali11}. The vibronic levels are denoted by the electronic state (left) and vibrational occupation (right) of the $A_1$ and $E$ modes that Gali et al calculated to have vibrational energies 77.0 meV and 63.3 meV associated with $^3A_2$ and energies 76.9 meV and 63.9 meV associated with $^3E$, respectively. The solid arrows indicate optical transitions originating from the ground vibronic levels of each electronic state that are allowed in the absence of the Jahn-Teller effect and the dashed arrows indicate the transitions that are allowed as a consequence of the effect. The transitions will result in a double one-phonon peak in both the absorption and luminescence bands in conflict with experiment \cite{davies76}. Gali et al did not identify the dashed transition in emission.}
\label{fig:galijahntellermistakereview}
\end{center}
\end{figure}

In spite of the conclusive empirical evidence of the presence of the Jahn-Teller effect in $^3E$, there has not yet been a thorough attempt at a Jahn-Teller explanation of the differences in the NV$^-$ optical absorption and luminescence bands. In their \textit{ab initio} calculation, Gali et al \cite{gali11} identified $A_1$ and $E$ pseudo-local modes with vibrational energies 76.9 meV and 63.9 meV associated with $^3E$, respectively, and suggested that if the Jahn-Teller effect was present in $^3E$, then optical transitions would be allowed from the ground vibronic level of $^3A_2$ to the first vibronic levels of $^3E$ corresponding to these modes (refer to figure \ref{fig:galijahntellermistakereview}), thereby yielding a double one-phonon peak in absorption. Whilst this conclusion was correct, Gali et al failed to identify that if this was the case then the modified selection rules would induce a similar doubling of the one-phonon peak in emission arising from the first vibronic levels of $^3A_2$ corresponding to the same $A_1$ and $E$ modes, which have ground state vibrational energies 77.0 meV and 63.3 meV, respectively. Gali et al's assertion is thus clearly incorrect. Another \textit{ab initio} calculation performed by Zhang et al \cite{zhang11} provided additional evidence for a weak Jahn-Teller effect in $^3E$ by finding that the $^3E$ electronic energy is reduced by $\sim$6.4 meV, if the symmetry of the centre is relaxed from $C_{3v}$ to $C_{1h}$. Zhang et al identified that the mode with a displacement coordinate equivalent to the symmetry lowering relaxation had an energy much greater than $\sim$6.4 meV. Consequently, this result implies a weak dynamic Jahn-Teller effect where the zero-point energy is greater than the Jahn-Teller distortion energy \cite{zhang11}. In summary, the empirical and \textit{ab initio} evidence indicates the presence of a weak dynamic Jahn-Teller in $^3E$ and the thorough theoretical treatment of such an effect remains an outstanding issue of the NV$^-$ centre that needs to be addressed.

\subsection{The NV$^-$ infrared vibronic band}

The infrared vibronic band of NV$^-$ differs to the optical band in several ways. Since the infrared band occurs between the intermediate singlet states ($^1A_1$ and $^1E$), it has so far only been observed in fluorescence via the excitation of the centre's optical transition and the non-radiative transfer of population to the upper singlet $^1A_1$ \cite{rogers08,manson10,acosta10b}. Consequently, the emission band has only been measured to date and there is no observation of the absorption band available for comparison. The emission band (refer to figure \ref{fig:infraredbandreview}) exhibits a one-phonon peak with an energy of 42.6 meV from the ZPL and a temperature independent total Huang-Rhys factor of $S\sim1$ \cite{rogers08}. The infrared emission band is thus very different to the optical band, where several phonon peaks occur at energies that are almost perfect multiples of the one phonon energy ($\sim67$ meV) \cite{davies76}. A further difference between the infrared and optical bands is the temperature independence of the infrared band \cite{rogers08,acosta10b}. Unlike the optical band, the features of the infrared emission band have been observed to be approximately temperature independent \cite{rogers08} and the infrared ZPL has been observed to not significantly broaden or shift in energy with temperature \cite{acosta10b}. Applying the symmetric mode model, the temperature independence of the infrared emission band together with its small Huang-Rhys factor imply that interactions with symmetric modes do not play a significant part in forming the infrared band. This conclusion is consistent with a simple molecular model, where since $^1A_1$ and $^1E$ belong to the same MO configuration $a_1^2e^2$, there is no axial shift of electronic charge involved in the infrared transition and therefore, no excitation of symmetric modes. Hence, it appears that the infrared band predominately arises due to interactions with non-symmetric modes.

\begin{figure}[hbtp]
\begin{center}
\includegraphics[width=0.575\columnwidth] {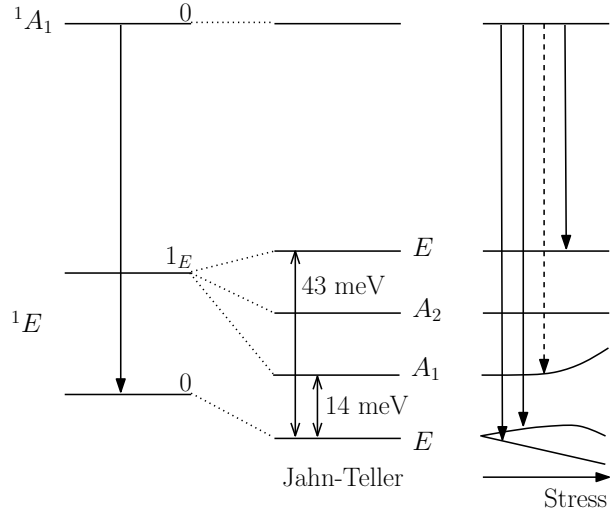}
\caption[The Jahn-Teller model of fluorescence depolarisation and homogeneous ZPL broadening of the NV$^-$ optical transition]{(a) The vibronic structure related to the infrared band of NV$^-$ \cite{manson10}. The ground ($0$) vibronic level of $^1A_1$ and the ground and first excited ($1_E$) vibronic levels of $^1E$ corresponding to the occupation of a pair of $E$ symmetric modes in the absence of the Jahn-Teller effect is depicted on the left hand side. The vibronic splittings due to linear and quadratic Jahn-Teller interactions are depicted in the centre and the interactions of the split vibronic levels in the presence of non-symmetric stress is depicted on the right hand side. Solid arrows originating from $^1A_1$ denote allowed infrared transitions in the absence of the Jahn-Teller effect (left) and in the presence of the effect (right). The dashed arrow denotes the additional infrared transition to the $A_1$ vibronic level that is allowed in the presence of stress. The solid arrows in the centre denote the known vibronic energies \cite{manson10}.}
\label{fig:1EvibronicLACreview}
\end{center}
\end{figure}

As discussed in the previous section and consistent with the preceding conclusion, the Jahn-Teller effect has been observed in $^1E$ via a vibronic LAC under uniaxial stress by Manson et al \cite{manson10}. The presence of the LAC can be explained by first identifying that in the absence of stress, the Jahn-Teller interaction of $^1E$ with a single pair of $E$ symmetric vibrational modes gives rise to vibronic levels of $E$, $A_1$ and $A_2$ symmetry, as depicted in figure \ref{fig:1EvibronicLACreview} \cite{manson10}. The ground vibronic state of $E$ symmetry splits in the presence of non-symmetric strain and interacts quadratically with the next vibronic level of $A_1$ symmetry \cite{manson10}. At large enough stress, the quadratic interaction between the vibronic levels will result in a LAC. Importantly, the $A_1$ vibronic level is not observed in the infrared spectra until the LAC is reached, since transitions from $^1A_1$ to the $A_1$ and $A_2$ vibronic levels of $^1E$ are forbidden unless there is significant stress mixing of those vibronic levels with the ground vibronic level of $^1E$. This behaviour enabled the identification of the $A_1$ vibronic level and furthermore, implied that the lower vibronic levels of $^1E$  are energetically well spaced, as there were no additional LACs observed \cite{manson10}.

From their stress measurements, Manson et al inferred that the $A_1$ vibronic level is $\sim 14.3$ meV above the ground vibronic level of $^1E$ at zero stress. The energy of the $E$ vibronic level above the ground vibronic level was directly identified from the emission band as being at $42.6$ meV, since the Jahn-Teller coupling of the lower $E$ vibronic levels of $^1E$ allows radiative transitions between them and $^1A_1$ with and without the application of stress \cite{manson10}. Consequently, the $A_2$ vibronic level must exist in the range $14.3-42.5$ meV  with an energy that is sufficiently greater than $14.3$ meV in order to not be observed in the uniaxial stress measurements. As concluded by Manson et al, the observed vibronic levels of $^1E$ can only occur if both linear and quadratic Jahn-Teller interactions are significant. In addition, since a vibronic LAC could only occur in $^1E$ (as $^1A_1$ cannot interact with non-symmetric modes) and that it must be on the high energy side of the $^1E$ ground vibronic level (as there are obviously no vibronic levels below the ground vibronic level), the observation of the vibronic LAC enabled the unambiguous ordering of the intermediate $^1A_1$ and $^1E$ singlets (as the LAC is observed in the IR emission band) \cite{manson10}.

\subsection{The NV$^0$ optical vibronic band}

The optical band of NV$^0$ has aspects that are similar to both the optical and infrared bands of NV$^-$. Since the NV$^0$ $^2E(a_1^2e)-$$^2A_1(a_1e^2)$ optical transition involves a similar change in MO configuration to the NV$^-$ optical transition, interactions with symmetric modes contribute significantly to the NV$^0$ optical band. The low temperature total Huang-Rhys factor of the NV$^0$ optical band is $S\sim3.3$ and the ZPL energy, intensity and width are dependent on temperature \cite{zaitsev}. Additionally, the Jahn-Teller effect is observed to be present in $^2E$ and it produces an analogous vibronic LAC as observed in $^1E$ of NV$^-$ under uniaxial stress \cite{davies79}. However, the energy ordering of the $A_1$ and $A_2$ vibronic levels of $^2E$ is unknown since the orbital symmetry of the optically excited $^2A$ of NV$^0$ has not been established. The lowest energy $A$ vibronic level of $^2E$ occurs at $\sim12$ meV, which is a very similar energy to the $A_1$ level of $^1E$ \cite{davies79}. Although there does not currently exist a detailed analysis of the NV$^-$ infrared band, from first impressions it is expected that there will be significant similarity between the Jahn-Teller interactions in $^2E$ of NV$^0$ and in $^1E$ of NV$^-$.

\subsection{The temperature dependence of the fine structure of the NV$^-$ $^3A_2$ level}

Apart from their presence in the optical and infrared bands of NV$^-$ and NV$^0$, potential vibronic effects have also been observed in the temperature dependence of the NV$^-$ ground state fine structure. In reference to the NV$^-$ ground state spin-Hamiltonian and potential for interactions with strain fields, Acosta et al \cite{acosta10,acosta11} observed that the fine structure splitting $D_{gs}$ between the $m_s =0$ and $m_s=\pm1$ spin sub-levels varied with temperature and that the strain parameter $\delta_\perp=\sqrt{\delta_x^2+\delta_y^2}$ that splits the $m_s=\pm1$ spin sub-levels in the presence of transverse strain did not vary with temperature. Acosta et al attempted to describe the temperature dependence of $D_{gs}$ by a model based purely on the thermal expansion of the crystal lattice, but were only partially successful. They ultimately concluded that additional vibronic effects beyond lattice expansion must be responsible for the temperature dependence of $D_{gs}$. The identification and description of these vibronic effects remain outstanding issues concerning NV$^-$.

By comparing the temperature dependence of $D_{gs}$ and that of the NV$^-$ optical ZPL energy it is evident that their dependence is very similar, which suggests that quadratic symmetric mode coupling could be responsible for the temperature dependent shift of $D_{gs}$, just as it was for the optical ZPL. This proposal is supported by the observation that the strain splitting $\delta_\perp$ does not change with temperature, thereby implying that interactions with non-symmetric modes are not sufficient to reduce the splitting as they do to the analogous splitting of $^3E$ in the process of orbital averaging. Consequently, it is clear that there is potentially a great deal of information concerning vibronic interactions in the NV centre available through the study of the NV$^-$ ground state temperature dependence.

This section has been principally focussed on the vibronic interactions that occur within a given electronic state and give rise to directly observable vibronic structure. In the next section, the key observations of the vibronic interactions between electronic states that give rise to non-radiative transitions will be discussed. As will be shown, these transitions have a significant role in the optical and spin dynamics of NV$^-$ and are central to understanding the properties of optical spin-polarisation and long lived spin coherence.

\section{Optical and spin dynamics}
\label{section:reviewopticalandspindynamics}

\begin{figure}[hbtp]
\begin{center}
\includegraphics[width=0.85\columnwidth] {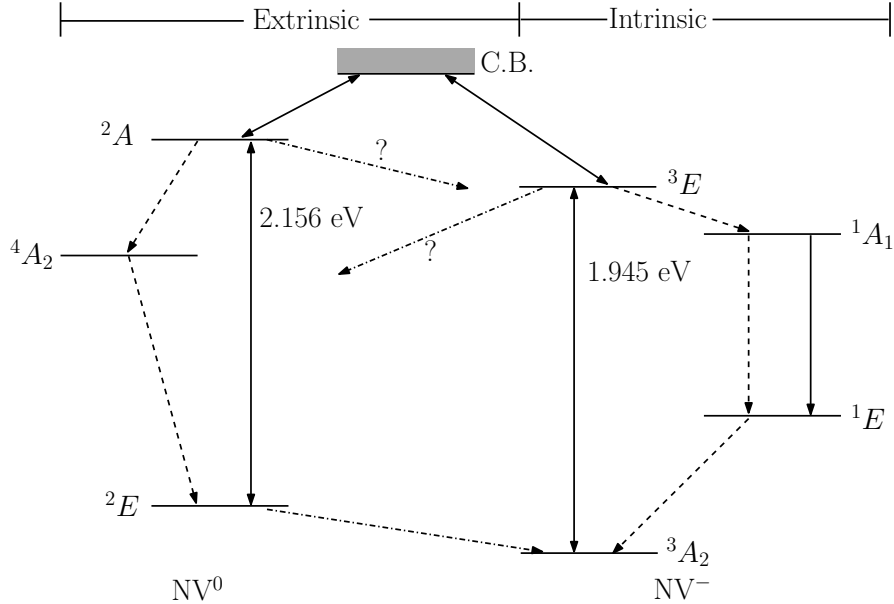}
\caption[The current understanding of the extrinsic and intrinsic aspects of the optical dynamics of NV$^-$]{The current understanding of the extrinsic and intrinsic aspects of the optical dynamics of NV$^-$. Solid arrows denote radiative transitions, dashed ($--$) arrows denote non-radiative transitions within a charge state and chain ($-\cdot-$) arrows denote non-radiative transitions between charge states. Electronic states of NV$^0$ are depicted on the left hand side, whereas electronic states of NV$^-$ are depicted on the right hand side. The optical ZPL energies of each charge state are as indicated. The upper most and central electronic state represents the continuum of states derived from the occupation of the conduction band. The conduction band states are believed to form an intermediate step in the quadratic photoconversion process between the NV charge states. In addition to the non-radiative transitions between the ground states of the two charge states, only non-radiative transitions that originate from their optically excited states are also depicted. Note that the final states of these transitions are unknown (as indicated by question marks) and that other non-radiative transitions between the charge states can potentially occur between any of the electronic states. }
\label{fig:opticadynamicsgeneralreview}
\end{center}
\end{figure}

Manson et al \cite{manson06,manson07} were the first to distinguish the fast and slow components of the NV$^-$ optical dynamics through dual optical pulse measurements.  The fast dynamics were attributed to intrinsic radiative and non-radiative transitions between the electronic states of NV$^-$ and the slow dynamics were attributed to extrinsic transitions involving photoconversion to NV$^0$. The current understanding of the intrinsic and extrinsic aspects of the optical dynamics of NV$^-$ are depicted in figure \ref{fig:opticadynamicsgeneralreview}. The intrinsic dynamics are responsible for optical spin-polarisation and readout. The extrinsic dynamics are responsible for the establishment of a dynamical equilibrium of the charge states when both are simultaneously excited optically as well as the photoconversion of NV$^-$ into NV$^0$ when just NV$^-$ is excited. Due to their significantly different timescales, the extrinsic and intrinsic optical dynamics of NV$^-$ can be treated individually in most conditions.

\subsection{The extrinsic optical dynamics of NV$^-$}

It is currently understood that the extrinsic dynamics involve both radiative and non-radiative transitions. The radiative transitions between the charge states are believed to occur between their optically excited levels via intermediate valence/conduction band derived levels \cite{waldherr11}. The radiative transitions between the charge states therefore correspond to photoconversion processes that are quadratic in the optical excitation intensity. Furthermore, since the radiative transitions do not explicitly involve electron donors/acceptors, but simply the inherent conduction/valence bands of the crystal, the radiative transitions are expected to be independent of the concentrations of donors/acceptors in the crystal. Differing from the radiative transitions, the non-radiative transitions occur directly between the electronic states of the charge states. Consequently, the non-radiative transitions must correspond to the tunneling of an electron between the NV centre and proximal electron donors/acceptors. Such tunneling will be highly dependent on the distance between the NV centre and the donors/acceptors, which will be itself statistically dependent on the concentrations of donors/acceptors. Apart from the non-radiative decay from the ground electronic state of the unstable charge state (NV$^0$ in the case of figure \ref{fig:opticadynamicsgeneralreview}) to the ground electronic state of the stable charge state (NV$^-$ in the case of figure \ref{fig:opticadynamicsgeneralreview}) in the absence of optical illumination \cite{gaebel06}, the other non-radiative transitions originate from the optically excited levels of each charge state. Thus, the non-radiative transitions correspond to photoconversion processes that are linear in optical excitation intensity \cite{manson05}.

The first indication that quadratic photoconversion processes occur in the NV centre was identified by Redman et al \cite{redman92b}, who observed a weak photoconductive current consistent with photoionisation occurring from a saturated excited electronic state of NV$^-$. This observation was later supported by the single centre studies \cite{beveratos00,hubbard07} that attempted to fit the observed NV$^-$ fluorescence autocorrelation function for different excitation intensities using the three level rate equation models that were discussed in section \ref{section:reviewelectronicstructure}. In order to correctly fit the autocorrelation function, the transition rates between $^3E$ and the long lived metastable state of the model had to be considered to be proportional to the optical excitation intensity. It is now understood that under the strong excitation intensities used in those studies, the long lived metastable state of the three level models effectively represented NV$^0$ and thus, the intensity dependent transitions to and from the metastable state were in fact representative of the radiative transitions between the charge states \cite{manson07}.  However, the conclusive evidence that quadratic photoconversion processes occur in the NV centre was provided by Waldherr et al \cite{waldherr11}, who observed that when NV$^-$ was continuously excited on resonance (1.945 eV) the fluorescence would decay exponentially and if, when the emission extinguished, an off-resonance (2.32 eV) excitation was instead applied, the NV$^-$ fluorescence would recover exponentially. Waldherr et al found that the decay and recovery rates were quadratic in excitation intensity for low intensities and linear at higher intensities. Furthermore, they identified by observing the changes in NMR spectra that under resonant/off-resonance optical excitation, the centre was being converted from NV$^-$ to NV$^0$ and back again. Hence, this identification and the dependence of the decay/recovery rates on excitation intensity conclusively demonstrated that radiative transitions between the charge states occur in each direction from the optical excited states of NV$^-$ and NV$^0$, such that the excited states become saturated at sufficiently high excitation intensities \cite{waldherr11}.

Direct observations of the linear photoconversion processes were made by Manson et al \cite{manson05}, who found that the ratio of the ZPL emission intensities of NV$^-$ and NV$^0$ follows a saturation curve when excited by an optical energy (2.32 eV) that is within the optical absorption bands of each charge state. The implication of the occurrence of linear processes is that the conversion between the charge states must proceed in both directions via a non-radiative decay from the optically excited states of the charge states. As discussed, such non-radiative decay requires the tunneling of an electron between the NV centre and proximal electron donors/acceptors, whose distances from the NV centre are statistically determined by the concentrations of donors/acceptors \cite{manson05}. Consistent with this conclusion, the ensemble measurements of Manson et al were performed using type Ib diamond samples, which contain high substitutional nitrogen (electron donor) concentrations. The conclusion also implies that linear processes are expected to occur in NV centres in type Ib nanodiamonds, where there are significant concentrations of both substitutional nitrogen and surface traps, however no studies have been conducted to date to confirm.

Although the extrinsic dynamics appear to be reasonably established, the understanding of their origins is certainly not. To date, there has not been a thorough microscopic model developed to explain the various radiative and non-radiative transitions between the charge states beyond simple speculations of the role of conduction and/or valence band states and electron donors/acceptors. There has also been no attempt to relate the dynamics to the concentrations of the charge states, spectral stability of the ZPLs or material properties. Consequently, the microscopic origins of the extrinsic dynamics and the stable charge state of the centre remain outstanding issues that need to be addressed.

\subsection{The intrinsic optical dynamics of NV$^-$}

\begin{figure}[hbtp]
\begin{center}
\includegraphics[width=0.6\columnwidth] {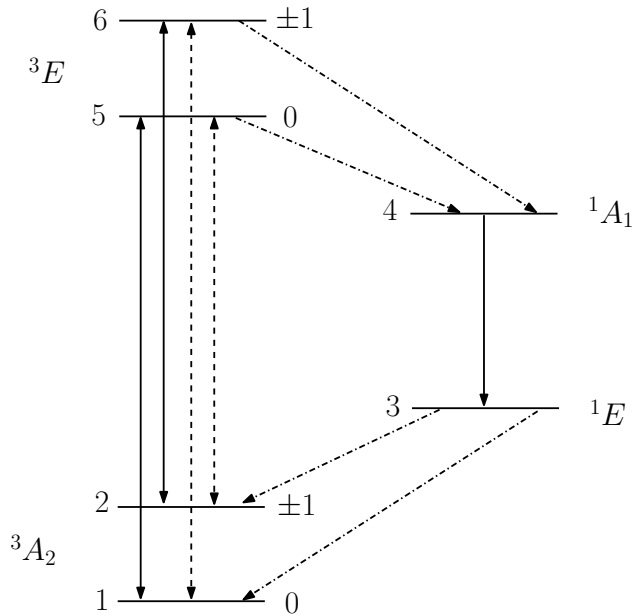}
\caption[The six level model of the intrinsic dynamics of NV$^-$]{The six level model of the intrinsic dynamics of NV$^-$. Electronic levels are labelled by their level index (1-6) on the left hand side and spin sub-level on the right hand side (triplet states only). The solid arrows denote transitions that conserve spin quantum numbers, dashed ($--$) arrows denote cross transitions between triplet spin sub-levels and chain ($-\cdot-$) arrows denote intersystem crossing transitions between triplet and singlet states.}
\label{fig:sixlevelmodelreview}
\end{center}
\end{figure}

Turning now to the intrinsic dynamics of NV$^-$, the intrinsic dynamics are most efficiently discussed using a six level model, such as the one depicted in figure \ref{fig:sixlevelmodelreview}. The six level model is an accurate representation of the NV$^-$ electronic structure at room temperature, however more complicated models are required when magnetic fields are applied such that the spin sub-levels interact or when operating at low temperatures with resonant excitation. Measurements of the transition rates of the six level model are contained in table \ref{tab:sixlevelmodelratereview}. Note that transitions between the spin sub-levels of each triplet state are not included in the six level model of figure \ref{fig:sixlevelmodelreview} since they are several orders of magnitude slower than the transitions of the optical dynamics. The transitions between the spin sub-levels of each triplet are considered to be aspects of the NV$^-$ spin dynamics, which will be discussed shortly. The other notable absences from the six level model of figure \ref{fig:sixlevelmodelreview} are the transitions directly from $^3E$ to $^1E$ and from $^1A_1$ to $^3A_2$. These transitions are of sufficiently high energy ($>1.2$ eV) that radiative decay will be expected to dominate any many-phonon non-radiative decay. Since there has been no detection of these transitions in the fluorescence spectra of NV$^-$ to date, they are believed to be negligible compared to the other transitions of the intrinsic optical dynamics.

\begin{table}
\caption[Experimental measurements of the NV$^-$ excited state lifetimes tabulated by reference]{\label{tab:sixlevelmodelratereview} Experimental measurements of the NV$^-$ excited state lifetimes tabulated by reference. The lifetimes are defined in terms of the transition rates $k_{ij}$ of the six level model depicted in figure \ref{fig:sixlevelmodelreview} by $T_i = 1/\sum_jk_{ij}$. Uncertainty ranges are included if provided by sources.}
\begin{tabular}{llllll}
\noalign{\smallskip}
\toprule
Reference & T (K) & $T_3$ (ns) & $T_4$ (ns) & $T_5$ (ns) & $T_6$ (ns) \\
\midrule
Collins$^e$ \cite{collins83} & 77-700 &  &  & $12.9\pm0.1$$^a$ &  \\
Lenef$^e$ \cite{lenef96b} & 77 &  &  & 12.96$\pm0.14$$^a$ &  \\
Manson$^e$ \cite{manson06} & Room & 300$^a$ &  &  &  \\
Batalov$^s$ \cite{batalov08} & Room &  &  & 12.0 & 7.8 \\
Acosta$^e$ \cite{acosta10} & 4.4-70 & 462$\pm10$$^b$ & 0.9$\pm0.5$ & 10.0$\pm0.3$$^a$ &  \\
Robledo$^s$ \cite{robledo10} & 8 &  &  & 10.9 &  \\
Robledo$^s$ \cite{robledo11} & 300 & 371$\pm6$$^b$ & & 13.7$\pm0.1$ & 7.3$\pm0.1$ \\
Beveratos$^{s,n}$ \cite{beveratos01} & Room &  &  & 25$^a$ &  \\
Beveratos$^{s,n}$ \cite{beveratos02} & Room &  &  & 23$^a$ &  \\
Neumann$^{s,n}$ \cite{neumann09} & Room &  &  & 23 & 12.7 \\
\bottomrule\noalign{\smallskip}
\multicolumn{6}{l}{$^a$did not distinguish triplet spin sub-levels/singlets; $^b$low T$<$5 K lifetime}\\
\multicolumn{6}{l}{$^e$ensemble; $^s$single centre; $^n$nanodiamond}
\end{tabular}
\end{table}

Collins et al \cite{collins83} were the first to measure the lifetime of $^3E$ and obtained 12.9$\pm$0.1 ns for NV$^-$ ensembles in natural type Ib diamond. Collins et al stated that the lifetime was independent of temperature in the range 77-700 K, which has been subsequently supported by another study \cite{acosta10b}, but seemingly contradicted by a recent third study performed by Toyli et al \cite{toyli12}. The inconsistency of Toyli et al's lifetime measurements has not been resolved and it is an important issue to be pursued. For the purposes of this discussion, the lifetime will be considered to be temperature independent. The first single centre studies \cite{gruber97,drab99} reported the observation of biexponential decay of the $^3E$ fluorescence with time instead of the monoexponential decays observed by Collins et al and other previous ensemble measurements \cite{lenef96b}. Using three level models, the single centre studies attributed the biexponential decay to the repopulation of $^3E$ from the long lived metastable state \cite{gruber97,drab99}. This is now known to be incorrect and the existence of a biexponential decay is the result of the spin sub-levels of $^3E$ having different lifetimes due to differing non-radiative decay rates to the intermediate singlet states. Batalov et al \cite{batalov08} were the first to measure the individual lifetimes of the $^3E$ spin sub-levels in bulk diamond. The lifetimes obtained by Batalov et al are largely consistent with the later measurements performed by Robledo et al \cite{robledo11}. The $^3E$ lifetimes approximately double in nanodiamond crystals that are much smaller than the fluorescence wavelength. This change is simply due to the reduction of the radiative emission rate induced by the decrease of the effective refractive index of the medium surrounding the NV centre. Beveratos et al \cite{beveratos01} were the first to measure the $^3E$ lifetime of a single centre in such a nanodiamond and observed a monoexponential decay corresponding to a lifetime of 25 ns. Neumann et al subsequently measured the individual lifetimes of the $^3E$ spin sub-levels for a single centre in a nanodiamond.

The equality of the $^3E$ lifetime obtained from observations of monoexponential fluorescence decay and the $m_s=0$ spin sub-level lifetime obtained from observations of biexponential decay, strongly suggests that the experimental conditions in which monoexponential decay was observed were such that the centre had been spin-polarised into the $m_s=0$ spin-projection prior to the recording of the fluorescence decay curve \cite{manson06}. Such preliminary spin-polarisation could have occurred if optical pulses that were too intense or long or insufficiently spaced were used to conduct the lifetime measurement. Furthermore, the temperature independence of the lifetime of the $m_s=0$ sub-level, even up to temperatures (700 K) where the highest energy phonons ($\sim$160 meV) of the diamond lattice are appreciably occupied, unambiguously implies that the non-radiative decay from the $m_s=0$ sub-level is negligible compared to the radiative decay. This conclusion is supported by the low temperature optical Rabi experiments of Batalov et al \cite{batalov08}, in which nearly pure radiative relaxation and dephasing were observed. In the absence of any strong spin-dependent interactions between other electronic states and $^3E$ and $^3A_2$, which have not been detected to date in any of the uniaxial stress, magnetic and electric field studies, the radiative decay rates of $^3E$ will be spin independent \cite{manson06}. Thus, the difference between the lifetimes of the $^3E$ $m_s=0$ and $m_s=\pm1$ spin sub-levels is due to the $m_s=\pm1$ sub-levels undergoing greater non-radiative decay, which ultimately results in the spin-dependent fluorescence of NV$^-$ \cite{manson06}. Using the observed sub-level lifetimes, a simple expression can be derived that relates the non-radiative contributions to the decay of each spin sub-level
\begin{equation}
\frac{T_6}{T_5} = \frac{1+\gamma_0}{1+\gamma_{\pm1}}
\end{equation}
where $\gamma_{m_s}$ is the ratio of the non-radiative decay rate of sub-level $m_s$ to the $^3E$ radiative decay rate. In the limit $\gamma_0\approx0$, the lifetime measurements of Batalov et al result in the estimation $\gamma_{\pm1}\approx0.54$, which increases if $\gamma_0>0$. Hence, it may be concluded that the non-radiative decay rate out of the $^3E$ $m_s=\pm1$ spin sub-levels is at least $50\%$ of the $^3E$ radiative decay rate \cite{manson06}. This conclusion is consistent with the lifetime measurements of single centres in nanodiamond crystals performed by Neumann et al.  Such a significant contribution of non-radiative transitions to the decay of the $^3E$ $m_s=\pm1$ spin sub-levels should ensure that their lifetime displays an observable temperature dependence. However, in the recent lifetime measurements of Toyli et al \cite{toyli12}, the lifetimes of the $m_s=\pm1$ spin sub-levels were observed to exhibit a small temperature reduction and the $m_s=0$ spin sub-levels were observed to undergo a much larger temperature reduction. These observations are inconsistent with the preceding discussion and pose an issue that needs to be resolved.

\begin{figure}[hbtp]
\begin{center}
\includegraphics[width=0.75\columnwidth] {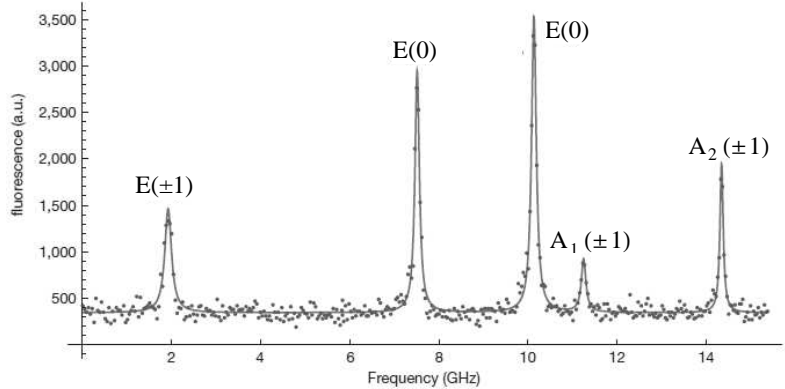}
\caption[Low temperature excitation spectra of the NV$^-$ $^3E$ fine structure]{Low temperature optical excitation spectra of the NV$^-$ $^3E$ fine structure obtained from a single centre using continuous microwave driving of the ground state magnetic resonance \cite{togan10}. The fine structure levels corresponding to the spectral lines are denoted by their spin-orbit symmetry and spin-projection. The splitting of the central $E(0)$ level into two lines signifies that there is a small strain present in the centre of $\sim3$ GHz. The lower energy $E(0)$ line is slightly less intense than the upper $E(0)$ line due to the small enhancement of the spin-spin coupling of the lower energy $E(0)$ branch with the $E(\pm1)$ level in the presence of strain.  The relative intensities of the upper two lines were used to infer their spin-orbit symmetry. The least intense line corresponds to the $A_1(\pm1)$ level since it has the most non-radiative decay paths. The highest energy line is the most intense out of the $m_s=\pm1$ lines and thus, corresponds to the $A_2(\pm1)$ level since it has the least number of non-radiative decay paths.}
\label{fig:toganfinestructurespectrareview}
\end{center}
\end{figure}

At low temperatures, each of the $^3E$ $m_s=\pm1$ fine structure levels are expected to have different lifetimes because of their different spin-orbit symmetry (refer to figure \ref{fig:NVexcitedstatefinestructurereview}). Simple symmetry arguments imply that the $E(\pm1)$ fine structure level can only decay to $^1A_1$ through the emission of $E$ symmetric phonons, whereas the $A_1(\pm1)$ level can decay directly to $^1A_1$  through the emission of $A_1$ phonons or in a two step process via the $E(\pm1)$ level through the emission of $E$ phonons. The $A_2(\pm1)$ fine structure level can only decay to $^1A_1$ in a two step process via the $E(\pm1)$ level through the emission of $E$ phonons. Consequently, based purely on the possible decay paths for each $m_s=\pm1$ fine structure level, $A_1(\pm1)$ is expected to have the shortest lifetime, $A_2(\pm1)$ the longest lifetime and $E(\pm1)$ an intermediate lifetime. As mentioned in section \ref{section:reviewelectronicstructure},  such simple lifetime arguments were used to determine the energy ordering of the $A_1$ and $A_2$ fine structure levels through the inspection of the low temperature excitation spectra of the $^3E$ fine structure (refer to figure \ref{fig:toganfinestructurespectrareview}) \cite{togan10}. At room temperature, orbital averaging is expected to lead to the single lifetime of the $m_s=\pm1$ spin sub-level in the six level model (refer to figure \ref{fig:sixlevelmodelreview}). A detailed model of the non-radiative decay mechanisms of the low temperature fine structure levels and their averaging at room temperature has not yet been developed and remains an outstanding issue concerning NV$^-$.

The radiative non spin-conserving transitions between spin sub-levels of the triplet states with different spin-projections occur due to the small mixing of the spin-projections in $^3E$ arising from spin-spin interaction (parameter $\lambda_{es}^\perp$ of the spin-Hamiltonian (\ref{eq:NVexcitedstateeffectiveHamiltonian})). Given the observed magnitude of the spin-spin interaction (refer to table \ref{tab:lowtempexcitedstatespinparametersreview}) being an order of magnitude smaller than the low temperature low strain fine structure splittings of $^3E$, the rates of the cross transitions are expected to be less than 1$\%$ of the spin-conserving radiative transition rate. This expectation has been supported by recent low temperature measurements using resonant excitation of the fine structure levels \cite{togan10}. As discussed in section \ref{section:reviewelectronicstructure}, in the presence of higher strain, the spin interaction causes the significant spin mixing in the lower branch of the $^3E$ low temperature fine structure that gives rise to optical spin-flip transitions. Therefore, the relative strength of the cross transitions are a function of the strain present at the NV centre. No model has yet been developed to describe the strain dependence of the cross transitions at room temperature. Regardless, it is expected that the radiative cross transitions play a minor role in the dynamics that lead to optical spin-polarisation, since the transitions are equal and thus, do not result in the preferential population of one particular spin-projection \cite{manson06,manson10}.

As discussed in section \ref{section:reviewelectronicstructure}, the lifetimes of the intermediate singlet states as well as their temperature dependence have been measured using NV$^-$ ensembles by Acosta et al \cite{acosta10}. The lifetime of $^1A_1$ was observed to be temperature independent in the temperature range $4.4-70$ K. Since it is currently believed that the non-radiative decay between the singlet states dominates the weak infrared decay \cite{rogers08}, the observed temperature independence of the lifetime of $^1A_1$ implies that the non-radiative decay out of $^1A_1$ occurs via high energy phonons, whose occupations do not appreciably change in the temperature range $4.4-70$ K. For example, the thermal occupation of a phonon with energy $67$ meV is $<3\times10^{-5}$ at 70 K. The lifetime of $^1E$, which is expected to undergo pure non-radiative decay to the ground $^3A_2$ state, on the other hand displays a significant temperature dependence \cite{acosta10,robledo11}. This signifies that the non-radiative decay between the singlet states occurs via different electron-phonon interactions than the decay from $^1E$ to $^3A_2$. Acosta et al found that the temperature dependence of the $^1E$ lifetime could be modeled by the emission of a phonon of energy $15(1)$ meV. The later single centre lifetime measurements of Robledo et al \cite{robledo11} were consistent with those of Acosta et al and obtained a phonon energy of $16.6\pm0.9$ meV.

Robledo et al concluded that the phonon energy was indicative of a third electronic state $\sim16$ meV below the lower energy electronic state involved in the infrared transition. Motivated by the \textit{ab initio} calculations of Ma et al \cite{ma10}, Robledo et al speculated that there existed three intermediate singlet states and that the lower energy electronic state involved in the infrared transition was $^1A_1$, which decayed via a $\sim16$ meV phonon to an even lower energy $^1E$. Given that Ma et al's calculation has been shown to be inconsistent with the observations of Manson et al \cite{manson10} and that the ordering of the singlets involved in the infrared transition has been established as depicted in figure \ref{fig:sixlevelmodelreview}, the speculation of Robledo et al is also brought into question. Given the detection by Manson et al of an $A_1$ vibronic level $\sim14.3$ meV above the ground vibronic level of $^1E$, the equivalence of the energy of the vibronic level and the observed phonon energies strongly suggests that the decay from $^1E$ involves the $A_1$ vibronic level instead of an unknown electronic state. Using this assertion, the non-radiative decay from $^1E$ to $^3A_2$ most likely originates from both the $^1E$ ground and $A_1$ (occupied in thermal equilibrium) vibronic levels and involves the emission of multiple high energy phonons that have no observable temperature dependence in the temperature ranges investigated by either Acosta et al or Robledo et al \cite{manson10}. This explanation is consistent with the vast majority of theoretical models and empirical observations to date, which do not indicate the presence of a third intermediate singlet state. However, this explanation is yet to be developed in any detail and must include the consideration of the Jahn-Teller effect in $^1E$ as well as the spin dependent electronic interactions that enable the non-radiative decay to $^3A_2$.

Using a sequence of ps optical pulses combined with the microwave control of the ground state electronic spin, Robledo et al also measured the ratio of the decay rates from $^1E$ to the spin sub-levels of $^3A_2$ and obtained $k_{31}/k_{32}=1.15\pm0.05$ for one centre and $1.6\pm0.4$ for a second centre. The small preference of the decay from the lower energy singlet $^1E$  to the $m_s=0$ spin sub-level implies that optical spin-polarisation into the $m_s=0$ spin-projection must result from the preferential decay from the $^3E$ $m_s=\pm1$ spin sub-levels to the higher energy singlet $^1A_1$, rather than the preferential decay from $^1E$ to the $^3A_2$ $m_s=0$ spin sub-level \cite{robledo11}. This implication can be demonstrated more quantitatively by applying the six level rate equation model to calculate the ground state spin-polarisation under continuous off-resonance excitation \cite{manson10}
\begin{equation}
{\cal P}_{gs} = \frac{P_0}{P_0+P_{\pm1}} = \frac{1+\gamma_0}{1+\gamma_0+\alpha(\beta+\gamma_0)}\approx1-\alpha\beta
\end{equation}
where $P_{m_s}$ are the populations of the ground state spin sub-levels under continuous off-resonance excitation, and $\alpha=k_{32}/k_{31}$ and $\beta=k_{54}/k_{64}$ are the ratios of the non-radiative decay rates from $^1E$ to the spin sub-levels of $^3A_2$ and from the spin sub-levels of $^3E$ to $^1A_1$, respectively. The approximation made in the above expression assumes that $\alpha\gamma_0\ll \alpha\beta \ll 1$. Consequently, in order for the spin-polarisation to be $\sim80\%$ (average quoted spin-polarisation from the vast range of spin-polarisations reported), $\alpha\beta$ must be $\sim0.2$. For example, given $\alpha\sim0.8$ as determined by Robledo et al, the above indicates that $\beta$ must be  $\sim0.2$ in order for ${\cal P}_{gs}\sim80\%$. Thus, it can be seen that reasonably high degrees of spin-polarisation can be achieved without the extreme differences in the decay rates from the spin sub-levels of $^3E$ to $^1A_1$ that would be expected if the transitions for one of the spin-projections were forbidden. It may be further concluded from the above expression that the dependence of spin-polarisation on the ratio of non-radiative transitions implies that spin-polarisation will exhibit a significant temperature dependence. This temperature dependence has been observed by Felton et al \cite{felton09}, who found using conventional EPR techniques that optical spin-polarisation decreased from 10-300 K. Spin-polarisation is therefore a reasonably subtle effect in NV$^-$, which may depend intricately on temperature, crystal strain and other extrinsic factors, thereby explaining the large variance in reported values. Hence, spin-polarisation requires a fully developed theoretical model for it to be studied in any detail.

\subsection{The spin dynamics of NV$^-$}

The spin dynamics of NV$^-$ include the transitions that occur between the spin sub-levels of a given triplet state that act to relax any population difference between them, as well as the fluctuating interactions that act to dephase the spin states. These two distinct processes are characterised (as for any spin system) by the spin relaxation rate $1/T_1$ and the spin homogeneous dephasing rate $1/T_2$. A further complication arises when measurements are performed on an ensemble of centres, where each centre has different local conditions, or when an ensemble of measurements of a single centre are performed, where each measurement is made in different conditions. For such ensemble measurements, the expectation value of a given observable must be averaged over the relevant distribution of conditions. This averaging leads to inhomogeneous spin dephasing that is characterised by the rate $1/T_2^\ast$, such that $T_2^\ast\leq T_2$. The spin relaxation and dephasing rates are sensitive to various factors including the densities and types of paramagnetic impurities present in the vicinity of the NV centre, the applied magnetic, electric and strain fields as well as electron-phonon interactions and temperature. Due to the many applications employing the spin of NV$^-$, its spin dynamics have been extensively studied and a sample of reported values are contained in table \ref{tab:spincoherencetimesreview}. Given the scope of this review being restricted to the intrinsic properties of the NV centre, only aspects of the dynamics that provide information about such properties will be discussed in detail.

\begin{table}
\caption[Experimental measurements of the NV$^-$ ground state spin relaxation and dephasing times tabulated by reference]{\label{tab:spincoherencetimesreview} Experimental measurements of the NV$^-$ ground state spin relaxation and dephasing times tabulated by reference. The magnetic field data is provided in units of Gauss or Tesla if known or operating frequency of EPR apparatus if utilised. }
\begin{tabular}{llllll}
\noalign{\smallskip}
\toprule
Reference & $T$ (K) & Mag. Field & $T_1$ (ms) & $T_2$ ($\mu$s) & $T_2^\ast$ ($\mu$s) \\
\midrule
Redman$^{e,b}$ \cite{redman91,glasbeek92} & 80 &  & 265 &  &   \\
Kennedy$^{e,b}$ \cite{kennedy02} & 1.5-100 & 35 GHz &  & 32 &   \\
Kennedy$^{e,b}$ \cite{kennedy03} & 300 & 35 GHz &  & 58 &   \\
Jelezko$^{s,n,b}$ \cite{jelezko04b} & Room &  &  & 1-2 &   \\
Jelezko$^{s,b}$ \cite{jelezko04} & Room &  &  & $<6$ &   \\
Popa$^{s,n,b}$ \cite{popa04} & Room & 0 &  & 0.9 &   \\
 &  & 100 G &  & 1.2 &   \\
Gaebel$^{s,a}$ \cite{gaebel06} & Room &  &  & 350 &   \\
Childress$^{s,a}$ \cite{childress06} & Room &  &  &  & 1.7 \\
Takahashi$^{e,b}$ \cite{takahashi08} & Room & 8 T & 7.7 & 6.7 &  \\
 & 40 & 8 T & 3.8$\times10^3$ &  &  \\
  & 20 & 8 T &  & 8.3 &  \\
   & 1.7 & 8 T &  & 250 &  \\
Balasub.$^{s,c}$ \cite{balasubramanian09} & Room &  &  & 1.8$\times10^3$ &  \\
Stanwix$^{e,a}$ \cite{stanwix10} & Room &  &  & 600 &  \\
\bottomrule\noalign{\smallskip}
\multicolumn{6}{l}{$^a$type IIa; $^b$type Ib; $^c$ isotopically pure type IIa}\\
\multicolumn{6}{l}{$^e$ensemble; $^s$single centre; $^n$nanodiamond}
\end{tabular}
\end{table}

\begin{figure}[hbtp]
\begin{center}
\mbox{
\subfigure[]{\includegraphics[width=0.55\columnwidth] {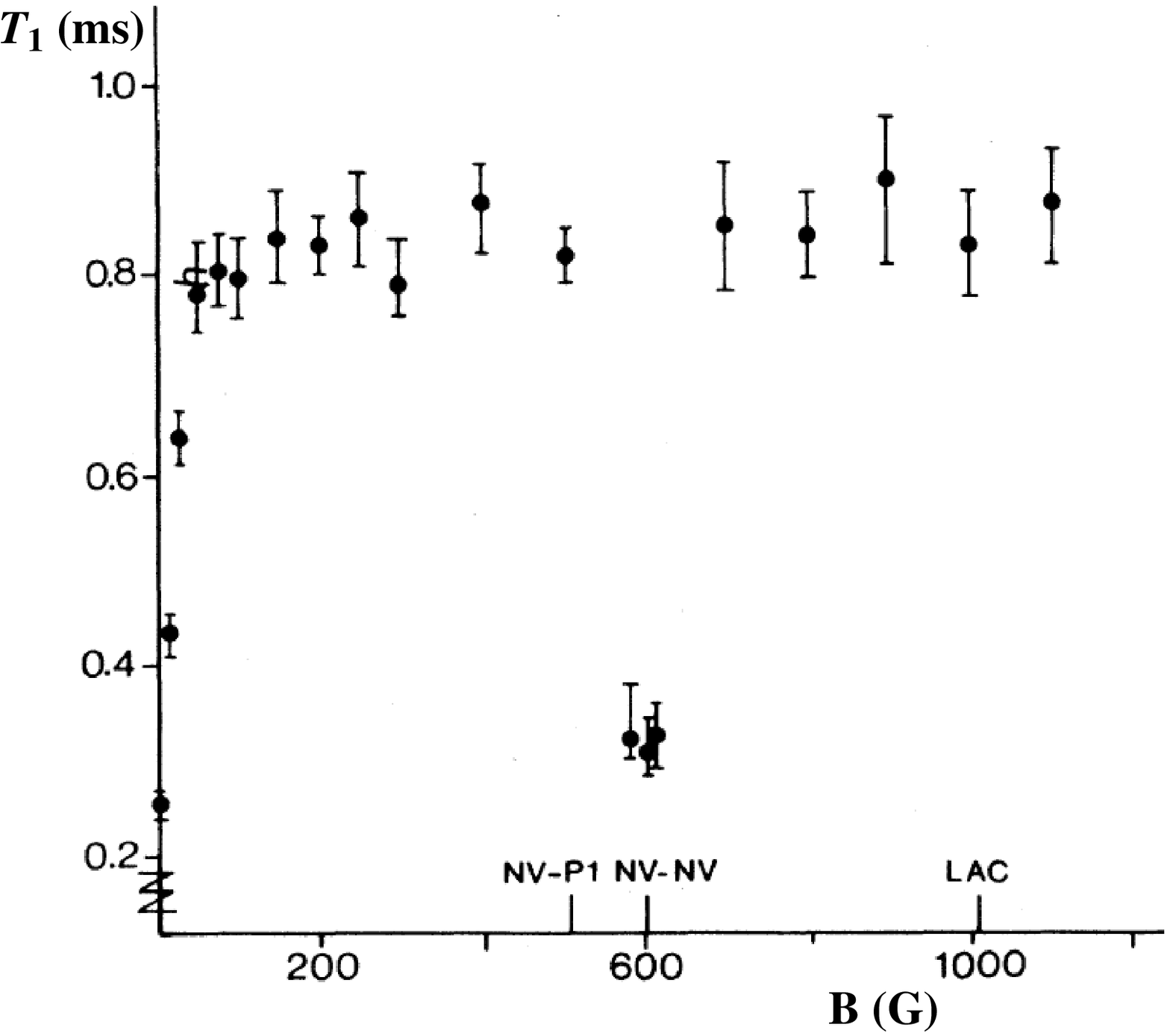}}
}
\mbox{
\subfigure[]{\includegraphics[width=0.45\columnwidth] {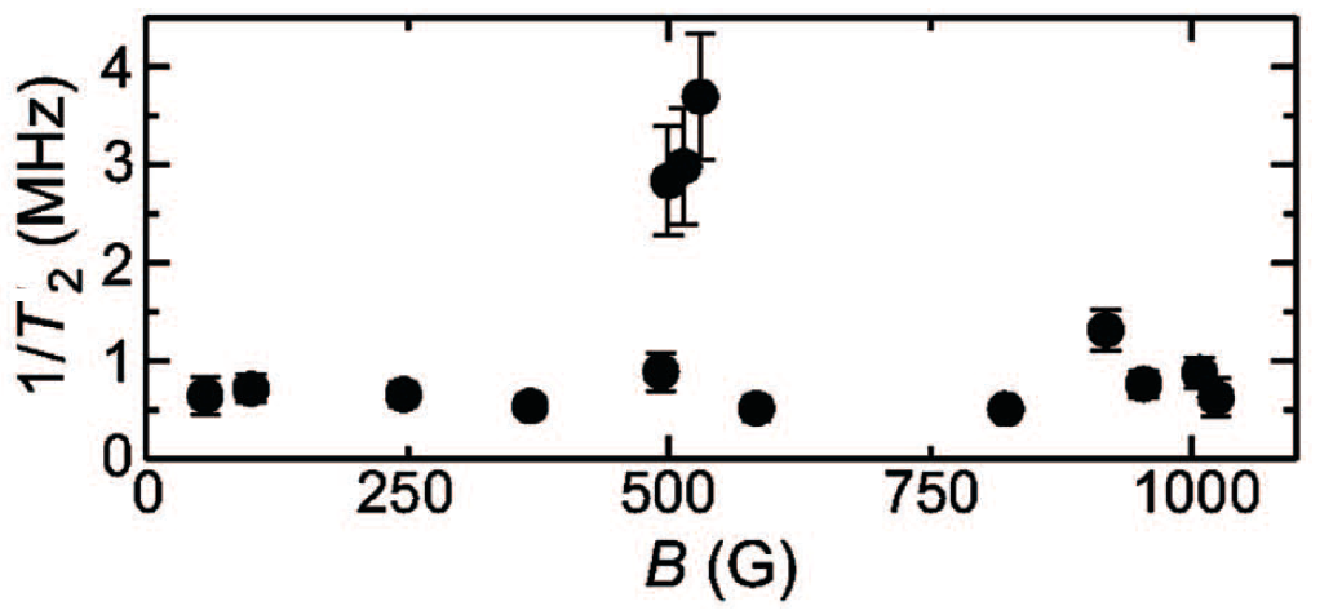}}
}
\mbox{
\subfigure[]{\includegraphics[width=0.5\columnwidth] {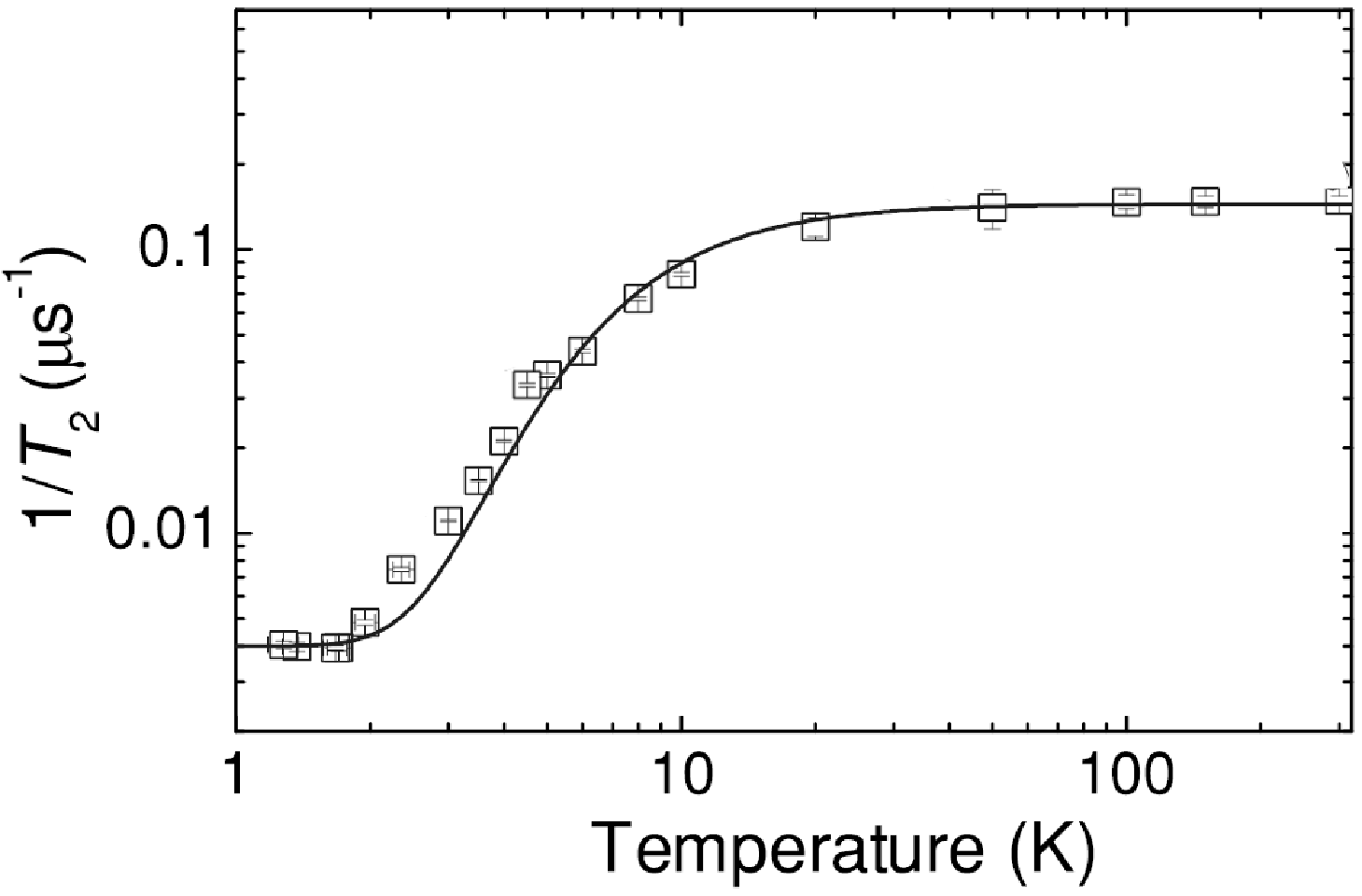}}
}
\caption[The observed magnetic field and temperature dependence of the NV$^-$ ground state spin relaxation and dephasing rates]{The observed axial magnetic field dependence of the NV$^-$ ground state (a) relaxation time $T_1$ \cite{vanOort89} and (b) homogeneous dephasing rate $1/T_2$ \cite{hanson06} at room temperature. (c) The observed temperature dependence of $1/T_2$ in an axial magnetic field of $\sim8$ T \cite{takahashi08}. Measurements were made in each case using NV$^-$ ensembles in type Ib diamond. The axial magnetic fields corresponding to the spin level resonances between the ground state NV$^-$ spin and the N$_s$ defect (NV-P1), two NV$^-$ sub-ensembles (NV-NV) and the ground state LAC are as marked in (a).}
\label{fig:T1magneticfielddependencereview}
\end{center}
\end{figure}

Van Oort and Glasbeek \cite{vanOort89} were the first to measure the dependence of the ground state spin relaxation time $T_1$ on an applied magnetic field (refer to figure \ref{fig:T1magneticfielddependencereview}). Van Oort and Glasbeek conducted their measurements using NV$^-$ ensembles in type Ib diamond and showed that the magnetic field dependence of the ground state $T_1$ was governed by the resonances of the ground state fine structure levels of NV$^-$ sub-ensembles (split by the applied magnetic field) and resonances between one sub-ensemble and the electronic spin of the $N_s$ centre. In the regions of these resonances, the NV$^-$ centre can undergo energy conserving spin flip-flops that redistribute any population difference in the spin sub-levels of one sub-ensemble with the other sub-ensembles or N$_s$ spins, thereby leading to a relaxation of any initial population difference \cite{vanOort91b,vanOort89}. As demonstrated by figure \ref{fig:T1magneticfielddependencereview}, the ground state $T_1$ is approximately magnetic field independent for fields that are away from any resonance. The observations of van Oort and Glasbeek can be extrapolated to single NV$^-$ centres by noting that the effects of any resonances with distant NV$^-$ centres will be negligible, leaving just resonances with N$_s$ centres to influence $T_1$. The N$_s$ centre is not special in this property as any paramagnetic impurity with an electronic g-factor comparable to NV$^-$ will become resonant in some magnetic field. This naturally precludes nuclear spins, such as those of $^{14}$N and $^{13}C$,  from inducing spin relaxation except in extreme field configurations.

Away from resonances with paramagnetic impurities, the ground state $T_1$ is governed by phonon induced transitions between the spin sub-levels. Redman et al \cite{redman91} were the first to measure the temperature dependence (in the range 100-500 K) of the ground state spin relaxation rate $1/T_1$ of an ensemble of NV$^-$ centres. They found that up to room temperature, the relaxation rate was well described by the expression $1/T_1\propto \omega^3/(e^{\hbar\omega/k_BT}-1)$ using $\omega=63$ meV. From the form of this expression and the equivalence of the fitted phonon energy with the dominant phonon energy of the optical band $\sim67$ meV, Redman et al concluded that relaxation occurred via a two-phonon process involving strongly interacting modes of energy $\sim63$ meV. Above room temperature, they found that the rate was well described by $1/T_1\propto T^5$, which as previously discussed in section \ref{section:reviewvibronicstructure}, is indicative of a two-phonon Raman process involving the continuum of low energy phonons. The $T_1$ measurements of Redman et al were later repeated using a different method involving very large axial magnetic fields ($\sim 8$ T) by Takahashi et al \cite{takahashi08}. In partial agreement with Redman et al, Takahashi et al found that $1/T_1\propto AT+BT^5$, where $A$ and $B$ are fitting parameters. The linear relaxation process is indicative of direct one-phonon transitions between the spin sub-levels, which although present in smaller magnetic fields, are typically insignificant \cite{stoneham75}. The direct transitions are likely to have gained significant enhancement due to the larger spin transition energies at $\sim8$ T, such that they became observable in Takahashi et al's measurements. Regardless, it is clear that further measurements of the temperature dependence of the ground state $T_1$ is required to confirm the phonon relaxation processes that are occurring. There is yet to be developed a detailed model of the possible phonon relaxation processes that would aid the analysis of any future measurements.

From their first ODMR experiments, van Oort et al \cite{vanOort88} have attributed the homogeneous dephasing of the NV$^-$ ground state spin to spin-spin interactions with the electronic and nuclear spin bath formed by other NV$^-$ sub-ensembles, the electronic and nuclear spins of the N$_s$ centre and the nuclear spins of the $^{13}$C isotopic impurities. Kennedy et al \cite{kennedy03,kennedy02} has shown that the interactions with the N$_s$ bath is the dominant dephasing mechanism in type Ib diamond and have correlated the ground state $T_2$ with the N concentration present in the diamond. This is supported by the measurements of Takahashi et al, who observed that the temperature dependence of $1/T_2$ was well described by the temperature dependence of the N$_s$ spin flip-flop rate, as well as the measurements of Hanson et al \cite{hanson06}, who observed the magnetic field dependence of $1/T_2$ to be governed by the resonances with N$_s$ in analogy with the magnetic field dependence of $1/T_1$ (see figure \ref{fig:T1magneticfielddependencereview}). Similarly, for single centres in type IIa diamond (undetectable N concentration), interactions with the $^{13}$C spin bath are the dominant dephasing mechanism and $T_2$ has been correlated with the $^{13}$C concentration \cite{balasubramanian09,acosta09,mizouchi09}. Indeed, in isotopically pure diamond, the ground state $T_2$ reaches $\sim$1.8 ms \cite{balasubramanian09} and the dynamics of the NV$^-$ spin become intricate due to coherent interactions with a few proximal $^{13}$C spins, instead of the incoherent interactions with an isotropic bath of $^{13}$C spins present in less pure diamond \cite{maze08b}. Consequently, only in isotopically pure diamond does the spin dephasing rate potentially have detectable contributions from phonon interactions. No contributions have been detected to date. Since the spin dephasing rate is governed in most cases by the spin-spin interactions between NV$^-$ and other electronic and nuclear spins, which could be equally described by the interactions of a canonical triplet spin system with other spins in the lattice, the spin dephasing rate does not provide any information about the intrinsic properties of the centre.

Apart from macroscopic inhomogeneities in the applied electric, magnetic and strain fields and in the preparation of the ensemble of NV$^-$ centres or single NV$^-$ centre, the inhomogeneous spin dephasing time $T_2^\ast$ is governed by the microscopic inhomogeneities of the local fields present at each NV$^-$ centre of the ensemble or the single centre induced by the slow fluctuations of other magnetic and electric defects in the lattice. The slow fluctuations of the defects alter the local fields at each centre of the ensemble for a given measurement or, for the case of repeated measurements of a single centre, alter the local fields for each measurement. The inhomogeneous dephasing due to paramagnetic impurities and its dependence on the applied magnetic field has been well observed and does not in itself yield any information on the intrinsic properties of NV$^-$. However, recently the dependence of the inhomogeneous dephasing on the electric and strain fields has been observed by Dolde et al \cite{dolde11} in their electrometry demonstration. They observed that $T_2^\ast$ increased dramatically when the magnetic and electric fields were such that their effects were of comparable magnitudes. Their observation emphasised the combined effects of the interactions of the NV$^-$ ground state spin with each of the different fields and suggested that the spin's sensitivity to the fluctuations of magnetic and electric defects could be controlled by applied fields. The combined effects of each of the fields on the NV$^-$ ground state spin is yet to be modeled in any detail and the observations of Dolde et al still require a full explanation.

The relaxation and dephasing of the spin associated with the $^3E$ excited state has only been investigated by Fuchs et al \cite{fuchs10} in their single centre excited state ODMR experiments at room temperature. Within a large axial magnetic field ($\sim1276$ G), Fuchs et al used an intense microwave field to manipulate the $^3E$ spin within the lifetime of $^3E$ and implemented Rabi, Ramsey and spin-echo pulse sequences. Fuchs et al measured the homogeneous dephasing time of the spin to be $T_2\sim5.8$ ns and found that the dephasing rate had almost equal contributions from both the decay out of $^3E$ as well as the spin-conserving phonon transitions between the $^3E$ fine structure levels that are responsible for orbital averaging at room temperature. The spin relaxation time of $^3E$ has not yet been measured. Fuchs et al modeled the phonon transitions between the fine structure levels by a fluctuating magnetic field, however a rigorous model of the dephasing and relaxation processes that also accounts for the orbital averaging and dynamic Jahn-Teller effects in $^3E$ is yet to be developed.

\section[Conclusion]{Conclusion}
\label{section:reviewkeycomponentsandoutline}

This review has, for the first time, assembled the body of facts that form our current understanding of the NV centre. Any future study of the centre should seek to resolve the remaining issues concerning the centre, which can be summarised as the following:
\begin{enumerate}
\item The microscopic origins of the equilibrium charge state, the photoconversion process and spectral diffusion

\item The temperature variations of the fine structures of the NV$^-$ ground and optically excited triplet states

\item A detailed analysis of the Jahn-Teller effect and its influence on the NV$^-$ optical and infrared vibronic bands

\item The rigorous characterisation of the NV$^-$ optical spin-polarisation and readout mechanisms
\end{enumerate}
It is evident that the resolution of the above issues will alleviate any barriers in the development of the current applications of NV$^-$ that are related to our understanding of the centre and will also enable the proposal and development of new applications.

The resolution of the first issue will facilitate greater control over the NV charge state during fabrication and implementation. The development of a model of the common microscopic origins of the charge related phenomena requires complementary \textit{ab initio} and experimental studies of the effects of the systematic manipulation of the microscopic distributions of charge donors and traps. Possible first experiments could simply involve the observation of NV ensembles in samples of varying macroscopic concentrations of isolated substitutional nitrogen defects. Alternatively, observations of single ion implanted centres in high purity diamond could be performed as functions of the ion implantation parameters. Given the significant crystal volume required to realistically model typical concentrations of donors and acceptors, the \textit{ab initio} techniques that have been previously applied to model the NV centre are unsuitable. More suitable techniques are likely to be found in charge transfer and reaction modeling techniques of chemistry.

The resolution of the second issue leads to an enhanced understanding of how to minimise the decoherence of the NV$^-$ spin due to temperature fluctuations in its current applications or how to optimise its future implementation as a thermometer. Whilst there exist comprehensive studies of the temperature shift of the ground state fine structure, there exists only one (ensemble) study of the temperature variation of the excited state fine structure from low to room temperatures. Future objectives should thus include the more detailed study of the temperature transformation of the NV$^-$ excited state fine structure as well as the development of theoretical models of the temperature variations based upon the non-radiative dynamics of the centre.

The resolution of the third issue will enable greater control over the optical transition and its coupling to photonic structures, as well as provide insights into how to extend spin-photon entanglement towards room temperature conditions. Experimental studies should seek to probe the polarisation, stress response and temperature dependence of different components of the vibronic bands. These experimental studies should be complemented by the systematic application of Jahn-Teller models of increasing complexity (i.e. linear vs quadratic vs anharmonic coupling and effective mode vs many modes). Other vibronic effects, such the vibronic coupling of energetically close electronic states should also be considered during the interpretation of the experimental observations.

The resolution of the last issue will provide the means to enhance the preparation and readout fidelity of the NV$^-$ ground state spin for its qubit and metrology applications. Given the predominately non-radiative nature of the NV$^-$ optical spin-polarisation mechanism, aspects of the mechanism can only be indirectly probed via the optical and infrared transitions. Consequently, the first steps in characterising the spin-polarisation mechanism is to extend the recent works \cite{robledo11,toyli12} that focused on measuring the temperature dependence of the lifetimes of all electronic levels involved in the optical cycle. The inherent spin dependence of the mechanism necessitates the measurement of the lifetimes of individual fine structure levels. It is therefore clear that the second, third and final unresolved issues are intimately linked. Theoretical work should thus attempt to identify the correlations between the vibronic/ non-radiative aspects of the these unresolved issues in order to ensure a consistent resolution of all three.

As a concluding remark, this review has demonstrated that the NV centre is rich in condensed matter physics, including important, but subtle, relativistic and vibronic effects. Indeed the depth of field of the physics exhibited by the NV centre has resulted in the above issues that have remained unresolved for almost 50 years. Yet, having stated this, much of the physics of the NV centre can be understood from the simple and intuitive molecular model proposed by Loubser and van Wyk \cite{loubser77, loubser78}. Consequently, the NV centre promises to be an excellent pedagogical tool for condensed matter physics and model system to aid in the understanding and identification of other defects in semiconductors with comparable or better properties for quantum technology applications. There is thus great motivation to pursue a more complete understanding of the NV centre.

\section*{Acknowledgements}
This work was supported by the Australian Research Council under the
Discovery Project scheme (DP0986635 and DP120102232), the EU commission (ERC grant SQUTEC), Specific Targeted Research Project  DIAMANT and the integrated project SOLID. M.W.D. wishes to acknowledge the David Hay Memorial Fund.

\section*{Appendix Properties of the $C_{3v}$ point group}

In this appendix, the key properties of the point symmetry group of the NV centre are detailed in order to provide a quick reference for those studying the centre. The group's symmetry elements, character table, direct-product table and Clebsch-Gordan coefficients are provided. The group projection operators, irreducible tensor operators and the Wigner-Eckart matrix element theorem are defined along with selection rules for the tensor operators.

\begin{figure}[hbtp]
\begin{center}
\mbox{
\subfigure[$e$]{\includegraphics[width=0.3\columnwidth] {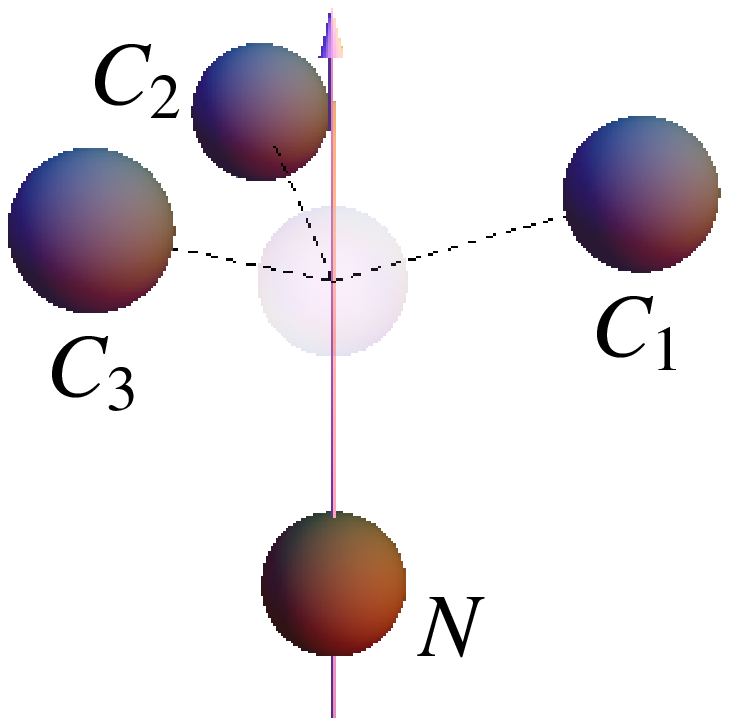}}
\subfigure[$C_3^+$]{\includegraphics[width=0.3\columnwidth] {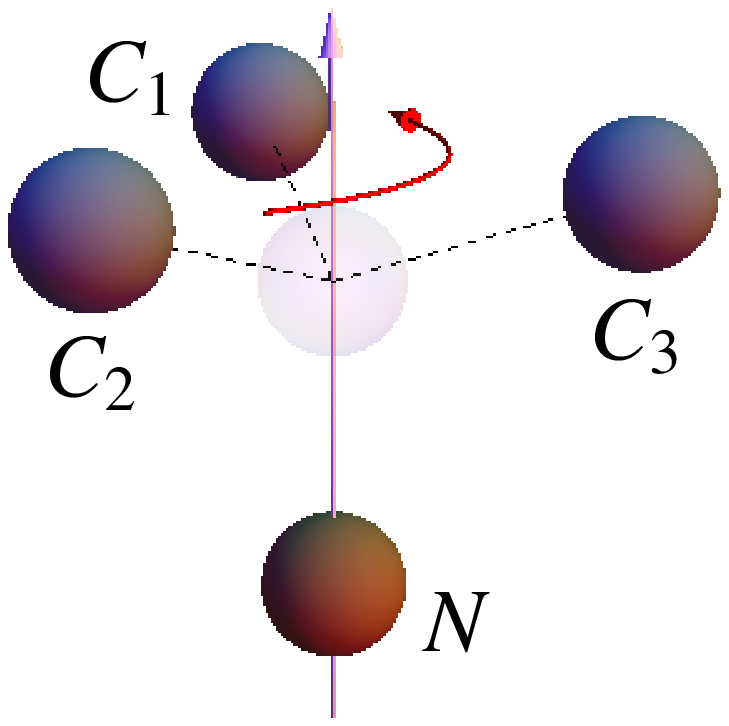}}
\subfigure[$C_3^-$]{\includegraphics[width=0.3\columnwidth] {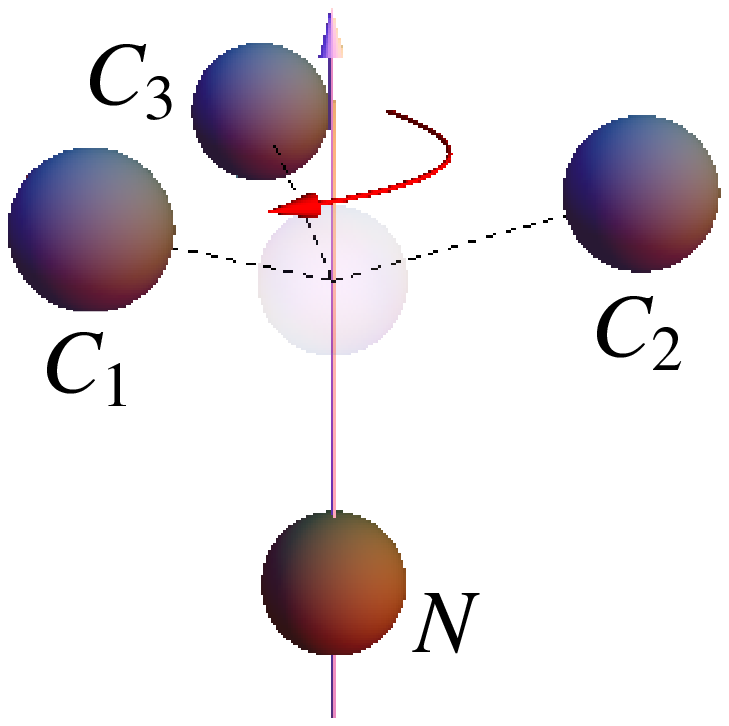}}}
\mbox{
\subfigure[$\sigma_1$]{\includegraphics[width=0.33\columnwidth] {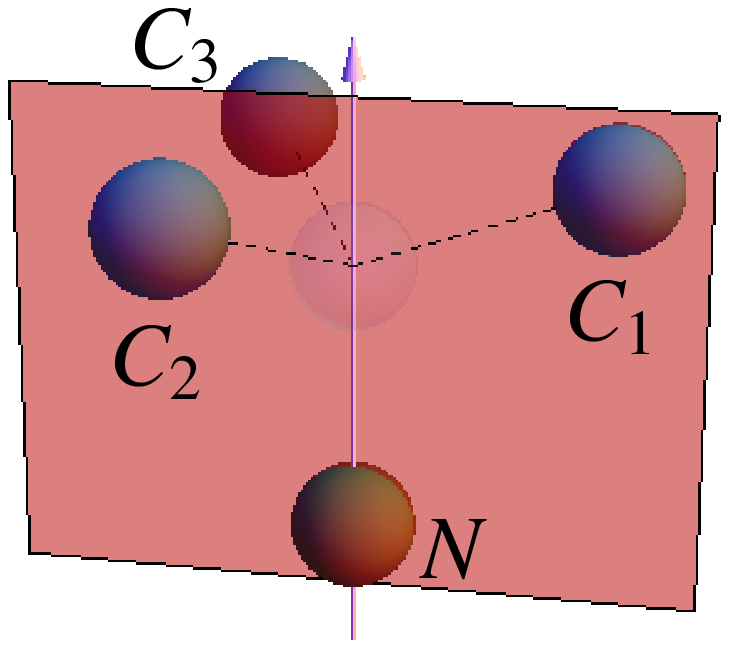}}
\subfigure[$\sigma_2$]{\includegraphics[width=0.3\columnwidth] {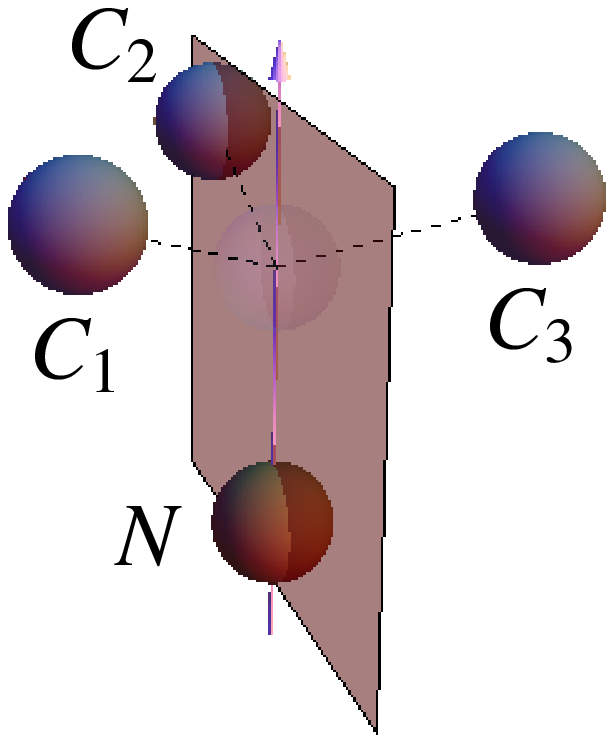}}
\subfigure[$\sigma_3$]{\includegraphics[width=0.33\columnwidth] {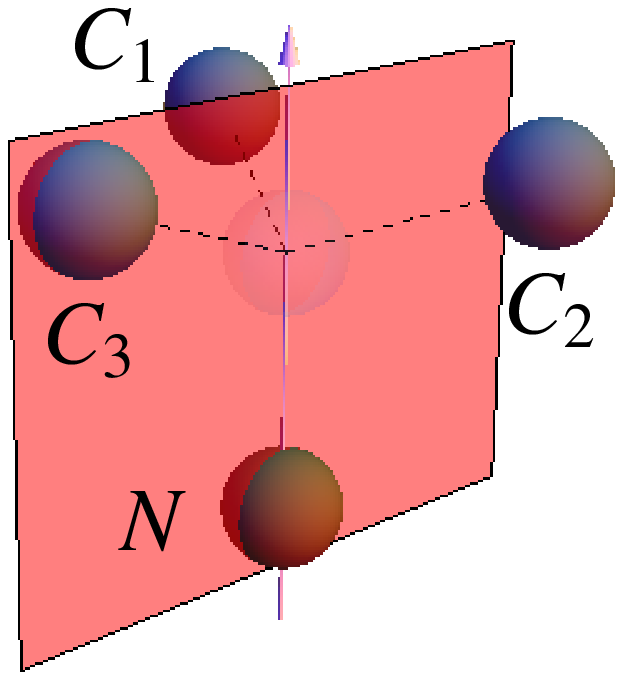}}
}
\caption[The operations of the $C_{3v}$ group elements on the structure of the NV centre]{The operations of the $C_{3v}$ group elements on the structure of the NV centre. The nearest neighbour nitrogen (brown) and carbon (grey) atoms to the vacancy (transparent) are as denoted. The operations of the group elements result in permutations of the carbon atom labels. The directions of rotation (red arrows) about the trigonal symmetry axis (white arrow) of the $C_3^+$ and $C_3^-$ group elements are depicted in (b) and (c). The planes of reflection (red squares) of the $\sigma_1$, $\sigma_2$ and $\sigma_3$ group elements are depicted in (d), (e) and (f).}
\label{fig:appendixBgroupoperations}
\end{center}
\end{figure}

The $C_{3v}$ group is a trigonal point symmetry group of order six. The elements of the group are defined with respect to the trigonal symmetry axis and form three classes, being the identity $e$, the $C_3$ rotations about the symmetry axis and the vertical reflections $\sigma_v$ through planes containing the symmetry axis. The operations of the six $C_{3v}$ group elements on the structure of the NV centre are depicted in figure \ref{fig:appendixBgroupoperations}. Given three classes, the $C_{3v}$ group possesses three irreducible representations $A_1$, $A_2$ and $E$, being of dimensions one, one and two, respectively.

The character and direct product multiplication tables of the $C_{3v}$ group are contained in tables \ref{tab:appendixBC3vcharactertable} and \ref{tab:appendixBC3vdirectproducttable}, respectively. Basis elements that transform as the $A_1$ irreducible representation are invariant under all of the group element operations. Basis elements that transform as the $A_2$ irreducible representation are invariant under the operations of the group elements of the $e$ and $C_3$ classes, but are antisymmetric under the operations of the group elements of the $\sigma_v$ class. Basis elements that transform as $E$ are non-symmetric. Defining the coordinate system such that the z axis is collinear with the trigonal symmetry axis, the irreducible representations that the coordinate components and coordinate rotations transform as are as identified in table \ref{tab:appendixBC3vcharactertable}.

Whilst the matrix representations of the one dimensional irreducible representations are as specified by the group character table, there is a degree of flexibility in the definition of the matrix representations of the $E$ irreducible representation. The $E$ irreducible representation is most simply defined via the operations of the group elements on the cartesian components $x,y$. However, the matrix representations of the $E$ irreducible representation may also be defined via the operations of the group elements on the spherical components $\xi_\pm=\mp\frac{1}{\sqrt{2}}(x\pm iy)$. The matrix representations (including both cartesian and spherical representations of $E$) of the $C_{3v}$ irreducible representations are contained in table \ref{tab:appendixBmatrixrepresentations}.

The $C_{3v}$ projection operator corresponding to the $\lambda$ row of the $j^{th}$ irreducible representation is defined as \cite{tinkham}
\begin{equation}
\hat{{\cal P}}_{\lambda\lambda}^{(j)}=\frac{l_j}{6}\sum_i\Gamma^{(j)}(g_i)_{\lambda\lambda}^\ast\hat{P}_{g_i}
\end{equation}
where $l_j$ is the dimension of the $j^{th}$ irreducible representation, $g_i$ is the $i^{th}$ group element, $\Gamma^{(j)}(g_i)_{\lambda\lambda}$ is the $\lambda^{th}$ diagonal matrix element of the matrix representation of the $g_i$ group element in the $j^{th}$ irreducible representation, and $\hat{P}_{g_i}$ is the unitary transformation operator corresponding to the group element $g_i$. Let $\psi_\lambda^j$ be a state that transforms as the $\lambda$ row of the $j^{th}$ irreducible representation, then $\psi_\lambda^j$ satisfies
\begin{equation}
\hat{{\cal P}}_{\kappa\kappa}^{(k)}\psi_\lambda^j=\delta_{\kappa,\lambda}\delta_{k,j}\psi_\lambda^j
\end{equation}
Similarly, an irreducible tensor operator $\hat{Q}_\lambda^j$ that transforms as the $\lambda$ row of the $j^{th}$ irreducible representation, is an operator that satisfies
\begin{equation}
\hat{P}_{g_i}\hat{Q}_\lambda^j\hat{P}_{g_i}^{-1}=\sum_\kappa^{l_j}\Gamma^{(j)}(g_i)_{\kappa\lambda}\hat{Q}_\kappa^j
\end{equation}
for every group element $g_i$ and every row $\lambda$. A matrix element $\bra{\psi_\gamma^r}\hat{Q}_\kappa^q\ket{\phi_\lambda^p}$ of an irreducible tensor operator can be expressed in terms of a set of reduced matrix elements $\bra{\psi^r}| \hat{Q}^q |\ket{\phi^p}$ and the Clebsch-Gordan coefficients $\left(\ldots \ \ldots|\ldots\right)$ of the $C_{3v}$ group via the Wigner-Eckart theorem
\begin{equation}
\bra{\psi_\gamma^r}\hat{Q}_\kappa^q\ket{\phi_\lambda^p} = \left(\begin{array}{cc|c} p & q & r \\ \lambda & \kappa & \gamma  \end{array}\right)^\ast\bra{\psi^r}| \hat{Q}^q |\ket{\phi^p}
\label{eq:appendixAwignereckart}
\end{equation}
Importantly, the reduced matrix elements are only dependent on the irreducible representations of the tensor operator and the basis states and not the row of the irreducible representations that they belong to. The set of reduced matrix elements of a tensor operator thus represents the minimum set of independent parameters (as determined by symmetry) that completely describe the matrix representation of a tensor operator. The Clebsch-Gordan coefficients corresponding to the cartesian and spherical representations of the $C_{3v}$ group are contained in tables \ref{tab:appendixBclebschgordancoefficients} and \ref{tab:appendixBsphericalclebschgordancoefficients}, respectively. These Clebsch-Gordan coefficients define the cartesian and spherical tensor operator selection rules contained in tables \ref{tab:cartesianselectionrules} and \ref{tab:sphericalselectionrules}, respectively.

\begin{table}[hbtp]
\begin{center}
\caption[Character table of the $C_{3v}$ point group]{Character table of the $C_{3v}$ point group \cite{tinkham}. The irreducible representations that the coordinate components ($x$, $y$, $z$),  quadratic coordinate components ($x^2$, $xy$, $xz$, $y^2$, $yz$, $z^2$) and coordinate rotations ($R_x$, $R_y$, $R_z$) transform as are as identified.}
\label{tab:appendixBC3vcharactertable}
\begin{tabular}{c|c|c|ccc}
\noalign{\smallskip}
\toprule
\multicolumn{3}{c|}{$C_{3v}$} & $e$ & $2C_3$ & $3\sigma_v$ \\
\midrule
$x^2+y^2$, $z^2$& $z$& $A_1$ & 1 & 1 & 1 \\
& $R_z$ & $A_2$ & 1 & 1 & -1 \\
$\left.\begin{array}{c}(x^2-y^2,xy) \\ (xz,yz) \\ \end{array}\right\}$ & $\left.\begin{array}{c} (x,y) \\ (R_x,R_y) \\ \end{array}\right\}$ & $E$ & 2 & -1 & 0 \\
\bottomrule
\end{tabular}
\end{center}
\end{table}
\begin{table}[hbtp]
\begin{center}
\caption[The $C_{3v}$ point group direct-product multiplication table]{The $C_{3v}$ point group direct product multiplication table \cite{mqm}.}
\label{tab:appendixBC3vdirectproducttable}
\begin{tabular}{c|ccc}
\noalign{\smallskip}
\toprule
$C_{3v}\otimes C_{3v}$ & $A_1$ & $A_2$ & $E$ \\
\midrule
$A_1$ & $A_1$ & $A_2$ & $E$ \\
$A_2$ & $A_2$ & $A_1$ & $E$ \\
$E$ & $E$ & $E$ & $A_1\oplus A_2\oplus E$\\
\bottomrule
\end{tabular}
\end{center}
\end{table}
\begin{sidewaystable}[hbtp]
\begin{center}
\caption[The matrix representations of the $C_{3v}$ group irreducible representations]{The matrix representations of the $C_{3v}$ group irreducible representations. Both the cartesian ($x$, $y$) \cite{lenef96} and spherical ($\xi_+$, $\xi_-$) \cite{altmann} representations of $E$ are detailed. $\epsilon=e^{2\pi i/3}$.}
\label{tab:appendixBmatrixrepresentations}
\begin{tabular}{c|cccccc|c}
\noalign{\smallskip}
\toprule
$C_{3v}$ & $e$ & $C_3^+$ & $C_3^-$ & $\sigma_1$ & $\sigma_2$ & $\sigma_3$ & \\
\midrule
$A_1$ & 1 & 1 & 1 & 1 & 1 & 1 & \\
$A_2$ & 1 & 1 & 1 & -1 & -1 & -1 & \\
$E$ & $\left(\begin{array}{cc} 1 & 0 \\ 0 & 1 \\ \end{array}\right)$ &
$\left(\begin{array}{cc} -\frac{1}{2} & \frac{\sqrt{3}}{2} \\ -\frac{\sqrt{3}}{2} & -\frac{1}{2} \\ \end{array}\right)$ &
$\left(\begin{array}{cc} -\frac{1}{2} & -\frac{\sqrt{3}}{2} \\ \frac{\sqrt{3}}{2} & -\frac{1}{2} \\ \end{array}\right)$ &
$\left(\begin{array}{cc} 1 & 0 \\ 0 & -1 \\ \end{array}\right)$ &
$\left(\begin{array}{cc} -\frac{1}{2} & -\frac{\sqrt{3}}{2} \\ -\frac{\sqrt{3}}{2} & \frac{1}{2} \\ \end{array}\right)$ &
$\left(\begin{array}{cc} -\frac{1}{2} & \frac{\sqrt{3}}{2} \\ \frac{\sqrt{3}}{2} & \frac{1}{2} \\ \end{array}\right)$ & $(x,y)$ \\
& $\left(\begin{array}{cc} 1 & 0 \\ 0 & 1 \\ \end{array}\right)$ &
$\left(\begin{array}{cc} \epsilon & 0 \\ 0 & \epsilon^\ast \\ \end{array}\right)$ &
$\left(\begin{array}{cc} \epsilon^\ast & 0 \\ 0 & \epsilon \\ \end{array}\right)$ &
$\left(\begin{array}{cc} 0 & -1 \\ -1 & 0 \\ \end{array}\right)$ &
$\left(\begin{array}{cc} 0 & -\epsilon  \\  -\epsilon^\ast & 0 \\ \end{array}\right)$ &
$\left(\begin{array}{cc} 0 & -\epsilon^\ast  \\  -\epsilon & 0 \\ \end{array}\right)$ &
$(\xi_+,\xi_-)$ \\
\bottomrule
\end{tabular}
\end{center}
\end{sidewaystable}
\begin{table}[hbtp]
\begin{center}
\caption[The Clebsch-Gordan coefficients corresponding to the cartesian matrix representations of the $C_{3v}$ irreducible representations]{The non-zero Clebsch-Gordan coefficients corresponding to the cartesian matrix representations of the $C_{3v}$ irreducible representations \cite{lenef96}. The indices $j$, $k$ denote the element of the matrix defining the Clebsch-Gordan coefficient in the basis $\{x$, $y\}$.}
\label{tab:appendixBclebschgordancoefficients}
\begin{tabular}{lcllcl}
\noalign{\smallskip}
\toprule
$\left(\begin{array}{cc|c} A_1 & A_1 & A_1 \\ 1 & 1 & 1  \end{array}\right)$ & = & 1 &
$\left(\begin{array}{cc|c} E & E & A_1 \\ j & k & 1  \end{array}\right)$ & = & $\frac{1}{\sqrt{2}}\left(\begin{array}{cc} 1 & 0 \\ 0 & 1 \end{array}\right)$ \\
$\left(\begin{array}{cc|c} A_1 & A_2 & A_2 \\ 1 & 1 & 1  \end{array}\right)$ & = & 1 & $\left(\begin{array}{cc|c} E & E & A_2 \\ j & k & 1  \end{array}\right)$ & = & $\frac{1}{\sqrt{2}}\left(\begin{array}{cc} 0 & 1 \\ -1 & 0\end{array}\right)$  \\
$\left(\begin{array}{cc|c} A_2 & A_2 & A_1 \\ 1 & 1 & 1  \end{array}\right)$ & = & 1 & $\left(\begin{array}{cc|c} E & E & E \\ j & k & x  \end{array}\right)$ & = & $\frac{1}{\sqrt{2}}\left(\begin{array}{cc} 1 & 0 \\ 0 & -1 \end{array}\right)$  \\
$\left(\begin{array}{cc|c} A_1 & E & E \\ 1 & j & k  \end{array}\right)$ & = & $\left(\begin{array}{cc} 1 & 0 \\ 0 & 1 \end{array}\right)$
& $\left(\begin{array}{cc|c} E & E & E \\ j & k & y  \end{array}\right)$ & = & $\frac{1}{\sqrt{2}}\left(\begin{array}{cc} 0 & -1 \\ -1 & 0 \end{array}\right)$ \\
$\left(\begin{array}{cc|c} A_2 & E & E \\ 1 & j & k  \end{array}\right)$ & = & $\left(\begin{array}{cc} 0 & 1\\ -1 & 0 \end{array}\right)$ & & & \\
\bottomrule
\end{tabular}
\end{center}
\end{table}
\begin{table}[hbtp]
\begin{center}
\caption[The Clebsch-Gordan coefficients corresponding to the spherical matrix representations of the $C_{3v}$ irreducible representations]{The non-zero Clebsch-Gordan coefficients corresponding to the spherical matrix representations of the $C_{3v}$ irreducible representations. The indices $j$, $k$ denote the element of the matrix defining the Clebsch-Gordan coefficient in the basis $\{\xi_+$, $\xi_-\}$.}
\label{tab:appendixBsphericalclebschgordancoefficients}
\begin{tabular}{lcllcl}
\noalign{\smallskip}
\toprule
$\left(\begin{array}{cc|c} A_1 & A_1 & A_1 \\ 1 & 1 & 1  \end{array}\right)$ & = & 1 &
$\left(\begin{array}{cc|c} E & E & A_1 \\ j & k & 1  \end{array}\right)$ & = & $\frac{1}{\sqrt{2}}\left(\begin{array}{cc} 0 & 1 \\ 1 & 0 \end{array}\right)$ \\
$\left(\begin{array}{cc|c} A_1 & A_2 & A_2 \\ 1 & 1 & 1  \end{array}\right)$ & = & 1 & $\left(\begin{array}{cc|c} E & E & A_2 \\ j & k & 1  \end{array}\right)$ & = & $\frac{1}{\sqrt{2}}\left(\begin{array}{cc} 0 & 1 \\ -1 & 0\end{array}\right)$  \\
$\left(\begin{array}{cc|c} A_2 & A_2 & A_1 \\ 1 & 1 & 1  \end{array}\right)$ & = & 1 & $\left(\begin{array}{cc|c} E & E & E \\ \xi_- & \xi_- & \xi_+  \end{array}\right)$ & = & 1  \\
$\left(\begin{array}{cc|c} A_1 & E & E \\ 1 & j & k  \end{array}\right)$ & = & $\left(\begin{array}{cc} 1 & 0 \\ 0 & 1 \end{array}\right)$
& $\left(\begin{array}{cc|c} E & E & E \\ \xi_+ & \xi_+ & \xi_-  \end{array}\right)$ & = & 1 \\
$\left(\begin{array}{cc|c} A_2 & E & E \\ 1 & j & k  \end{array}\right)$ & = & $\left(\begin{array}{cc} 1 & 0 \\ 0 & -1 \end{array}\right)$ & & & \\
\bottomrule
\end{tabular}
\end{center}
\end{table}
\begin{table}[hbtp]
\begin{center}
\caption{The selection rules for cartesian tensor operators as derived using the Clebsch-Gordan coefficients of table \ref{tab:appendixBclebschgordancoefficients}. The elements of the four tables below correspond to the Clebsch-Gordan coefficients of the matrix elements $\me{\psi_\gamma^r}{\hat{O}_\kappa^q}{\phi_\lambda^p}$ as defined by the Wigner-Eckart theorem.}
\label{tab:cartesianselectionrules}
\begin{tabular}{c|cccc|c|cccc}
\noalign{\smallskip}
\toprule
$\hat{O}^{A_1}$ & $\phi^{A_1}$ & $\phi^{A_2}$ & $\phi_x^E$ & $\phi_y^E$ & $\hat{O}^{A_2}$ & $\phi^{A_1}$ & $\phi^{A_2}$ & $\phi_x^E$ & $\phi_y^E$ \\
\midrule
$\psi^{A_1}$ & 1 & 0 & 0 & 0 & $\psi^{A_1}$ & 0 & 1 & 0 & 0 \\
$\psi^{A_2}$ & 0 & 1 & 0 & 0 & $\psi^{A_2}$ & 1 & 0 & 0 & 0 \\
$\psi_x^E$ & 0 & 0 & 1 & 0 & $\psi_x^E$ & 0 & 0 & 0 & -1 \\
$\psi_y^E$ & 0 & 0 & 0 & 1 & $\psi_y^E$ & 0 & 0 & 1 & 0 \\
\midrule
$\hat{O}_x^{E}$ & $\phi^{A_1}$ & $\phi^{A_2}$ & $\phi_x^E$ & $\phi_y^E$ & $\hat{O}_y^{E}$ & $\phi^{A_1}$ & $\phi^{A_2}$ & $\phi_x^E$ & $\phi_y^E$ \\
\midrule
$\psi^{A_1}$ & 0 & 0 & $\frac{1}{\sqrt{2}}$ & 0 & $\psi^{A_1}$ & 0 & 0 & 0 & $\frac{1}{\sqrt{2}}$ \\
$\psi^{A_2}$ & 0 & 0 & 0 & $\frac{-1}{\sqrt{2}}$ & $\psi^{A_2}$ & 0 & 0 & $\frac{1}{\sqrt{2}}$ & 0 \\
$\psi_x^E$ & $\frac{1}{\sqrt{2}}$ & 0 & $\frac{1}{\sqrt{2}}$ & 0 & $\psi_x^E$ & 0 & $\frac{1}{\sqrt{2}}$ & 0 & $\frac{-1}{\sqrt{2}}$ \\
$\psi_y^E$ & 0 & $\frac{-1}{\sqrt{2}}$ & 0 & $\frac{-1}{\sqrt{2}}$ & $\psi_y^E$ & $\frac{1}{\sqrt{2}}$ & 0 & $\frac{-1}{\sqrt{2}}$ & 0 \\
\bottomrule
\end{tabular}
\end{center}
\end{table}
\begin{table}[hbtp]
\begin{center}
\caption{The selection rules for spherical tensor operators as derived using the Clebsch-Gordan coefficients of table \ref{tab:appendixBsphericalclebschgordancoefficients}. The elements of the four tables below correspond to the Clebsch-Gordan coefficients of the matrix elements $\me{\psi_\gamma^r}{\hat{O}_\kappa^q}{\phi_\lambda^p}$ as defined by the Wigner-Eckart theorem.}
\label{tab:sphericalselectionrules}
\begin{tabular}{c|cccc|c|cccc}
\noalign{\smallskip}
\toprule
$\hat{O}^{A_1}$ & $\phi^{A_1}$ & $\phi^{A_2}$ & $\phi_{\xi_+}^E$ & $\phi_{\xi_-}^E$ & $\hat{O}^{A_2}$ & $\phi^{A_1}$ & $\phi^{A_2}$ & $\phi_{\xi_+}^E$ & $\phi_{\xi_-}^E$ \\
\midrule
$\psi^{A_1}$ & 1 & 0 & 0 & 0 & $\psi^{A_1}$ & 0 & 1 & 0 & 0 \\
$\psi^{A_2}$ & 0 & 1 & 0 & 0 & $\psi^{A_2}$ & 1 & 0 & 0 & 0 \\
$\psi_{\xi_+}^E$ & 0 & 0 & 1 & 0 & $\psi_{\xi_+}^E$ & 0 & 0 & 1 & 0 \\
$\psi_{\xi_-}^E$ & 0 & 0 & 0 & 1 & $\psi_{\xi_-}^E$ & 0 & 0 & 0 & -1 \\
\midrule
$\hat{O}_{\xi_+}^{E}$ & $\phi^{A_1}$ & $\phi^{A_2}$ & $\phi_{\xi_+}^E$ & $\phi_{\xi_-}^E$ & $\hat{O}_{\xi_-}^{E}$ & $\phi^{A_1}$ & $\phi^{A_2}$ & $\phi_{\xi_+}^E$ & $\phi_{\xi_-}^E$ \\
\midrule
$\psi^{A_1}$ & 0 & 0 & 0 & $\frac{1}{\sqrt{2}}$ & $\psi^{A_1}$ & 0 & 0 &  $\frac{1}{\sqrt{2}}$ & 0 \\
$\psi^{A_2}$ & 0 & 0 & 0 & $\frac{-1}{\sqrt{2}}$ & $\psi^{A_2}$ & 0 & 0 & $\frac{1}{\sqrt{2}}$ & 0  \\
$\psi_{\xi_+}^E$ & 0 & 0 & 0 & 0 & $\psi_{\xi_+}^E$ & $\frac{1}{\sqrt{2}}$ & $\frac{1}{\sqrt{2}}$ & 0 & 1 \\
$\psi_{\xi_-}^E$ & $\frac{1}{\sqrt{2}}$ & $\frac{-1}{\sqrt{2}}$ & 1 & 0 & $\psi_{\xi_-}^E$ & 0 & 0 & 0 & 0 \\
\bottomrule
\end{tabular}
\end{center}
\end{table}

\end{document}